\documentclass[aps,prx,twocolumn,groupedaddress,showpacs,amsmath,amssymb,longbibliography, amsfonts,10pt]{revtex4-2}

\usepackage{amsmath}
\usepackage{amsfonts}
\usepackage{amssymb}
\usepackage{graphicx}
\usepackage{xcolor}
\usepackage{bbold}
\usepackage[citecolor=blue,urlcolor=blue]{hyperref}
\hypersetup{colorlinks}
\usepackage{mathtools}
\usepackage{standalone}
\usepackage{tikz}
\usepackage[compat=1.1.0]{tikz-feynman}
\usetikzlibrary{arrows.meta}
\usepackage[nolist,nohyperlinks]{acronym}
\newacro{RG}[RG]{renomalization group}
\newacro{fdr}[FDR]{fluctuation-dissipation relation}
\newacro{EP}[EP]{exceptional point}
\usepackage[normalem]{ulem}
\usepackage{physics}
\usepackage[makeroom]{cancel}
\usepackage{slashed}
\usepackage[shortlabels]{enumitem}

\newacro{CEP}[CEP]{critical exceptional point}
\newacro{frg}[FRG]{functional RG}
\newacro{DSE}[DSE]{Dyson-Schwinger equation}
\newacro{MSRJD}[MSRJD]{Martin-Siggia-Rose-Janssen-DeDominicis}
\newacro{BKT}[BKT]{Berezinskii–Kosterlitz–Thouless}
\newacro{KPZ}[KPZ]{Kardar–Parisi–Zhang}
\newacro{EFT}[EFT]{effective field theory}
\newacro{Nlsm}[NL$\sigma$M]{non-linear $\sigma$ model}
\usepackage{lipsum}  
\newcommand{\xx}{\mathbf{x}}
\newcommand{\RR}{\mathbf{\tiny{R}}}
\newcommand{\yy}{\mathbf{y}}
\newcommand{\rr}{\mathbf{r}}
\newcommand{\res}{\widetilde{\boldsymbol{\phi}}}

\newcommand{\ee}{\mathrm{e}}
\newcommand{\kk}{\mathbf{k}}
\newcommand{\BB}{\mathbf{B}}
\newcommand{\Svec}{\mathbf{S}}

\newcommand{\pp}{\mathbf{p}}
\newcommand{\Avec}{\mathbf{A}}
\newcommand{\vecpi}{\boldsymbol\Pi}
\newcommand{\vecxi}{\boldsymbol\xi}
\newcommand{\vecphi}{\boldsymbol{\phi}}
\newcommand{\vecvarphi}{\boldsymbol\varphi}
\newcommand{\sgn}{{\rm{sgn}}}
\newcommand{\SP}{\mathbf{S}}
\newcommand{\B}{\mathcal{B}}
\newcommand{\A}{\mathcal{A}}
\newcommand{\id}{\mathbb{1}}
\newcommand{\vp}{\vec{p}}
\newcommand{\vq}{\vec{q}}
\newcommand{\vk}{\vec{k}}
\newcommand{\vx}{\vec{x}}

\newcommand{\fc}[1]{{\color{red} [\textsf{#1}]}}
\newcommand{\ale}[1]{{\color{blue} [\textsf{#1}]}}
\newcommand{\vecx}{\mathbf{x}}
\newcommand{\vecy}{\mathbf{y}}
\newcommand{\vecq}{\mathbf{q}}
\newcommand{\vecp}{\mathbf{p}}
\newcommand{\veck}{\mathbf{k}}
\newcommand{\vecs}{\mathbf{s}}
\newcommand{\I}{\mathcal{I}}
\newcommand{\Ev}[1]{\langle #1 \rangle}
\newcommand{\Str}{\text{Str}}
\newcommand{\STr}{\text{STr}}
\newcommand{\mi}{\mathrm{i}}
\newcommand{\me}{\mathrm{e}}

\newcommand{\jlc}[1]{\textcolor{magenta}{\textit{[JL: #1]}}}
\newcommand{\jlm}[1]{\textcolor{magenta}{#1}}
\newcommand{\jls}[1]{\textcolor{magenta}{\ifmmode\text{\cancel{\ensuremath{#1}}}\else\sout{#1}\fi}}
\newcommand{\SD}[1]{\textcolor{blue}{\textit{[SD: #1]}}}

\newcommand{\RM}[1]{\textcolor{orange}{\textit{[RM: #1]}}}

\newcommand{\RMs}[1]{\textcolor{orange}{\ifmmode\text{\cancel{\ensuremath{#1}}}\else\sout{#1}\fi}}
\newcommand{\tz}[1]{\textcolor{olive}{\textit{[TZ: #1]}}}
\newcommand{\tzs}[1]{\textcolor{olive}{\ifmmode\text{\cancel{\ensuremath{#1}}}\else\sout{#1}\fi}}
\newcommand{\tzm}[1]{\textcolor{olive}{#1}}
\newcommand\numberthis{\addtocounter{align}{1}\tag{\theequation}}

\definecolor{alabaster}{RGB}{11,96,172}
\definecolor{acidgreen}{rgb}{0.5,0.8,0.2}
\definecolor{burgundy}{RGB}{231, 0, 14}

\tikzset
{
   ->-/.style={decoration={markings,mark=at position 0.5 with {\arrow{Stealth}}},
               postaction={decorate}}
}

\begin{document}
\title{Fermion quantum criticality far from equilibrium}

\author{Rohan Mittal}
\author{Tom Zander} 
\author{Johannes Lang} 
\author{Sebastian Diehl}

\affiliation{Institut f\"ur Theoretische Physik, Universit\"at zu K\"oln, 50937 Cologne, Germany}

\begin{abstract}

Driving a quantum system out of equilibrium while preserving its subtle quantum mechanical correlations on large scales presents a major challenge, both fundamentally and for technological applications. At its core, this challenge is pinpointed by the question of how quantum effects can persist at asymptotic scales, analogous to quantum critical points in equilibrium. In this work, we construct such a scenario using fermions as building blocks. These fermions undergo an absorbing-to-absorbing state transition between two topologically distinct and quantum-correlated dark states. Starting from a microscopic, interacting Lindbladian, we derive an effective Lindblad-Keldysh field theory in which critical fermions couple to a bosonic bath with hydrodynamic fluctuations associated with particle number conservation. A key feature of this field theory is an emergent symmetry that protects the purity of the fermions' state even in the presence of the thermal bath. We quantitatively characterize the critical point using a leading-order expansion around the upper critical dimension, thereby establishing the first non-equilibrium universality class of fermions. The symmetry protection mechanism, which exhibits parallels to the problem of directed percolation, suggests a pathway toward a broader class of robust, universal quantum phenomena in fermionic systems.

\end{abstract}

\date{\today}

\maketitle
\section{Introduction}

The critical behavior near a second-order phase transition is one of the strongest manifestations of universality, where macroscopic observables lose memory of microscopic details. Universality classes are typically distinguished by dimensionality and internal symmetries, but additional categories exist.

One key differentiation is between classical and quantum critical behavior, distinguishing between transitions that occur in mixed and pure states.
Quantum fluctuations are more subtle than classical statistical ones and are asymptotically overwritten at any finite temperature in equilibrium systems~\cite{Sachdev2023,Loehneysen2007,Sachdev2008}. Another distinction arises from particle statistics, {\it i.e.\/}, whether bosons or fermions carry the long-distance fluctuations, rooted in their drastically different spectral and occupation properties~\cite{Boyack_2021,Gegenwart2008}. Importantly, fermions contribute to the asymptotic long-wavelength behavior only in pure states. At finite temperature, they freeze out, meaning they lack a nontrivial classical limit. Finally, an essential aspect is whether the transition occurs in or out of equilibrium. Often, a small perturbation of the system out of equilibrium on the microscopic scale levels out, and the universal behavior falls into one of the known equilibrium classes. But there are also genuine non-equilibrium universality classes with no equilibrium counterpart. Prominent examples, which we will connect to in this work, are so-called absorbing state transitions, which belong to the directed percolation universality class~\cite{Hinrichsen2000,OdorRev,Tauber2014a}; but recent work has demonstrated further instances~\cite{Young2020,Daviet2024,Nahum_2020}.

The quest for novel universal structures out of equilibrium is a particularly active area of research. This is fueled by rapid experimental progress in diverse platforms from light-matter solid state systems~\cite{Carusotto2013,Noh_2016,Huebener2021,Cavalleri2018} to cold atoms and ions~\cite{Mivehvar_2021,Mueller_Qsim_2012,Harrington_2022,Browaeys_2020,SaffmanRev}, as well as functionalized matter devices in the noisy intermediate-scale quantum
(NISQ) regime~\cite{Vool_2017,Blais_2021,Preskill2018}. These systems combine coherent unitary dynamics with dissipative processes such as decoherence and loss on an equal footing~\cite{Sieberer2016,Weimer2021RMP,Sieberer2023}. Despite being 'made of quantum ingredients', such instances of driven open quantum matter tend to show classical equilibrium or non-equilibrium criticality, with a finite temperature replaced by a finite Markovian noise level~\cite{Sieberer2014, Mitra2006, overbeck2017}.

However, as a matter of principle, the quantum nature of these systems allows for exceptions to this rule. Indeed, such cases have been identified. They rely on the existence of noiseless critical modes, which can be realized with carefully engineered non-Markovian~\cite{dalla2010quantum} or even Markovian~\cite{Marino2016a} dissipation. Quantum scaling features have also been observed numerically in quantum generalizations of directed percolation models~\cite{Carollo_2019}.
Recently, a mechanism for genuine quantum directed percolation has been proposed~\cite{Thompson2024}. This demonstrates that non-equilibrium quantum universality classes can exist beyond mere classical analogs. All these instances share one trait: they represent bicritical points where double fine-tuning is needed to reach criticality: One of them ensures criticality, the other one the purity of the state.

Another route towards universal quantum non-equilibrium phenomena in open systems has been taken in~\cite{Nahum_2020}, studying a diffusion-annihilation process of Majorana defects.  Here, the topological nature of the defects ensures that the system is protected from decoherence, leading to scaling behavior distinct from classical diffusion-annihilation models.

Quantum non-equilibrium phenomena have also been found in pre-thermal dynamics of isolated quantum systems subject to a sudden quench of the system parameters. These systems can show universal behavior as a transient phenomenon at short times~\cite{Chiocchetta_2015,Chiocchetta_2017}, in part exhibiting scaling behavior without equilibrium counterpart when fermions are involved~\cite{Jian_2019}.

In this work, we establish a novel scenario of stationary quantum criticality far from equilibrium. It builds on a topological phase transition, which unfolds in the pure stationary state of Lindbladian dynamics. It differs from previous open system instances in two fundamental ways. First, it is based on fermionic -- rather than bosonic -- matter as a resource for quantum criticality. The fermions are the carriers of the critical fluctuations, a situation without a classical counterpart. Second, it relies on symmetry protection such that only a single fine-tuning is needed to reach criticality. In this respect, the new scenario more closely resembles zero temperature quantum phase transitions of fermions, as well as the original variant of directed percolation: In the former, thermal equilibrium itself acts as the protecting symmetry~\cite{sieberer2015prb, Haehl2017,  Crossley2017}, while in the latter, protection arises from a so-called rapidity inversion symmetry~\cite{Tauber2014a}. Despite these analogies, our setting crucially departs from both precedents: first, the critical theory manifestly violates detailed balance; second, the carriers of criticality are fermion excitations on top of a pure, entangled state. 

We now summarize the main elements of our construction and key results before turning to their detailed derivation. Our approach starts from a microscopic lattice model featuring a many-body dark state, derives an intermediate scale `mesoscopic' fermion-boson theory both based on  diagrammatic calculations and complementary symmetry arguments, and then connects to the macroscopic quantum critical behavior.

\subsection{Synopsis and key results} 

\textit{Microscopic model: Quantum absorbing-to-absorbing state transition ---} 
In this work, we investigate phase transitions occurring in a pure state of matter as a function of a single tuning parameter $\theta$, stabilized by a Lindbladian generator of dynamics. A key concept is that of a unique many-body dark state $\ket{\text{DS}(\theta)}$, a pure-state dynamical fixed point of the Lindbladian, into which the system evolves regardless of its initial density matrix. 
A dark state is a quantum analog of an absorbing state in classical physics: it represents a 'vacuum state' where all dynamics halts once reached. A key difference, however, is that the dark states considered here represent nontrivial vacua, filled with particles and equipped with nontrivial quantum correlations, while in the classical realm, the vacuum is trivial, given by an uncorrelated absorbing state \footnote{A more precise analogy is this: the degrees of freedom of the critical long wavelength theory annihilate a non-trivial quantum vacuum in our case, while the critical theory of absorbing state transitions in the directed percolation universality class act on a trivial vacuum. }. Specifically, the dark states in our analysis correspond to chiral topological insulators.

The competition driving the transition is realized by Lindblad operators schematically defined by (cf. Sec.~\ref{sec:MicroModel})
\begin{align}\label{eq:X}
\hat{X}(\theta)=\cos(\theta+\pi/4)\hat{L}+\sin(\theta+\pi/4)\hat{R}\,,
\end{align}
with tuning parameter $\theta \in [-\pi/4,\pi/4]$. $\hat L$ and $\hat R$ stabilize dark states that are qualitatively distinct by the value of their topological index -- below, a chiral winding number $n =\pm 1$. At $\theta=0$, where both Lindblad operators contribute equally, a critical point emerges, where the fermions become gapless, and marking the transition between the two topological phases. We also incorporate a Hamiltonian that features $\ket{\text{DS}(\theta)}$ as an eigenstate, thus not modifying the stationary state and preserving its purity.
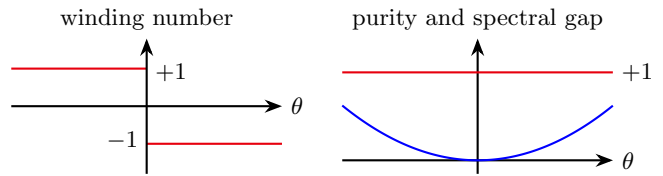
\begin{figure}
\centering
     \begin{tikzpicture}
\def\y{1.8}
\def\z{0.8}
\draw[samples=1000,color=burgundy,domain=-\y:0, thick] plot (\x,{0.5});
\draw[samples=1000,color=burgundy,domain= 0:\y, thick] plot (\x,{-0.5});
\draw[-Stealth, thick] (-\y,0) -- (\y,0) node[right] {$\theta$};
\draw[-Stealth, thick] (0,-0.5*\y) -- (0,0.5*\y) node[above] {winding number};
\node at (0,0.25*\y) [right] {$+1$};
\node at (0,-0.25*\y) [left] {$-1$};

\draw[-Stealth, thick] (\y+\z,-0.4*\y) -- (\z+3*\y,-0.4*\y) node[right] {$\theta$};
\draw[-Stealth, thick] (2*\y+\z,-0.5*\y) -- (2*\y+\z,0.5*\y) node[above] {purity and spectral gap};
\draw[samples=1000,color=burgundy,domain=\y+\z:\z+3*\y, thick] plot (\x,{0.25*\y});
\draw[samples=1000,color=blue,domain=\y+\z:\z+3*\y, thick] plot (\x,{0.125*\y*(\x-\z-2*\y)^2-0.4*\y)});
\node at (\z+3*\y,0.25*\y) [right] {$+1$};

\end{tikzpicture}
    \caption{Quantum absorbing-to-absorbing state transition. The Lindblad generator has a unique topological dark state for any value of the tuning parameter $\theta\neq 0$, see Eq.~\eqref{eq:X}. The phase transition is between fermionic dark states, which are topologically distinct by the value of a winding number $n$. The fermionic spectral gap, given by the smallest dissipation rate in the problem, closes at the transition, giving rise to divergent length and time scales characteristic of a critical point. The fermions remain in a pure state across the phase transition, emphasizing the analogy to quantum critical phenomena.}
    \label{fig:transition}
\end{figure}
Two ingredients are key for the scenario which we will develop: First, the transition proceeds far from equilibrium. The Lindblad generator violates the symmetry connected to thermodynamic equilibrium~\cite{sieberer2015prb, Haehl2017,  Crossley2017}, including at $T=0$~\cite{sieberer2015prb}, anticipating the emergence of novel critical behavior. A second key feature is particle number conservation. Regarding the fermionic dynamics sketched above, physical realizations with this property have been proposed, including superfluid~\cite{Bardyn2013,Iemini2016,Yi2012,Hoening2012,Diehl2010,Diehl2011,Bardyn2012} and insulating~\cite{Goldstein2019, Shavit2020, Huang2022, Tonielli2020, Nosov_2023, lyublinskaya2023diffusive} dark states. Particle number conservation leads to the presence of a diffusive mode coupling to the critical fermions. The diffusive dynamics induced by the purely fermionic model turns out to be too weak to give rise to non-Gaussian critical behavior; this finding is model specific and can be traced back to chiral symmetry. Coupling additional number conserving bosons held at a finite temperature instead gives rise to an interacting quantum critical point. 

\textit{Mesoscopic fermion-boson theory and fermionic dark state symmetry ---} 
For the practical evaluation, in Sec.~\ref{app:HS} we map the Lindblad equation into an equivalent Lindblad-Keldysh functional integral. We perform a Hubbard-Stratonovich decoupling of the interacting fermion problem guided by the exact knowledge of the dark state. This approach allows us to systematically track the coupling to additional bosonic modes. In this framework, we distill an effective intermediate scale action of gapless, critical fermions coupled via a cubic term to bosonic density fluctuations. The latter result from the coupling to the number-conserving bosonic bath, exhibiting diffusive density fluctuations that remain gapless throughout the transition.  

We then perform a systematic symmetry analysis of this intermediate scale action. A key insight of this work is the identification of an emergent anti-unitary parity symmetry that protects the purity of the fermion state across the transition. It can be written as a concatenation of time inversion and an additional discrete $\mathbb{Z}_2$ transformation of fermionic and bosonic fields. Crucially, it exploits the fermionic nature of the problem, transforming the fermionic Grassmann fields and their conjugates in distinct, independent ways. We refer to it as fermionic dark state symmetry (FDS). 

The FDS is identified for the mesoscopic action based on a coarse graining procedure of the microscopic theory. Once identified, it provides a powerful tool constraining the physics at even larger distances, no matter the precise microscopic physics. In particular, FDS enforces a locking between dissipation and fluctuations (in a way distinct from thermodynamic equilibrium), ensuring the presence of a pure fermion dark state at all values of $\theta$. Applying this reasoning to the critical region, 
this symmetry principle protects quantum criticality far from equilibrium. 

\textit{Macroscopic scale: Symmetry protected universality class of non-equilibrium quantum critical fermions ---} In Sec.~\ref{sec:RG}, we quantitatively extract the novel universality class associated with the non-equilibrium quantum critical point in a renormalization group (RG) analysis. Again, the FDS is key: translated into an RG language, due to the locking of spectral and noise sectors, it prohibits the generation of a relevant coupling (see Fig.~\ref{fig:RGpicture}). Consistent with the intuition gained from the microscopic model, only a single fine-tuning is needed to reach the critical point. This is in stark contrast to previous instances of non-equilibrium quantum criticality, which correspond to bicritical points~\cite{dalla2010quantum,dalla2012dynamics,Marino2016a,Marino2016b, DallaTorre2012PRB, Marino2016PRB}, necessitating the fine-tuning of two parameters.

\begin{figure}
    \input{pure_state_manifold} 
    \caption{Structure of the RG flow and implications of the fermionic dark state symmetry (FDS). In $d = 4-\epsilon$ dimensions, there is an unstable Gaussian fixed point (black dot at the origin) and an interacting Wilson-Fisher fixed point (red dot) with a single relevant direction, describing fermions far from equilibrium made critical by fine-tuning a single parameter. The FDS confines the flow to a hypersurface embedded into a larger coupling space where the symmetry is generically absent, which includes an additional relevant direction at the fixed point. The symmetry thus protects the system from activating this relevant direction. This bears analogies to the problem of directed percolation (see Sec.~\ref{sec:percolation}).}
    \label{fig:RGpicture}
\end{figure}

Within the manifold of FDS-preserving couplings, the fixed point structure comprises a noninteracting Gaussian and an interacting Wilson-Fisher fixed point below the upper critical dimension $d_c=4$. Interactions are relevant at the Gaussian fixed point, while the non-Gaussian fixed point is stable. The lower critical dimension is $d_l = 2$. We work in a dimensional expansion, which is controlled by $\epsilon = 4-d$. At the leading, one-loop order, the corresponding universality class is characterized by the exponents 
\begin{align}
    \nu =  1/2-\epsilon/8, \quad z = 2+\epsilon/2, \quad \eta_Z=\epsilon/2,
\end{align}
which describe the divergence of the correlation length, the fermion dynamical exponent, and the anomalous dimension of the fermionic field, respectively. 
All exponents receive large anomalous corrections in the physically relevant case $d=3$, compared to their canonical values $\nu = 1/2, z=2, \eta_Z=0$, mirroring the interacting Wilson-Fisher fixed point governing the universal behavior.

\textit{Outline ---}
The remainder of this paper follows the structure outlined above. In Sec.~\ref{sec:MicroModel}, we introduce the microscopic Lindblad model and provide the qualitative picture for the phase transition. The coupling to an interacting thermal bosonic bath is introduced in Sec.~\ref{sec:diffusivemodes}. In Sec.~\ref{sec:symmetries} we discuss the microscopic and emergent symmetries of the model and derive the mesoscopic effective action. The macroscopic critical behavior is then analyzed using an RG approach in Sec.~\ref{sec:RG}. In Sec.~\ref{sec:percolation}, we discuss how the new universality class relates both to equilibrium quantum critical points and to the non-equilibrium directed percolation class. We offer additional discussion and an outlook in Sec.~\ref{sec:outlook}.

\section{Microscopic model: Dissipative topological insulator and basic picture for the phase transition } \label{sec:MicroModel}
In this section, we derive an interacting, fermionic Lindblad equation with a unique and pure topological dark state $\ket{\text{DS}(\theta)}$ that can be tuned across a topological phase transition by varying the parameter $\theta$. Building on the general principles outlined in~\cite{Bardyn2013,Huang2022}, we construct a system in odd spatial dimensions $d$ equipped with chiral symmetry, corresponding to class AIII of the Altland-Zirnbauer classification~\cite{Altland1997,Altland2021}.

\subsection{Dark state Lindbladian}\label{sec:Lindblad}

The Lindbladian reads
\begin{align}\label{eq:Lindblad}
    \partial_t \hat \rho =\int_\vecx - \mi [\hat H ,\hat \rho] + \sum_\alpha \big(\hat L_\alpha\hat \rho \hat L_\alpha^\dag - \tfrac{1}{2} \{\hat L_\alpha^\dag \hat L_\alpha,\hat \rho \} \big) ,\end{align}
with Hamiltonian density $\hat H = \hat H [\hat{\psi}^\dag(\boldsymbol{x}),\hat{\psi}(\boldsymbol{x})]$ and Lindblad (or jump)  operators $\hat L_\alpha = \hat L_\alpha [\hat{\psi}^\dag(\boldsymbol{x}),\hat{\psi}(\boldsymbol{x})]$ depending on the canonical fermion operators $\hat{\psi}^\dag,\hat{\psi}$. The index $\alpha$ labels different types of Lindblad operators, which will be specified below. Furthermore, we have introduced the shorthand notation $\int_\vecx\equiv\int \dd^dx$.

A dark state $\ket{\text{DS}}$ is a pure state stationary solution of the Lindblad master equation defined as
\begin{align}\label{eq:dark_state_property}
	\hat{\mathcal{L}}\left(\ket{\text{DS}}\bra{\text{DS}}\right) = 0\,,
\end{align}
where $\hat{\mathcal{L}}$ denotes the right-hand side of Eq.~\eqref{eq:Lindblad}. This condition is fulfilled if
\begin{align}\label{eq:DS_HandL}
	\hat{H}\ket{\text{DS}}=E\ket{\text{DS}}\quad \hat{L}_\alpha\ket{\text{DS}} = 0\; \forall \alpha \,.
\end{align}
We consider a setting in which the dark state is unique—that is, no other stationary pure states exist under periodic boundary conditions—and exhibits topologically nontrivial properties. Furthermore, the specific dark states studied in this work are Gaussian, even though the underlying Lindblad generator is interacting.

We begin our discussion with purely dissipative dynamics, {\it i.e.\/}, without a Hamiltonian. The latter is reintroduced at the end of this section.

Concretely, we consider number-conserving Lindblad operators of the form~\cite{Diehl2011, Bardyn2013}
\begin{align}\label{eq:jumpop1}
    \begin{split}
		\hat{L}^{(U)}_{\alpha,\beta}(\vecx) &= \hat{\psi}^\dag_\alpha(\vecx) \hat{f}_\beta(\vecx),  \quad \beta\in U,\ \alpha\in I\,, \\
		\hat{L}^{(L)}_{\alpha,\beta}(\vecx) &= \hat{\psi}_\alpha(\vecx)\hat{f}^\dag_\beta(\vecx), \quad \beta\in L,\ \alpha\in I\,.
    \end{split}
\end{align}
Here, the operators $\{\hat{\psi}_\beta(\vecq)\}_{\beta\in I}$ denote fermionic creation and annihilation operators with quasi-momentum in the $d$-dimensional first Brillouin zone, $\vecq\in \text{BZ}^d$. The index set $I$, of cardinality $N \equiv |I|$, labels the complete set of bands. We partition it into upper ($U$) and lower ($L$) bands. Requiring chiral symmetry, $N$ must be even, and each subset must contain $N/2$ sub-bands; the rate premultiplying all Lindblad operators must be identical. Without loss of generality, we set this global dissipation rate to unity.

The operators $\{\hat{f}_\beta(\vecq)\}_{\beta\in I}$ are defined via a linear transformation 	
\begin{align}
		\hat{f}_\beta(\vecq) = V_{\beta,\delta}(\vecq)\hat{\psi}_\delta(\vecq), \quad \beta,\delta\in I\,,
\end{align}
with the (almost) unitary matrix
\begin{align}\label{eq:Vmat}
	V(\vecq) = \alpha_\mu(\vecq)\gamma_\mu, \quad \mu =1,\hdots, d+1\,,
\end{align}
{\it i.e.\/}, it satisfies $V^\dagger V\sim \mathbb{1}_N$, and $\gamma_\mu$ are the generators of the Clifford algebra satisfying $\{\gamma_\mu,\gamma_\nu\}=2\delta_{\mu\nu}$~\cite{Huang2022}. Here and in the following, repeated indices are summed over, unless stated otherwise.

The operators~\eqref{eq:jumpop1} can be understood as follows. If a fermion occupies the upper band $\beta$ at position $\mathbf{x}$ the application of $\hat{L}^{(U)}_{\alpha,\beta}(\mathbf{x})$ transports it to a new state. Depending on the choice of $\alpha$, this state can be from an upper or a lower band. Similarly, a hole in a lower band is transported by $\hat{L}^{(L)}_{\alpha,\beta}$. In the absence of holes in the lower bands and without particles in the upper bands, the operators~\eqref{eq:jumpop1} act trivially.
A dark state therefore is the stationary state annihilated by all $\hat{f}_{\beta\in U}$ and $\hat{f}^\dagger_{\beta \in L}$. Such a state is given by
\begin{align}\label{eq:darkstate}
	\ket{\text{DS}} = \mathcal{N}\prod_{\beta\in L,\vecq}\hat{f}_\beta^\dag(\vecq)\ket{0}, \quad L\subset I\,,
\end{align}
where $\ket{0}$ represents the vacuum and $\mathcal{N}$ is a normalization constant. Using the canonical anticommutation relations (CAR) it can be verified that $\hat{L}^{(X)}_\beta\ket{\text{DS}} = 0\ \forall \beta\in I\,, \forall X\in\{U,L\}$, which implies that $\hat{\mathcal{L}}\left(\ket{\text{DS}}\bra{\text{DS}}\right) = 0$. Hence, following~\cite{Kraus2008}, one finds that the pure dark state~\eqref{eq:darkstate} is indeed the unique stationary state of the evolution. 

The Lindblad operators~\eqref{eq:jumpop1} possess a strong $U(1)$ symmetry generated by the total particle number 
\begin{align}\label{eq:Charge}
\hat{Q}=\int_\mathbf{x}\hat{\psi}^\dag_\alpha(\mathbf{x})\hat{\psi}_\alpha(\mathbf{x})\,,
\end{align}
{\it i.e.\/}, they satisfy
\begin{align}\label{eq:Lcommutator}
[\hat{L}^{(X)}_{\alpha,\beta}, \hat{Q}] =0 \quad \forall \alpha,\beta \in I\,.
\end{align}
Physically, this means that the Lindbladian dynamics conserve the total particle number.  In Sec.~\ref{sec:micro-sym}, we give a more detailed discussion of the implications of this global conservation law.

As mentioned earlier, we are interested in systems with chiral symmetry, as for these, the critical theory of the dissipative topological transition is particularly simple, as we will see in Sec.~\ref{sec:RG}. 
Chiral symmetry, together with the requirement of nontrivial topological phases, restricts our discussion to odd spatial dimensions~\cite{Schnyder2008, Kitaev2009, Ryu_2010, Chui2015, Moessner_book}. There, we can define a matrix $\gamma_{d+2}$ with the property $\{\gamma_{d+2},\gamma_\mu\}=0$ for $\mu =1,\hdots d+1$. 

For concreteness, we choose a basis where $\gamma_{d+2}=\sigma_x\otimes\mathbb{1}_{N/2}$. In this representation, the chiral transformation is given by
\begin{align}\label{eq:chiral_sym}
    \mathcal{S}: \quad \hat{\psi} &\rightarrow \gamma_{d+2}\hat{\psi}^\dag, \quad \mi\rightarrow -\mi\,,
\end{align} under which the fermions transform as
\begin{align} \label{eq:chiral_transformation}
	&\forall \alpha\in U:\  \hat{\psi}_\alpha \rightarrow \hat{\psi}_\beta^\dag\ \text{ and }\  \hat{f}_\alpha\rightarrow -\hat{f}_\beta^\dag\ \text{ with }\ \beta\in L,
\end{align}
and similarly with exchanged upper and lower bands. Consequently, the Lindblad operators transform as 
\begin{align}
	&\forall \alpha\in Y:\  \hat{L}^{(U)}_{\alpha,\beta} \rightarrow -\hat{L}^{(L)}_{\delta,\epsilon}\ \text{ with }\ \delta\in Y^c,\epsilon\in U,
\end{align}
where $Y \in \{L,U\}$ and $Y^c$ its complement. Again, the same is true with $U$ and $L$ exchanged.
The prefactor $-1$ is inconsequential, as the jump operators always appear quadratically in the Lindblad equation. The chiral transformation effectively relabels the upper and lower bands. Since the Lindblad equation involves a summation over all bands, it remains invariant. Hence, the system belongs to class AIII~\cite{Altland2021} and the dark states constructed in Eq.~\eqref{eq:darkstate} can be characterized by a chiral winding invariant. 

We proceed to show how the model can be tuned between topological phases with distinct winding numbers while preserving the purity of the dark state. This is achieved by the following construction:
The definition of the fermionic operators $\hat{f}^\dagger_\alpha$ is quite general. In particular, they can create excitations with arbitrary definite chiral winding numbers. We use this to split the set of operators $\hat{f}_\alpha$ further in $\{\hat{l}_\alpha,\hat{r}_\alpha\}$, each stabilizing a topologically distinct dark state ({\it i.e.\/}, of definite chiral winding number). Specifically, $\hat{l}_\alpha$ and $\hat{r}_\alpha$ stabilize left-winding states with $n=+1$ and right-winding states ($n=-1$), respectively, with $n$ defined below in Eq.~\eqref{eq:winding}. We then introduce competition between these topological phases by defining the following coherent superposition
\begin{align}\label{eq:coherent_sup_x}
\begin{split}
		\hat{x}_\beta(\theta) &= \cos(\theta+\pi/4) \hat{l}_{\beta} + \sin(\theta+\pi/4)\hat{r}_{\beta} \\&= W_{\beta\delta}(\theta)\hat{\psi}_\delta, \quad \beta,\delta\in I\,,
\end{split}
\end{align}
where $\theta$ is a tuning parameter $\theta\in[-\pi/4,\pi/4]$ that drives the transition between the two topological phases (see Fig.~\ref{fig:transition} for an illustration). For the transition to occur in a pure state, a sufficient condition is that $\forall\ \theta\in\mathbb{R}$, $W(\theta)$ is an almost unitary matrix~\cite{Bardyn2012,Altland2021}. This criterion ensures that the operator $\hat{x}_\alpha$ satisfies the CAR up to a normalization constant.

For fermionic operators $\hat{l}(\vecq) = V_l(\vecq)\hat{\psi}(\vecq)$, $\hat{r}(\vecq) = V_r(\vecq)\hat{\psi}(\vecq)$, where $V_{l,r}(\vecq)$ are almost unitary matrices, the matrix $W$ is also almost unitary if and only if
\begin{align}
    \left(V^\dag_lV_r + V^\dag_rV_l\right) = g(\vecq)\mathbb{1}_N\,,
\end{align}
where $g(\vecq)\in \mathbb{R}$ is a smooth function. For the general construction discussed here, with $V_{l,r}=\alpha(\vecq)^{l,r}_\mu\gamma_\mu$ (cf.~\eqref{eq:Vmat}) this is the case and $g(\vecq) = 2\alpha(\vecq)_\mu^l\alpha_\mu^r(\vecq)$~\footnote{This follows from straightforward manipulations of $\gamma$ matrices, 
\begin{align*}
    \left(V^\dag_lV_r + V^\dag_rV_l\right)&=\alpha^l_\mu\alpha^r_\nu\gamma_\mu\gamma_\nu + \alpha^r_\nu\alpha^l_\mu\gamma_\nu\gamma_\mu\,,\\
    &=2\alpha^l_\mu\delta_{\mu\nu}\alpha^r_\nu + \alpha^l_\mu\alpha^r_\nu \sigma_{\mu\nu} + \alpha^r_\nu\alpha^l_\mu \sigma_{\nu\mu}\\
    &=2\alpha^l_\mu\delta_{\mu\nu}\alpha^r_\nu + \left(\alpha^l_\mu\alpha^r_\nu - \alpha^r_\nu\alpha^l_\mu \right)\sigma_{\mu\nu}\\
    &=2\alpha^l_\mu\delta_{\mu\nu}\alpha^r_\nu\,,
\end{align*}
where $\sigma_{\mu\nu} = \frac{1}{2}[\gamma_\mu,\gamma_\nu]$.}.  Therefore, the matrix $W(\theta)$ can be recast as
\begin{align}\label{eq:WfU}
W(\vecq)=\sqrt{\lambda(\vecq)}\ U(\vecq)\,,
\end{align}
with unitary $U(\vecq)$ and
\begin{align}\label{eq:lambda}
\lambda(\vecq) &= z_\mu(\theta,\vecq) z_\mu(\theta,\vecq)\geq 0, \\
    z_\mu(\vecq) &= \cos(\theta+\pi/4)\alpha(\vecq)^l_\mu + \sin(\theta+\pi/4)\alpha(\vecq)^r_\mu.
\end{align} 
Thus, the operators $\hat{x}_\beta$ satisfy the canonical algebra. Returning to our earlier discussion, the number-conserving quasilocal jump operators
\begin{align}\label{eq:jumpop2}
\begin{split}
	\hat{L}^{(U)}_{\alpha,\beta}(\theta) &= \hat{\psi}^\dag_\alpha \hat{x}_\beta(\theta),  \quad \beta\in U,\ \alpha\in I, \\
	\hat{L}^{(L)}_{\alpha,\beta}(\theta) &= \hat{\psi}_\alpha \hat{x}^\dag_\beta(\theta), \quad \beta\in L,\ \alpha\in I
\end{split}
\end{align}
therefore stabilize the pure dark state
\begin{align}\label{eq:darkstate2}
    \ket{\text{DS}(\theta)} = \mathcal{N}'\prod_{\vecq,\beta\in L} \hat{x}_\beta^\dag(\theta)\ket{0}\,,
\end{align}
{\it i.e.\/}, $\hat{\mathcal{L}}\left(\ket{\text{DS}(\theta)}\bra{\text{DS}(\theta)}\right) = 0$. For the two limiting cases $\theta=\pm\pi/4$, the Lindbladian system with jump operators~\eqref{eq:jumpop2} have dark states $\ket{\text{DS}_{l,r}}$ defined by $\hat{l}\ket{\text{DS}_l} = 0$, $\hat{r}\ket{\text{DS}_r} = 0$ respectively. 

By construction, it is impossible to deform these two states into one another without changing the chiral winding number. Thus, the system must undergo a quantum phase transition in a pure state at the critical value $\theta=\theta_c=0$, where both the operators $\hat{l},\hat{r}$ are in an equal superposition, and the dark space is two-dimensional. 

Until now, we have discussed purely dissipative dynamics. However, we can add a Hamiltonian density that satisfies Eq.~\eqref{eq:DS_HandL}. To illustrate this point, we enrich our model with the simplest choice
\begin{align}\label{eq:Hamiltonian}
    \hat{H}(\vecq)= h \hat{x}^\dagger_\alpha(\vecq)\hat{x}_\alpha(\vecq)=h \hat{\psi}^\dagger_\gamma(\vecq) W_{\gamma \alpha}(\theta)^\dagger W_{\alpha\beta}(\theta) \hat{\psi}_\beta(\vecq) \,.
\end{align}
In fact, we will see below that this is the only quadratic Hamiltonian that can be added, which is consistent with all symmetries of the system. It does not affect the dark state but contributes a coherent term to the dynamics. 
\begin{figure}[!th] 
\def\cube at (#1,#2){\filldraw[fill=black!10,draw=black] (#1,#2) rectangle (#1+\w,#2+\w);
\filldraw[fill=black!10,draw=black] (#1+0,#2+\w) -- (#1+\wd,#2+\w+\wd) -- (#1+\w+\wd,#2+\w+\wd) -- (#1+\w,#2+\w) -- (#1+0,#2+\w);
\filldraw[fill=black!10,draw=black] (#1+\w,#2+0) -- (#1+\w+\wd,#2+\wd) -- (#1+\w+\wd,#2+\w+\wd) -- (#1+\w,#2+\w) -- (#1+\w,#2+0);}

\def\sigmax at (#1,#2){\draw[burgundy,very thick] (#1+\w+0.5*\wd,#2+0.5*\w+0.5*\wd) -- (#1+\a,#2+0.5*\w+0.5*\wd);}

\def\sigmay at (#1,#2){\draw[blue,very thick] (#1+\w+\wd,#2+\w+\wd) -- (#1+\ad*\a+0.5*\w,#2+\ad*\a+0.5*\w);}

\def\sigmaz at (#1,#2){\draw[acidgreen,very thick] (#1+0.5*\w+0.5*\wd,#2+\w+0.5*\wd) -- (#1+0.5*\w+0.5*\wd,#2+\a);}

\begin{tikzpicture}


\pgfdeclareimage[height=5cm, width=5cm]{image1}{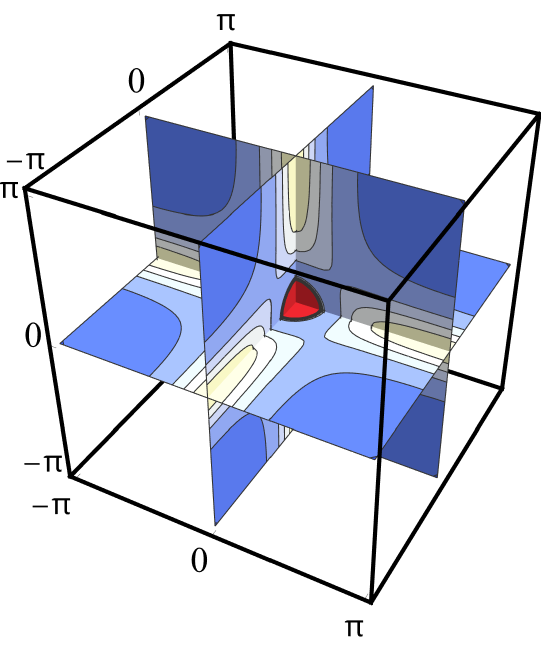}
\pgftext[left,base, at={\pgfpoint{0.75cm}{-11.05cm}} ]{\pgfuseimage{image1}}

\def\b{1.4} 
\def\h{1} 
\def\x{7} 
\def\y{3.8} 
\def\z{2.2} 
\def\w{0.3} 


\draw[thick] (0,0) -- (\x,0);
\draw[thick] (0,-\z) -- (\x,-\z);
\draw[thick] (0,-\y-\z) -- (\x,-\y-\z);

\node at (0,\y) [below right] {(a)};
\node at (0,0) [below right] {(b)};
\node at (0,-\z) [below right] {(c)};
\node at (0,-\y-\z) [below right] {(d)};

\node at (0.75*\x, -0.25*\z) [above] {$\check\tau_x$};
\node at (0.75*\x, -0.5*\z-0.5*\w) [above] {$\check\tau_y$};
\node at (0.75*\x, -0.75*\z-\w) [above] {$\check\tau_z$};

\node at (6,-0.5*\y-\z) [] {$\cdots$};
\node at (0.75,-0.5*\y-\z) [] {$\cdots$};
\node at (6.4,2.8-\y-\z) [rotate=45] {$\cdots$};
\node at (0.3,1.1-\y-\z) [rotate=45] {$\cdots$};
\node at (0.5*\x-0.15,-\y-\z+0.4) [rotate=90] {$\cdots$};
\node at (0.5*\x-0.15,-\z-0.35) [rotate=90] {$\cdots$};

\node at (0.333*\x-0.5*0.333*\b, 0.5*\y-\h) [below] {$\vecx$};
\node at (2*0.333*\x+0.5*0.333*\b, 0.5*\y-\h) [below] {$\vecx+\mathbf{e}_i$};



\filldraw[fill=burgundy!30,draw=black, thick] (0.333*\x-0.333*2*\b,0.5*\y-\h) rectangle (0.333*\x+0.333*\b,0.5*\y);
\filldraw[fill=blue!25,draw=black, thick] (0.333*\x-0.333*2*\b,0.5*\y) rectangle (0.333*\x+0.333*\b,0.5*\y+\h);
\filldraw[fill=burgundy!30,draw=black, thick] (0.333*2*\x-0.333*\b,0.5*\y-\h) rectangle (0.333*2*\x+0.666*\b,0.5*\y);
\filldraw[fill=blue!25,draw=black, thick] (0.333*2*\x-0.333*\b,0.5*\y) rectangle (0.333*2*\x+0.666*\b,0.5*\y+\h);

\foreach \t in {1,2,3,4,5} {
\draw (0.333*\x-0.333*\b, 0.5*\y-\h+0.333*\t*\h) -- (0.333*\x, 0.5*\y-\h+0.333*\t*\h);
}
\foreach \t in {1,2,3,4,5} {
\draw (0.333*2*\x+0.333*\b, 0.5*\y-\h+0.333*\t*\h) -- (0.333*2*\x, 0.5*\y-\h+0.333*\t*\h);
}

\draw[-Stealth,black] (0.333*\x-0.222*\b , 0.5*\y-\h+0.333*5*\h) -- (0.333*\x-0.222*\b , 0.5*\y-\h+0.333*2*\h);
\draw[-Stealth,black] (0.333*\x-0.111*\b , 0.5*\y-\h+0.333*4*\h) -- (0.333*\x-0.111*\b , 0.5*\y-\h+0.333*\h);
\draw[-Stealth,black] (0.333*2*\x+0.222*\b , 0.5*\y-\h+0.333*5*\h) -- (0.333*2*\x+0.222*\b , 0.5*\y-\h+0.333*2*\h);
\draw[-Stealth,black] (0.333*2*\x+0.111*\b , 0.5*\y-\h+0.333*4*\h) -- (0.333*2*\x+0.111*\b , 0.5*\y-\h+0.333*\h);

\node at (0.333*\x-0.222*\b , 0.5*\y-\h+0.333*5*\h) [below right = -0.1] {$\mathbb{1}$};
\node at (0.333*\x-0.111*\b , 0.5*\y-\h+0.333*4*\h) [below right = -0.1] {$\mathbb{1}$};
\node at (0.333*2*\x+0.222*\b , 0.5*\y-\h+0.333*5*\h) [below left = -0.1] {$\mathbb{1}$};
\node at (0.333*2*\x+0.111*\b , 0.5*\y-\h+0.333*4*\h) [below left = -0.1] {$\mathbb{1}$};




\draw[Stealth-Stealth, black] (0.333*\x+0.333*\b,0.5*\y-\h) to [out=315,in=225] node[below] {$\mathbb{1},\check\tau_i$} (0.333*2*\x-0.333*\b,0.5*\y-\h);
\draw[Stealth-Stealth, black] (0.333*\x+0.333*\b,0.5*\y+\h) to [out=45,in=135] node[above] {$\mathbb{1},\check\tau_i$} (0.333*2*\x-0.333*\b,0.5*\y+\h);











\draw[-Stealth, black] (0.333*2*\x-0.333*\b,0.5*\y-0.45*\h) to (0.333*\x+0.333*\b,0.5*\y+0.55*\h);
\draw[-Stealth, black] (0.333*\x+0.333*\b,0.5*\y+0.45*\h) to (0.333*2*\x-0.333*\b,0.5*\y-0.55*\h);

\draw[-Stealth, black] (0.333*\x+0.333*\b,0.5*\y-0.45*\h) to (0.333*2*\x-0.333*\b,0.5*\y+0.55*\h);
\draw[-Stealth, black] (0.333*2*\x-0.333*\b,0.5*\y+0.45*\h) to (0.333*\x+0.333*\b,0.5*\y-0.55*\h);


\node at (0.333*\x+0.333*\b,0.5*\y-0.5*\h) [right = 0.1] {$\mathbb{1}$};
\node at (0.333*\x+0.333*\b,0.5*\y+0.5*\h) [right = 0.1] {$\check\tau_i$};
\node at (0.333*2*\x-0.333*\b,0.5*\y-0.5*\h) [left = 0.1] {$\mathbb{1}$};
\node at (0.333*2*\x-0.333*\b,0.5*\y+0.5*\h) [left = 0.1] {$\check\tau_i$};

\node at (0.333*\x-2*0.333*\b , 0.5*\y+\h) [below right] {$U$};
\node at (0.333*\x-2*0.333*\b , 0.5*\y) [below right] {$L$};
\node at (2*0.333*\x-4*0.333*\b+\b , 0.5*\y+\h) [below right] {$U$};
\node at (2*0.333*\x-4*0.333*\b+\b , 0.5*\y) [below right] {$L$};

\def\bNew{1}
\def\hNew{0.7}

\filldraw[fill=burgundy!30,draw=black, thick] (0.25*\x-0.5*\bNew,-0.5*\z-\hNew) rectangle (0.25*\x+0.5*\bNew,-0.5*\z);
\filldraw[fill=blue!25,draw=black, thick] (0.25*\x-0.5*\bNew,-0.5*\z) rectangle (0.25*\x+0.5*\bNew,-0.5*\z+\hNew);
\node at (0.3*\x+0.5*\bNew,-0.5*\z) [] {$\equiv$};

\foreach \t in {1,2,3,4,5} {
\draw (0.25*\x-0.5*0.333*\bNew,-0.5*\z+0.333*\t*\hNew-\hNew) -- (0.25*\x+0.5*0.333*\bNew,-0.5*\z+0.333*\t*\hNew-\hNew);
}


\def\wd{\w*0.3} 
\def\a{1.1} 
\def\ad{0.4} 
\def\bpx{0.5*\x-2.5*\a+0.5*\w+0.5*\wd} 
\def\bpy{-0.5*\y-\z-\a-0.5*\w-0.5*\wd} 

\cube at (0.45*\x-0.5*\w-0.5*\wd,-0.5*\z-0.5*\w)

\foreach \t in {1,2,3} {
\cube at (2*0.333*\x-0.5*\w-0.5*\wd, -\t*0.25*\z-0.5*\t*\w)
\cube at (2*0.333*\x+\a-0.5*\w-0.5*\wd, -\t*0.25*\z-0.5*\t*\w)
}
\sigmax at (2*0.333*\x-0.5*\w-0.5*\wd, -0.25*\z-0.5*\w)

\draw[blue,very thick] (2*0.333*\x+0.5*\w,-0.5*\z+0.5*\wd-0.5*\w) -- (2*0.333*\x-0.5*\w-0.5*\wd+\a,-0.5*\z+0.5*\wd-0.5*\w);

\draw[acidgreen, very thick] (2*0.333*\x-0.5*\w-0.5*\wd+\w+0.5*\wd,-0.75*\z+0.5*\wd-\w) -- (2*0.333*\x-0.5*\w-0.5*\wd+\a,-0.75*\z+0.5*\wd-\w);


\foreach \i in {0,1,2,3,4} {
\foreach \j in {0,1,2} {
    \cube at (\bpx+\i*\a,\bpy+\j*\a)
}}
\foreach \i in {0,1,2,3,4} {
\foreach \j in {0,1} {
    \sigmaz at (\bpx+\i*\a,\bpy+\j*\a)
}}
\foreach \i in {0,1,2,3} {
\foreach \j in {0,1,2} {
    \sigmax at (\bpx+\i*\a,\bpy+\j*\a)
}}
\foreach \k in {0,1,2,3,4} {
\foreach \l in {1} {
    \cube at (\bpx+\k*\a+\ad*\a,\bpy+\l*\a+\ad*\a)
    \cube at (\bpx+\k*\a-\ad*\a,\bpy+\l*\a-\ad*\a)
    \sigmay at (\bpx+\k*\a-\ad*\a,\bpy+\l*\a-\ad*\a)
    \sigmay at (\bpx+\k*\a,\bpy+\l*\a)
}}


\end{tikzpicture}
    \caption{
    Microscopic model for a topological dark state transition on a 3D cubic lattice. (a) Two adjacent unit cells, each hosting upper ($U$, blue) and lower ($L$, red) bands, with two sub-bands per band. All (sub-)bands are coupled by the processes in Eqs.~(\ref{eq:Vmat},\ref{eq:Aops}) across sites; $\sigma$ matrices (acting between $U$, $L$) are implicit, while $\check\tau$ and $\mathbb{1}$ (acting in sub-bands) are shown. Note that some couplings are unidirectional and drive the system into the dark state. (b) The onsite Hilbert spaces are packaged into cubes to visualize the lattice model. (c) Pauli matrices $\sigma_i$ connect unit cells along spatial direction $i$, which is characteristic of chiral models. (d) The winding density (integrand of Eq.~\eqref{eq:winding}) is concentrated near $\vecq=0$.}
\label{fig:model}
\end{figure}
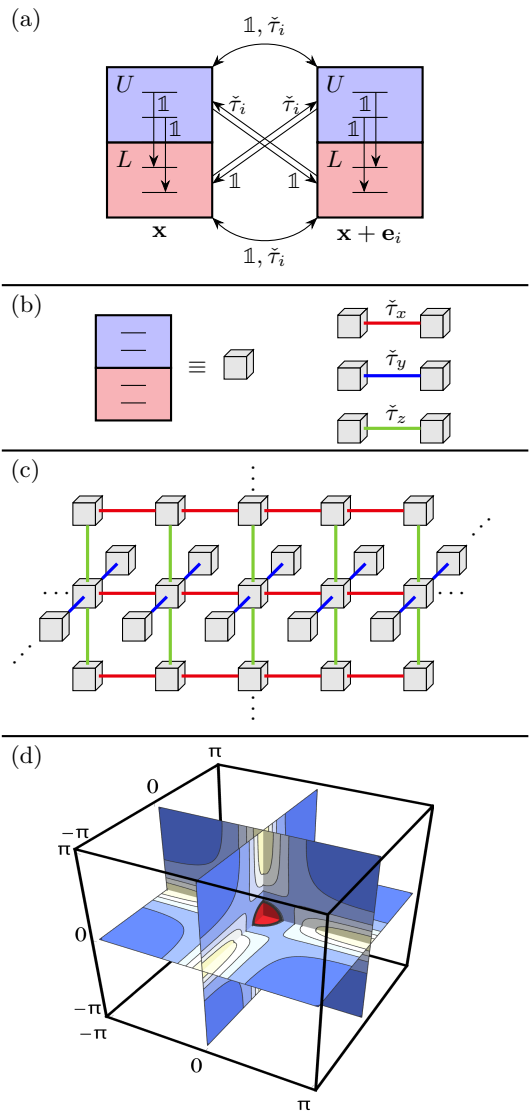

\textit{Microscopic model ---} In this work, our primary interest lies in the critical behavior of the system, and we therefore mostly treat the system as a continuum model, focusing on long wavelengths in the following~\footnote{Moving from the lattice to the continuum changes the topology of the base manifold. However, all symmetry classes can be captured in continuum Dirac models~\cite{Ryu_2010}, see~\cite{Huang2022} for a discussion of Dirac stationary states. In particular, the microscopic model considered here has a continuum limit preserving the topological properties, as the winding density is concentrated around zero momentum.}. Nonetheless, to illustrate the connection across scales—from the microscopic lattice structure to macroscopic emergent behavior—we present a lattice model that realizes the aforementioned general properties for a dissipative class AIII insulator in $d=3$. It is defined by 
\begin{align}\label{eq:lattice_model}
\begin{split}
    W(\vecq) =& \sum_{i=1}^3\sin(q_i a)\sigma_y\otimes\check\tau_i \\&+ \left(\sin(\theta)-3+\sum_{i=1}^3\cos(q_i a)\right)\sigma_z\otimes\mathbb{1}_2\,,
\end{split}
\end{align}
where $a$ denotes the lattice spacing and the hopping amplitude is adjusted using the single parameter $\theta\in\mathbb{R}$. The real space form of $W(\vecq)$ reads
\begin{align}\label{eq:Aops}
    W(\vecx-\mathbf{y}) = \; &\sin(\theta)\sigma_z \otimes \mathbb{1}_2 \delta_{\vecx,\mathbf{y}} \\
    &- \frac{1}{2}\left(\mi \sigma_y \otimes \check\tau_i - \sigma_z \otimes \mathbb{1}_2\right) \delta_{\vecx,\mathbf{y}-a \mathbf{e}_i} \notag \\
    &+ \frac{1}{2}\left(\mi \sigma_y \otimes \check\tau_i + \sigma_z \otimes \mathbb{1}_2\right) \delta_{\vecx,\mathbf{y}+a \mathbf{e}_i}\,. \notag
\end{align}
Here, $\sigma_i$ are Pauli matrices acting in the space of upper ($U$) and lower bands ($L$), each of which contains two sub-bands in three dimensions. These are acted on by the Pauli matrices $\check\tau_i$. The $\mathbf{e}_i$ indicate unit vectors in the three spatial directions. Further, one finds that the model stabilizes a dark state with chiral winding number given by $n=\text{sign}(\sin(\theta))+1$ (see Eq.~\eqref{eq:winding}). For $\sin(\theta)=0$ the dissipative gap (see Eq.~\eqref{eq:diss_gap}) closes at $\vecq=0$.  Expanding in small momenta, one recovers the continuum model we discuss in the remainder of this paper.

\subsection{Dynamical mean field theory}\label{sec:dynamical_MF}

As mentioned previously, the system is strongly interacting. Accordingly, a comprehensive field-theoretic analysis--presented in Secs.~\ref{sec:symmetries},~\ref{sec:RG} and App.~\ref{app:HS}--is necessary to characterize the universality class of the transition.
In this section, we introduce the field theoretical framework, and provide a qualitative mean-field picture to build intuition and introduce key quantities, including the dissipative and noise gaps, as well as the topological invariant that jumps at the critical point.

The fermionic Lindblad equation \eqref{eq:Lindblad} can be mapped into a Keldysh functional integral \cite{kamenev_book, Sieberer2023}, 
\begin{align}\label{eq:Kpart}
    \mathcal{Z}= \int \mathcal{D}[\bar{\psi},\psi]\me^{\mi S[\bar\psi,\psi]}\,,
\end{align}
where the integration is over Grassmann fields $\psi$ and $\bar{\psi}$. For the Lindblad operators from Eq.~\eqref{eq:jumpop2}, the microscopic action can be presented in a way that lends itself to a subsequent decoupling into a boson-fermion theory, 
\begin{align}\label{eq:micro_action}
\begin{split}
    S[\bar\psi,\psi]=&S_0[\bar\psi,\psi]+S_L[\bar\psi,\psi]\,,\\    S_0[\bar\psi,\psi]=&\sum_{\nu}\nu\int_{\mathbf{x},t}\,\left(\bar{\psi}^\nu_{\alpha}(\vecx,t)\mi \partial_t\psi^\nu_{\alpha}(\vecx,t)- H_\nu[\bar{\psi},\psi]\right)\,,\\
    S_L[\bar\psi,\psi]=&\int_{\mathbf{x},t}\, X_{\alpha}^{\nu\rho}(\vecx,t)\Psi_{\alpha}^{\rho\nu}(\vecx,t)=\text{Tr}(X\Psi)\,.
\end{split}
\end{align}
Here, we have introduced the short-hand notation $\int_{\mathbf{x},t}\equiv \int\dd^d x\dd t$ for space-time integration, as well as $\text{Tr}$ representing the integral over space-
time and trace over all indices. The upper index, $\nu,\rho$ here, taking values $+$ or $-$, is the contour index that labels the branches of the Keldysh contour~\cite{kamenev_book, Sieberer2016}.

Furthermore, we have defined Grassmann bilinears
\begin{align}\label{eq:bilinears}
\begin{split}
    X_{U}^{\nu\rho}(\vecx,t)&= \mi \sum_{\alpha \in U}\bar{x}_{\alpha}^\nu(\vecx,t+\nu 0^+)x_{\alpha}^\rho(\vecx,t)\,,\\
    X_{L}^{\nu\rho}(\vecx,t)&= \mi \sum_{\alpha \in L}x_{\alpha}^\nu(\vecx,t+\nu 0^+)\bar{x}_{\alpha}^\rho(\vecx,t-\rho 0^+)\,,\\
    \Psi_{U}^{\nu\rho}(\vecx,t)&= c^{\rho\nu}\sum_{\alpha\in I}\psi_{\alpha}^\rho(\vecx,t+\rho 0^+)\bar{\psi}_{\alpha}^\nu(\vecx,t)\,,\\
    \Psi_{L}^{\nu\rho}(\vecx,t)&= c^{\rho\nu}\sum_{\alpha\in I}\bar{\psi}_{\alpha}^\rho(\vecx,t)\psi_{\alpha}^\nu(\vecx,t)\,,
\end{split}
\end{align}
where the contour index is not summed over. The matrix
\begin{align}\label{eq:c_matrix}
c=\begin{pmatrix}
    1&0\\-2&1
\end{pmatrix}
\end{align} 
weighs and couples the Keldysh contours, with a structure characteristic of Lindblad dynamics. Since the Lindblad operators and their products in the evolution equation~\eqref{eq:Lindblad} are not normal ordered, we have introduced infinitesimal time shifts that track the proper ordering~\cite{Sieberer2014}. The Grassmann fields $x$, $\bar{x}$ correspond to the fermionic creation and annihilation operators $\hat{x}$, $\hat{x}^\dagger$ given by (cf. Eq.~\eqref{eq:coherent_sup_x} in the operator formalism)
\begin{align}\label{eq:W_pos}
    x_\alpha(\vecx,t) = \int_\vecy W_{\alpha\beta}(\vecx-\vecy)\psi_\beta(\vecy,t)\,,
\end{align}
where $W_{\alpha\beta}(\vecx-\vecy)$ is the position space representation of Eq.~\eqref{eq:WfU}. 
The quadratic Hamiltonian \eqref{eq:Hamiltonian} takes the form 
\begin{align}
    H_\nu[\bar\psi,\psi] = h\bar x^\nu_\alpha(\vecx,t) x^\nu_\alpha(\vecx,t)\,.
\end{align}
Using the knowledge of the exact stationary state~\eqref{eq:darkstate2}, we can linearize the interacting Lindbladian within a mean-field approach. The mean-field decoupling can be done either at the operator level~\cite{Bardyn2013}, or equivalently via a Lindblad-Keldysh path integral~\cite{Tonielli2020}, in a Hartree-Fock approach. In this work, we introduce a third strategy, based on a Hubbard-Stratonovich decoupling of the quartic terms, which effectively replaces fermion bilinears by matrix fields $X^{\nu\rho}\rightarrow \boldsymbol{\eta}^{\nu\rho}, \Psi^{\nu\rho}\rightarrow \boldsymbol{\phi}^{\nu\rho}$. The mean field theory is then obtained from a saddle point condition for $\boldsymbol{\eta},\boldsymbol{\phi}$, cf. App.~\ref{app:HS}. A crucial advantage of this approach is its ability to capture beyond mean field effects in terms of bosonic fluctuations. This is key to assess the importance of gapless diffusive modes imposed by particle number conservation. 

Applied to the model introduced in Sec.~\ref{sec:Lindblad}, both approaches replace the quartic terms in action with bilinears in the fermionic fields $\bar{x}_\beta$, i.e 
\begin{align}
    \begin{split}
        X^{\rho\nu}_{U}\Psi_U^{\nu\rho}(\vecx,t)\rightarrow  X^{\rho\nu}_{U}\Ev{\Psi_U^{\nu\rho}(\vecx,t)} = \varphi_0 X_U^{\rho\nu}(c^\top)^{\nu\rho}\, , \\
        X^{\rho\nu}_{L}\Psi_L^{\nu\rho}(\vecx,t)\rightarrow  X^{\rho\nu}_{L}\Ev{\Psi_L^{\nu\rho}(\vecx,t)} = \varphi_0  X_L^{\rho\nu}(c^\top)^{\nu\rho}\, ,
    \end{split}
\end{align}
where $\varphi_0= 1/2$ is the mean particle density at half filling. All other expectation values involving the fields $x_\alpha$ and $\psi_\alpha$ vanish due to the dark state property (cf. Sec.~\ref{sec:MicroModel}). Using Eq.~\eqref{eq:coherent_sup_x} to express the fields $x_\alpha^\pm$ in terms of $\psi_\alpha^\pm$ and performing a rotation of the fermionic fields into the classical-quantum basis defined by
\begin{align}\label{eq:cq}
    \psi^c=\frac{1}{\sqrt{2}}(\psi^++\psi^-),\quad \psi^q=\frac{1}{\sqrt{2}}(\psi^+-\psi^-)\,,
\end{align}
yields a quadratic fermionic action 
\begin{align}\label{eq:bare_action}
    \begin{split}
        S_0 &= \int_{\mathbf{q},\omega}\ \bar{\psi}(\omega,\vecq)\mqty(0 & P^A\\P^R & P^K)\psi(\omega,\vecq)\\
        &=\int_{\mathbf{q},\omega}\ \bar{\psi}(\omega,\vecq)\mqty(0 &\omega - K(\vecq) \\ \omega -K^\dagger(\vecq)  & 2\mi P(\vecq) )\psi(\omega,\vecq)\,,
    \end{split}
\end{align}
where we have taken the infinite lattice size limit and moved to a Fourier representation in terms of the continuous variables $(\omega,\vecq)$, and used the notation $\int_{\mathbf{q},\omega}\equiv \int \dd\omega\dd^d q$.  
The matrices $K(\vecq)$ and $P(\vecq)$ are defined as 
\begin{align}\label{eq:matrices}
\begin{split}
    K(\vecq) &=H(\vecq)+\mi D(\vecq)=h(\vecq)\mathbb{1}_N+\mi D(\vecq),\\
    D(\vecq)&= W(\vecq)^\dag W(\vecq) = \lambda(\vecq)\mathbb{1}_N\,, \\
    P(\vecq) &= W(\vecq)^\dag\tau_z W(\vecq)= \lambda(\vecq) U(\vecq)^\dag \tau_z U(\vecq)\,, 
\end{split}
\end{align}
where we have introduced $h(\vecq)=h \lambda(\vecq)$, $\lambda(\vecq)$ is specified in Eq.~\eqref{eq:lambda} and the sign of $\tau_z=\sigma_z\otimes\mathbb{1}_{N/2}$ distinguishes between empty upper $(+1)$ and filled lower bands $(-1)$. All matrices $H, D$ and $P$ are Hermitian.
The fields $\psi$ and $\bar{\psi}$ in Eq.~\eqref{eq:bare_action} are $2N$-component Grassmann fields of the form
\begin{align} \label{eq:Grassmann_fields}
\begin{split}
    \psi(\omega,\vecq) &= \mqty(\psi^c(\omega,\vecq)\\\psi^q(\omega,\vecq))\,, \quad \psi^{c,q}= \mqty(\psi^{c,q}_U(\omega,\vecq)\\ \psi^{c,q}_L(\omega,\vecq))\,, \\
    \bar{\psi}(\omega,\vecq) &= \mqty(\bar{\psi}^c(\omega,\vecq)\\\bar{\psi}^q(\omega,\vecq))\,, \quad \bar{\psi}^{c,q}= \mqty(\bar{\psi}^{c,q}_U(\omega,\vecq)\,,\\ \bar{\psi}^{c,q}_L(\omega,\vecq))\,,
\end{split}
\end{align}
and $\psi_{U/L}^{c/q}$ are $N/2=2^{(d-1)/2}$-component Grassmann fields in band space, with the labels $c$ and $q$ denoting classical and quantum fields, introduced by the doubling of degrees of freedom on the Keldysh contour \cite{kamenev_book} 
Both have vanishing expectation values, $\langle \psi^{c,q}\rangle = 0$. 
Their role, however, becomes clear when considering the Green's functions
\begin{align}\label{eq:operatorgreen}
\begin{split}
    G^R(\vecx,t;\vecx',t') &= -\mi\Theta(t-t')\langle\{\hat{\psi}(\vecx,t),\hat{\psi}^\dag(\vecx',t') \}\rangle\\
    &=-\mi\Theta(t-t')\langle \psi^c(\vecx,t)\bar{\psi}^q(\vecx',t') \rangle\,,\\
    G^A(\vecx,t;\vecx',t') &= \mi\Theta(t'-t)\langle\{\hat{\psi}(\vecx,t),\hat{\psi}^\dag(\vecx',t') \}\rangle\\
    &=-\mi\Theta(t'-t)\langle \psi^q(\vecx,t)\bar{\psi}^c(\vecx',t') \rangle\,,\\
    G^K(\vecx,t;\vecx',t') &= -\mi\langle[\hat{\psi}(\vecx,t),\hat{\psi}^\dag(\vecx',t')]\rangle\\
    &=-\mi\langle\psi^c(\vecx,t)\bar{\psi}^c(\vecx',t')\rangle\,.
\end{split}
\end{align}
Thus, correlation functions containing quantum fields encode information about the system's response to an external perturbation. Conversely, the correlation functions comprising only classical fields encode information about the fluctuations. 
The retarded and advanced response functions $G^R$ and $G^A$ are uniquely determined by each other via $[G^A]^\dag(\vecx',t';\vecx,t)=G^R(\vecx,t;\vecx',t')$, where the dagger denotes Hermitian conjugation in band space. In thermal equilibrium, response and Keldysh Green's functions are related by fluctuation-dissipation relations. Out of equilibrium, no such rigid relation exists in general.

So far, our discussion of Green's functions has been general. For the linearized model, they can be obtained by inverting the matrix in Eq.~\eqref{eq:bare_action}, yielding
\begin{align}\label{eq:bare_propagators}
    \begin{split}
        G^R(\omega,\vecq) &= (P^R(\omega,\vecq))^{-1} = \left(\omega - K^\dag(\vecq)\right)^{-1}\,, \\
        G^K(\omega,\vecq) &= -(P^R(\omega,\vecq))^{-1} P^K(\vecq) (P^A(\omega,\vecq))^{-1}\,.
    \end{split}
\end{align}

At the level of the Gaussian theory, one can identify three important scales of dimension energy, which are typically independent of each other. We briefly review these below before moving to the long-wavelength limit of dissipative AIII topological insulators near criticality.

(i) The bare \textit{dissipative gap} is defined as
\begin{align}\label{eq:diss_gap}
    \Delta_{d}(\theta) = \min_{|\vecq|}{(\text{eigvals}(D(\vecq)))}\,. 
\end{align} 
By construction, the matrix $D(\vecq)$ is positive definite and $\Delta_{d}(\theta) \geq 0$ provides a minimal damping/relaxation rate (or a characteristic time scale)
\begin{align}
    \tau_c(\theta) \sim \Delta_{d}^{-1}(\theta)\,,
\end{align}
at which single particle excitations decay into the stationary state. Conversely, if the dissipative gap vanishes at some $\theta = \theta_c$, this leads to a singular behavior of $G^R(\omega,\vecq)$ as $\omega,\vecq\rightarrow 0$. In turn, this signals an infrared divergence associated with the physics of continuous phase transitions. Hence,  $\Delta_{d}(\theta)$ is analogous to a `mass term' that is tuned to zero to reach criticality.

(ii) A second important gap is the \textit{noise gap}
\begin{align}
    \Delta_n(\theta) =\min_{|\vecq|}|\text{eigvals}(P(\vecq))|\,.
\end{align}
The noise gap has been introduced here as an auxiliary quantity, which, out of equilibrium, is independent of the dissipative gap. Its physical meaning will become clear in the following discussion, once the purity gap has been introduced below.
For the class of models presented above, one easily verifies the \emph{fermionic dark state condition} (FDC)
\begin{equation}\label{eq:FDC}
    [D,P]=[H,P]=[H,D]=0, \quad P^2=D^2
\end{equation} 
for all $\vecq$. Consequently, both these matrices can be simultaneously diagonalized with $\text{eigvals}(P) =\pm \text{eigvals}(D)$. This implies a locking of the dissipative and noise gap. Due to this locking, they must vanish simultaneously at $\theta=\theta_c$, which implies that a single fine-tuning is sufficient to access the critical point. We will elaborate on this, including in the presence of interactions, in more detail in Sec.~\ref{sec:symmetries}.

(iii) Finally, one can define the \textit{purity gap}, which characterizes the purity of the state. The density matrix of a noninteracting or Gaussian theory can be fully characterized by the Hermitian covariance matrix
\begin{align}\label{eq:covariance}
\begin{split}
    \Gamma_{\alpha\beta}(\vecq) &= \langle [\hat{\psi}_\alpha(\vecq,t),\hat{\psi}_\beta^\dag(\vecq,t)]\rangle\\
    &=\frac{\mi}{2\pi}\int_\omega\ G_{\alpha\beta}^K(\omega,\vecq). 
\end{split}
\end{align}
In its diagonal basis, it describes the mode occupation of fermionic quasiparticles, which constrains its eigenvalues $\text{eigvals}(\Gamma)\in [-1,1]$ due to Pauli's principle.
With this, one defines the purity gap as
\begin{align}\label{eq:pure_gap}
\Delta_p=\min_{|\vecq|}{|\text{eigvals}(\Gamma(\vecq))|}\,.
\end{align}
In the present case, the purity gap is maximal, {\it i.e.\/} $\Delta_p = 1$: Due to the commutativity of the involved matrices, one finds the simple form $\Gamma(\vecq) = D^{-1}(\vecq) P(\vecq)$, and thus, using $D^2=P^2$, $\Gamma(\vecq)^2 = \mathbb{1}$. Thus, as the name suggests, the FDC is a sufficient criterion for the purity of the states; in particular, this confirms our previous assertion that the phase transition occurs between pure states.

To elucidate its physical significance, we now discuss the interpretation of the noise gap. For driven dissipative Markovian systems, the noise matrix $P(\vecq)$ is independent of frequency and encodes information about the topology of the state. Since $\Gamma(\vecq)=D^{-1}(\vecq)P(\vecq)$, the eigenvalues of $P(\vecq)$ determine the occupation of fermionic quasiparticles. The vanishing of the noise gap, by itself, as $\theta\rightarrow\theta_c$ is not associated with an infrared divergence, and thereby does not by itself signal criticality. For fermions, however, the bound $|\text{eigvals}(\Gamma(\vecq))|\leq 1$ as $|\vecq|\rightarrow 0$ enforces that the dissipative gap can close only together with the noise gap. Conversely, a vanishing noise gap may occur while the dissipative gap remains finite, still satisfying the condition $|\Gamma|\leq 1$. This situation arises in mixed–state phase transitions, where a closing noise gap indicates a topological transition without criticality; see \cite{Sieberer2023,Huang2022}.

In equilibrium, the noise gap reflects thermal mode occupations and is tied to the dissipative gap by the fluctuation–dissipation relation, enforcing a fixed link between response and fluctuations. For spinless fermions, $\Delta_n = 2i\Delta_d F(\omega)$, where $F(\omega)=1-2\xi n_F(\omega)$, with $n_F$ the Fermi distribution function~\cite{kamenev_book}. Out of equilibrium, by contrast, the noise and dissipative gaps are independent, which implies the same for the correlation and response functions.

\paragraph*{Topological invariant ---} Chiral symmetry (as defined in Eq.~\eqref{eq:chiral_transformation}) implies in particular $\{\Gamma(\vecq),\gamma_{d+2}\}=0$, which in turn means that $\Gamma(\vecq)$ can be brought into a chiral form~\footnote{To do this explicitly, consider the basis where $\gamma_{d+1}=\sigma_z\otimes\mathbb{1}$, then the chiral representation can be obtained by considering a unitary transformation $\Lambda = \exp(\mi \frac{\pi}{4}\sigma_y\otimes\mathbb{1})$ acting on the $\gamma$ matrices
\begin{align*}
\begin{split}
&\gamma_{d+1}\rightarrow\Lambda^{-1}\gamma_{d+1}\Lambda=\sigma_x\otimes\mathbb{1},\\
&\gamma_i \rightarrow \Lambda^{-1}\gamma_i\Lambda = \gamma_\alpha\ \forall i=1,\dots,d,\\
&\gamma_{d+2} \rightarrow \Lambda^{-1}\gamma_{d+2}\Lambda = -\sigma_z\otimes\mathbb{1}\,.
\end{split}
\end{align*}
}
\begin{equation*}
    \Gamma(\vecq) = \mqty(\admat{Q(\vecq),Q^\dag(\vecq)}),\quad QQ^\dag = \mathbb{1}\,.
\end{equation*}
Using this we can define the chiral winding number (in odd spatial dimensions $d=2k+1$) as~\cite{Ryu_2010} 
\begin{align}\label{eq:winding}
    n = \int_{\vecq}\frac{(-1)^k k!}{d!}\left(\frac{\mi}{2\pi}\right)^{k+1} \Tr\left[\left(Q^\dag\dd Q\right)^{\wedge d}\right]\,,
\end{align}
where $\dd Q = \partial_{q_j}Q(\vecq)\ \dd q_j$, and, deviating from the remainder of this paper, the trace $\Tr$ runs only over band space.

\paragraph*{Near-critical dissipative topological insulators ---} Near the critical point, topological insulators undergoing a dark state phase transition can be described by the continuum limit obtained by expanding the coefficients $\alpha(\vecq)^{l/r}_\mu$ in Eq.~\eqref{eq:Vmat} to linear order in $\vecq$ and absorbing the lattice constant $a$ into the couplings. 
For the left-winding ($n=+1$) and right-winding fermions ($n=-1$), one finds the dimensionless matrix
\begin{align}\label{eq:unitary_dirac}
    V_{l/r}(\vecq) &= q_\alpha\gamma_\alpha + \text{sign}(n)\gamma_{d+1}\,.
\end{align}
Consequently, the dark state shares the topological properties of the ground state of the Dirac equation in $d$ dimensions~\cite{Shen_2017}. Utilizing Eq.~\eqref{eq:lambda}, we identify 
\begin{align}
\begin{split}
    z_\alpha(\theta,\vecq) &= \sqrt{2}\cos(\theta)q_\alpha,\quad \alpha=1,...,d\,,\\
    z_{d+1}(\theta,\vecq) &= -\sqrt{2}\sin(\theta)\,,
\end{split}
\end{align}
which yield the $D$ and $P$ matrices
\begin{align}\label{eq:D_P_effective}
\begin{split}
    D(\vecq) &=\left( \vecq^2 + 1 + \left(\vecq^2- 1\right)\cos(2\theta)\right)\mathbb{1}\,,\\
    P(\vecq) &= -\left[(\vecq^2 + 1)\cos(2\theta) +\vecq^2 - 1\right]\gamma_{d+1} \\
    &\quad-4\cos(\theta)\sin(\theta)q_\alpha\gamma_\alpha\,.
\end{split}
\end{align}
From $D(\vecq)$ follows the bare dissipative gap 
\begin{align}
    \Delta_{d} = 2\sin^2(\theta)\,,
\end{align}
which is positive for all values of $\theta$, except at $\theta_c = 0$ where it vanishes together with the noise gap, marking the bare value of the phase transition. At the critical point, both $D\sim q^2$ and $P\sim q^2$, which is referred to as quantum scaling~\cite{Sieberer2023}. 

From this, the mean-field critical exponents can be extracted. The dynamical exponent, defined by $\omega_c(q) \sim q^z$, is $z=2$ on the mean-field level, while the static correlation length $\xi \sim \Delta_{s}^{-\nu}$ gives $\nu = 1/2$. 

Finally, in $d=3$, the topological invariant is
\begin{align}\label{eq:winding_3d}
    n(\theta) = \text{sign}(\sin(\theta)),\quad \theta\neq0\,,
\end{align}
with $\theta\in[-\pi/4,\pi/4]$. It switches from $n=-1$ to $n=1$ at $\theta_c=0$, confirming that the phase transition occurs between two topologically distinct insulators. This has been illustrated in the left panel in Fig.~\ref{fig:transition}.

Although our construction focused on microscopic Class AIII topological insulators realizing a pure–to–pure state transition, it readily generalizes to other symmetry classes. The Dirac description captures a broad family of microscopic models—mirroring the equilibrium case, where each symmetry class supporting topological insulators (or superconductors) in $d$ spatial dimensions admits a representative low–energy Dirac Hamiltonian, and a "dimension-reduction" procedure yields the corresponding representatives in lower dimensions~\cite{Ryu_2010}. Technically, Eq.~\eqref{eq:unitary_dirac} corresponds to the lowest–order gradient expansion of the matrix $V(\vecq)$. We highlight two directions for further study: (a) setting one of the components of the spatial momenta to zero, reduces Dirac model for Class AIII in $d=3$ to a dissipatively stabilized Chern insulator in $d=2$ \cite{Tonielli2020,Huang2022} which belongs to Class A; (b) then replacing the field operators $\hat{l},\hat{r}$ with Bogoliubov operators stabilizing a BCS $p$–wave state yields Class D topological superconductors~\cite{Bardyn2012}.

In Sec.~\ref{sec:symmetries}, we adopt a complementary viewpoint, where building upon the long wavelength Dirac description, we present the necessary ingredients for a systematic symmetry-based construction of a mesoscopic field theory describing the critical theory of pure-to-pure state transitions.

\subsection{Diffusive modes}\label{sec:diffusivemodes}
The microscopic model described in Sec.~\ref{sec:Lindblad} conserves the total particle number. Following the general construction of Noether currents, cf. Sec.~\ref{sec:symmetries}, we expect density fluctuations to be gapless. Such gapless modes can couple to the fermions and therefore affect the critical behavior~\cite{Hohenberg1977}. However, under chiral symmetry \eqref{eq:chiral_sym}, the density behaves as $\hat{n}(x)\to N - \hat{n}(x)$ and therefore exhibits no fluctuations, {\it i.e.\/} $n(x)=N/2$. Consequently, the low-energy effective theory is the mean-field action derived in Sec.~\ref{sec:dynamical_MF}. Importantly, the dark-state critical fixed point of the strongly interacting fermionic model introduced in Sec.~\ref{sec:Lindblad} is therefore Gaussian. This general argument is complemented and confirmed via an explicit calculation of the bosonic diffusive action in App.~\ref{app:Hydro_mode} -- the vanishing of the bosonic noise level implies the Gaussianity of the fixed point.

However, the Gaussianity of the fixed point is a very peculiar feature of the self-interacting fermions. Here we extend the microscopic model in a way that preserves the purity of the  fermion state -- more operationally, the property  $\Gamma(\vecq)^2=\mathbb{1}$, including in the interacting problem (cf. Sec.~\ref{sec:FDS}) -- while featuring a finite bosonic noise level. In turn, this leads to a non-Gaussian fixed point without any additional fine-tuning. 

To this end, we couple our microscopic dissipative fermion model to number-conserving identical bosons $\varphi_i$ with a large number of field components $i = 1,\dots,M$ held at a finite temperature $T$ (above their condensation threshold). Finite temperature is important to ensure a finite noise level for the bosonic density diffusion, while large $M$ is required to preserve fermion purity. Simultaneously, the diffusive mode and its interaction with the fermions is unaffected by the large-$M$ limit.

We consider the simplest possible interaction term,
\begin{align}\label{eq:fermi-boson_int}
    \hat{H}_\text{int}=\frac{g}{\sqrt{N}\sqrt{M}}\sum_{i}^M\sum_{j}^N\int_{\vecx} \hat{\psi}_j^\dagger(\vecx)\hat{\psi}_j(\vecx)\hat{\varphi}_i^\dagger(\vecx)\hat{\varphi}_i(\vecx)\,.
\end{align}
The bare action of the $U(M)$ symmetric complex bosons near thermal equilibrium at finite temperature $T$ is given by
    \begin{align}
        \begin{split}
            S_0=\sum_i^M\int_{\omega,\vecq} \varphi_i^\dag(\omega,\vecq)\begin{pmatrix}0 & D^A(\omega,\vecq)\\ D^R(\omega,\vecq) & D^K(\omega,\vecq)\end{pmatrix}\varphi_i(\omega,\vecq)\,,
        \end{split}
    \end{align}
where $\varphi_i(\omega,\vecq)=\begin{pmatrix}\varphi_c(\omega,\veck)&\varphi_q(\omega,\veck)\end{pmatrix}_i^\top$ is a $2$ component complex field or each $i$ and the inverse propagators are given by
\begin{align}
    \begin{split}
    D^R(\omega,\vecq) &= (\omega + \tau - K\vecq^2 + \mi \epsilon)\, ,\\
    D^K(\omega,\vecq) &= \coth(\frac{\omega}{2T})\left(D^R(\omega,\vecq) - D^A(\omega,\vecq)\right)\,.
    \end{split}
\end{align}
We further allow for interactions between bosons
\begin{align}\label{eq:boson_vertex}
    \hat{H}_\text{B}=\frac{g_B}{M}\sum_{i,j}^M\int_\vecx \hat{\varphi}_j^\dagger(\vecx)\hat{\varphi}_j(\vecx)\hat{\varphi}_i^\dagger(\vecx)\hat{\varphi}_i(\vecx)\,,
\end{align}
where $i,j=1,\dots,M$ label the components of the bosonic operators and $g_B$ is a real coupling constant. 
\begin{figure}[t!]
  \scalebox{0.95}{\begin{tikzpicture}
\def\r{1}
\def\sx{3+0.5}
\def\sy{1.5}
\tikzset{snake it/.style={decorate, decoration=snake}}

    \path (-1.3,0.5) node[node font=\Large,black] {$\Sigma_F =$};
    \path (-0.45,0.5) node[node font=\Large, black] {$\frac{1}{2}$};
    \begin{feynman}
			\vertex (a0) at (0,0);
			\vertex (a1) at (1,0);
            \vertex (a2) at (2,0);
			\vertex (d1) at (1,1);
            \vertex (d2) at (1,1.0+\r);
			\diagram*{
				(a0) -- [fermion ,thick] (a1),
				(a1)-- [ fermion, thick] (a2),
                (a1)-- [scalar, thick] (d1),
                (d1) -- [photon,half left] (d2), (d2) --[photon, half left] (d1)
			};
    \end{feynman}
    \filldraw[black] (d1) circle (1pt);
    \draw [arrows={-Latex[length=6.5pt]}] (1.0,1.0+\r)--(0.9,1.0+\r);
    \filldraw[black] (a1) circle (1pt) node[anchor=north,font=\large,black] {$g$};

    \draw[xshift=-2mm] (2.5,0.5) node[node font=\Large, black] {$+$};
    \draw[xshift=-2mm] (3.1,0.5) node[node font=\Large, black] {$\frac{1}{N}$};
    \begin{feynman}[xshift=-2mm]
			\vertex (a0) at (0+\sx,0);
			\vertex (a1) at (1+\sx,0);
            \vertex (a2) at (2.5+\sx,0);
            \vertex (a3) at (3.5+\sx,0);
			\vertex (d1) at (1+\sx,1.0);
            \vertex (d2) at (2.5+\sx,1.0);
			\diagram*{
				(a0) -- [fermion ,thick] (a1),
				(a1)-- [ fermion, thick] (a2),
                (a2)-- [ fermion, thick] (a3),
                (a1)-- [scalar, thick] (d1),
                (a2)-- [scalar, thick] (d2),
			};
    \end{feynman}
    \filldraw[black,xshift=-2mm] (d1) circle (2pt);
    \filldraw[black,xshift=-2mm] (d2) circle (2pt);
    \draw[black,line width=1mm,xshift=-2mm] (1+\sx-0.03,1.0) -- (2.5+\sx+0.03,1.0);
    
    \draw[black,line width=1mm] (-2.0,-\sy) -- (-1.0,-\sy) node[anchor=west, black,node font=\Large] {$=$};
    \filldraw[black] (a1) circle (1pt) node[anchor=north,font=\large,black] {$g$};
    \filldraw[black] (a2) circle (1pt) node[anchor=north,font=\large,black] {$g$};
    
    \draw[draw=black, thick, snake it,decoration={amplitude=1.5,segment length=7}] (0,0-\sy) to [out=45,in=145, relative]  (2,0-\sy);
    \draw[draw=black, thick, snake it, decoration={amplitude=1.5,segment length=7}] (0,0-\sy) to [out=-45,in=-145, relative]  (2,0-\sy);

    \draw [arrows={-Latex[length=6.5pt]}] (1.0,-\sy+0.43) --(1.15,-\sy+0.44);
    \draw [arrows={-Latex[length=6.5pt]}] (1.0,-\sy-0.42) --(0.85,-\sy-0.41);
    
    \def\s{0.25}
    \path (2.5+0.1,-\sy) node[node font=\Large, black] {$+$};
        \draw[draw=black, thick, snake it,decoration={amplitude=1.5,segment length=7}] (0.0+\sx-\s,-\sy) to [out=45,in=145, relative]  (2.0+0.0+\sx-\s,-\sy);
    \draw[draw=black, thick, snake it, decoration={amplitude=1.5,segment length=7}] (0.0+\sx-\s,-\sy) to [out=-45,in=-145, relative]  (2.0+\sx-\s,-\sy);
        \filldraw[black] (2.0+\sx-\s,-\sy) circle (2pt) node[anchor=north west,font=\large] {$g_B$};
        \draw[black,line width=1mm] (2.0+\sx-\s,-\sy) -- (3.5+\sx-\s,-\sy);
        \draw [arrows={-Latex[length=6.5pt]}] (1.0+\sx-0.2,-\sy+0.43) --(1.15+\sx-0.2,-\sy+0.44);
    \draw [arrows={-Latex[length=6.5pt]}] (1.0+\sx-0.2,-\sy-0.42) --(0.85+\sx-0.2,-\sy-0.41);
\end{tikzpicture}}
  \caption{ 
  Fermionic self-energy to next-to-leading order in $1/N$. Fermions correspond to straight lines, and complex bosons correspond to wavy lines. The dotted lines denote the fermion-boson interaction vertex $g$ as shown in ~\eqref{eq:fermi-boson_int}. The second line depicts the Dyson equation for the bosonic density fluctuations to leading order in $1/N$; in particular, there is no suppression in $1/M$.}
  \label{fig:Dyson}
\end{figure}
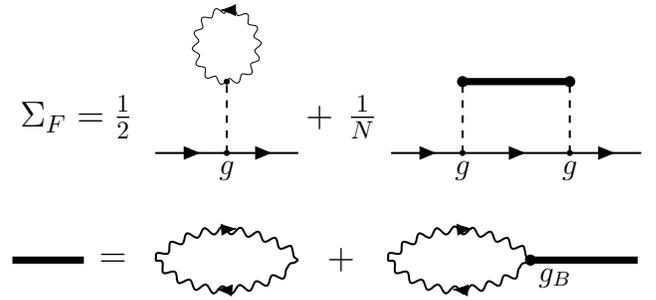
\begin{figure}
   \input{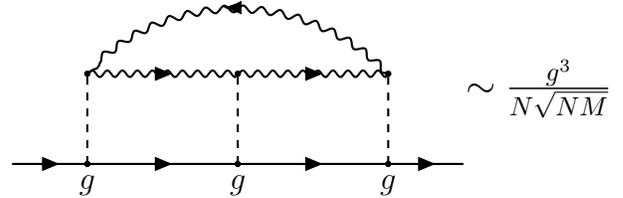}
   \caption{ All contributions to the fermionic self-energy other than those caused by hydrodynamic fluctuations are suppressed in the large $M$ limit. Here we show the leading omitted correction to the fermion self-energy from the interaction term~\eqref{eq:fermi-boson_int}, which scales as $M^{-1/2}$.}
   \label{fig:Extension}
\end{figure}
Diagrammatically in Fig.~\ref{fig:Dyson}, the first diagram corresponds to an irrelevant constant energy shift, while the second is typically associated with thermalization. All further contributions to $\Sigma_F$ are at most $\order{M^{-1/2}}$; the leading neglected diagram appears in Fig.~\ref{fig:Extension}. These are parametrically small at large $M$ and can be safely neglected.

The second line of Fig.~\ref{fig:Dyson} represents the propagator of the bosonic density,
\begin{equation}\label{eq:density_bosonic}
    \hat{n}(\vecx) = \frac{1}{M}\sum_i^M\hat\varphi_i^\dag(\vecx)\hat\varphi_i(\vecx)\,.
\end{equation}
While bosonic effects on fermions are negligible at large $M$, the Dyson equation for density fluctuations (Fig.~\ref{fig:Dyson}) involves no $1/M$ suppression. This necessitates a non-perturbative treatment and gives rise to a collective mode whose long-wavelength behavior is fully determined by symmetry.

Since the coupled system conserves the total number of bosons while (strong) translation invariance is explicitly broken by the dissipative fermionic lattice model, the hydrodynamic density fluctuations are described by gapless diffusive real bosons at temperature $T$\cite{Liu:2018Uz},\footnote{In the absence of dissipation and/or presence of strong translation invariance, the collective excitations of the bosons are instead described by propagating sound modes.}. The effective action of the coupled system is therefore given by
\begin{align}\label{eq:action}
    S=S_F+S_B+S_Y,
\end{align}
with the bare fermionic action $S_F$ stated in Eq.~\eqref{eq:bare_action}.
The fermions interact with the bosonic density fluctuations
{\small
\begin{align}\label{eq:Hydro_action}
	S_B=\frac{1}{2} \int_{\vecx,t}  \phi^\top(\vecx,t)\mqty(0 & -\partial_t - D\nabla^2\\ \partial_t - D\nabla^2 & -2 \mi T D \nabla^2) \phi(\vecx,t)\,,
\end{align}}
via Yukawa interactions
\begin{align}\label{eq:Yukawa_action_1}
	S_Y=\frac{g}{\sqrt{N}}\!\int_{\mathbf{x},t}\!\! \left[\bar{\psi}^q_i(\mathbf{x},t)\psi^c_i(\mathbf{x},t)\!+\bar{\psi}^c_i(\mathbf{x},t)\psi^q_i(\mathbf{x},t)\!\right]\!n_c(\mathbf{x},t).
\end{align}
In Eq.~\eqref{eq:Hydro_action}, $D\in\mathbb{R}^+$ is the diffusion constant, and we have introduced $\phi(\vecx,t)=(n_c(\vecx,t),\theta_q(\vecx,t))^\top$. All interactions involving $\theta_q$ are RG irrelevant, as will be shown in Sec.~\ref{sec:RG}, and have thus been omitted here.

The action~\eqref{eq:action} together with Eqs.~\eqref{eq:bare_action},\eqref{eq:Hydro_action},\eqref{eq:Yukawa_action_1} constitutes the interacting system to be analyzed in the following. In Sec.~\ref{sec:FDS}, we show that the coupling between dark state fermions and a real diffusive mode at finite temperature preserves the purity of the fermions, and in Sec.~
\ref{sec:RG} that it induces a novel interacting fixed point.

\section{Symmetry analysis and effective action}\label{sec:symmetries}

Microscopic models provide valuable intuition for the concrete realization of critical phenomena. Yet, universal properties are fixed by symmetry and dimensionality of the long-wavelength action rather than microscopic details. Symmetries determine the allowed perturbations and thereby the fixed points and relevant directions of the renormalization flow. In light of this, we now adopt a complementary viewpoint to Sec.~\ref{sec:MicroModel} by shifting from explicit microscopic constructions to a symmetry–based field–theoretic perspective. Our goal is to identify the unitary and anti-unitary symmetries that constrain the long–wavelength action governing pure–to–pure state transitions in Class AIII topological insulators. The long-wavelength critical theory is then constructed based on a gradient expansion, governed by symmetry principles alone.

The effective action can be constructed from the fundamental symmetries of the Keldysh action: probability conservation, Hermiticity, and chiral symmetry of the underlying Lindblad generator. In addition, we uncover a weak unitary symmetry associated with the $\mathrm{Pin}(d)$ group, arising in our near–critical Dirac stationary state. We show how this symmetry, together with chiral symmetry, restricts the action and fixes the number of relevant directions that must be tuned to reach criticality.

Beyond these microscopic symmetries, the long–wavelength theory can also display emergent symmetries, not explicit microscopically but arising  at long wavelengths after coarse graining. In our case, the dark–state condition enforcing purity of the stationary state yields a new emergent symmetry, which we call \textit{fermionic dark state symmetry}, ensuring the purity of the steady state at criticality.

Equations~\eqref{eq:hermiticity},~\eqref{eq:prob_consv},~\eqref{eq:chs} and ~\eqref{eq:Pin_spinorfields} thus assemble the complete symmetry framework needed for the construction of an effective field theory for pure-to-pure state transitions in AIII insulators. More specifically, it will allow us to systematically construct all possible relevant perturbations in the vicinity of the Gaussian fixed point.

\subsection{Symmetries of the microscopic model}\label{sec:micro-sym}

(i) \textit{Hermiticity ---} A Hermitian density matrix requires the Lindbladian to satisfy $\hat{\mathcal{L}}[\hat{\rho}]^\dag \overset{!}{=}\hat{\mathcal{L}}[\hat{\rho}]$. In the path integral formulation, this implies that under 
conjugation of the fields, defined by
\begin{align}\label{eq:hermiticity}
\begin{split}
\mathcal{C}:\psi^\sigma(\vecx,t)&\to\bar{\psi}^{-\sigma}(\vecx,t)\,,\; \bar{\psi}^\sigma(\vecx,t)\to-\psi^{-\sigma}(\vecx,t)\,,\\
\phi^\sigma(\vecx,t)&\to{\phi^*}^{-\sigma}(\vecx,t)\,, \;{\phi^*}^\sigma(\vecx,t)\to\phi^{-\sigma}(\vecx,t)\,,\\
\mi&\to-\mi\,,
\end{split}
\end{align}
the Keldysh action satisfies 
\begin{align}
    &\mathcal{C}(S) = -S.
\end{align}
This enforces the matrices $H$, $D$, and $P$ to be Hermitian
\begin{align}\label{eq:D,P_Hermiticity}
    H(\vecq)^\dag=H(\vecq),\quad D(\vecq)^\dag = D(\vecq),&\quad P(\vecq)^\dag = P(\vecq)\,,
\end{align}
consistent with Eq.~\eqref{eq:matrices}.

(ii) \textit{Probability conservation ---} The trace-preserving nature of the Lindbladian, $\partial_t\Tr(\hat\rho)=\Tr(\hat{\mathcal{L}}(\hat{\rho})) = 0$, implies probability conservation. In the path integral formulation, this corresponds to the constraint~\cite{kamenev_book}
\begin{align}\label{eq:prob_consv}
    &S[\Phi_c,\overline{\Phi}_c,\Phi_q=0,\overline{\Phi}_q=0] = 0\,.
\end{align}
Here $\Phi_c,\Phi_q$ collect all field variables, including bosons. Probability conservation ensures that the classical-classical $(c-c)$ component in the bare action~\eqref{eq:bare_action}, and therefore all correlation functions containing only $q$ fields, vanish~\cite{kamenev_book}. The related property of causality implies that the matrix $D$ is positive definite, such that the poles of the retarded Green's function lie in the lower half of the complex plane.

(iii) \textit{Chiral symmetry ---} An important constraint on the number of independent parameters of the low-energy theory follows from the chiral symmetry, cf. Eq.~\eqref{eq:chiral_transformation}. 
In the field theory representation, the anti-unitary chiral symmetry transformation $\mathcal{S}$ acts trivially in Keldysh space and as 
$\gamma_{d+2}$ in band space (cf. Eq.~\eqref{eq:chiral_sym}), {\it i.e.\/}, 
\begin{align}\label{eq:chs}
    \mathcal{S}:\quad \psi(\vecx,t)&\rightarrow \left(\mathbb{1}_2\otimes \gamma_{d+2}\right)\bar{\psi}(\vecx,t), \quad
    \mi\rightarrow -\mi.  
\end{align}

The action is invariant under $\mathcal{S}$ if and only if $N_U = N_L$ (cf. Sec.~\ref{sec:MicroModel}) and the matrices $H, D,P$ satisfy 
\begin{align}\label{eq:chsimp}
\begin{split}
    &\gamma_{d+2} H(\vecq) \gamma_{d+2} = + H(\vecq), \\
    &\gamma_{d+2} D(\vecq) \gamma_{d+2} = + D(\vecq), \\
    &\gamma_{d+2} P(\vecq) \gamma_{d+2} = - P(\vecq). 
\end{split}
\end{align}
By construction, the quadratic fermionic action ~\eqref{eq:bare_action} is symmetric under chiral transformations. This follows from the fact that $\gamma_{d+2}$ anti-commutes with $W(\vecx)$ and $\gamma_{d+1}$
\footnote{Since the product of the generators $\gamma_{d+2}\sim \mathbb{1}$ is trivial in odd dimensions, the quadratic action in odd dimensions can also be interpreted as a representative of near-critical topological insulators in Class A.}.

(iv) \textit{Fate of strong $U(1)$ symmetry for fermions ---}
As mentioned in Sec.~\ref{sec:diffusivemodes}, Class AIII topological insulators do not exhibit density fluctuations due to chiral symmetry, which locks the local particle density to $1/2$; this prohibits hydrodynamic density fluctuations. The precise mechanism is explored in  App.~\ref{app:Hydro_mode}, where we perform a diagrammatic fluctuation analysis of the microscopic action~\eqref{eq:micro_action} around the mean field solution. By a suitable parameterization of the collective modes, we derive a low-energy effective theory for the hydrodynamic mode associated with the strong $U(1)$ symmetry. We find that the hydrodynamic action features a vanishing noise coefficient, implying that all density fluctuations are "frozen" out (analogously to a thermal bath at zero temperature). Consequently, for $d>2$, any coupling of fermion bilinears to the local conserved density $\sim \bar\psi_c^\top \psi_c$, is irrelevant near the quantum critical point, and it suffices to treat the fermions in a mean field description in the remainder of the text.

(v) \textit{Weak Pin$(d)$ symmetry and constraints of the long wavelength theory ---}  
The Lindbladian~\eqref{eq:Lindblad} possesses a weak symmetry under the action of a symmetry group $G$ if and only if it commutes with the symmetry generator in the following sense~\cite{Sieberer2023},\cite{deGroot2022}
\begin{equation}\label{eq:weaksymm}
    \mathcal{L}(\hat U_g^\dag \rho \hat U_g) \overset{!}{=} \hat U_g^\dag \mathcal{L}(\rho) \hat U_g\,,\ \forall g\in G\,,
\end{equation}
where $\hat U_g$ is a unitary operator implementing the action of the symmetry group on the Fock space. 
To construct Lindbladians with weak symmetries, it is useful to reformulate the above statement in terms of the jump operators $L_{a}$, with the multi-index $a=(\alpha,\beta)$ (cf. Eq.~\eqref{eq:jumpop1},Eq.~\eqref{eq:jumpop2}). The Lindbladian satisfies the weak symmetry condition if and only if the jump operators are charged under the action of the symmetry group, \textit{i.e.} \cite{deGroot2022},
\begin{equation}\label{eq:weaksymm_jump}
    \hat{U}_g \hat{L}_{} \hat U_g^\dag = \sum_{b} R(g)_{a}\hat{L}_{b}\,,
\end{equation}
where $R(g)$ is a unitary representation of the group $G$. For Abelian groups, $R(g) = \exp(\mi\alpha(g))$ is one-dimensional, while for non-Abelian symmetry groups, $R(g)$ is given by a rank-$4$ tensor. For the operators $x_\alpha$ which transform as a singlet under the symmetry group, $R(g)$ is a unitary matrix.

If the jump operators carry a trivial charge, \textit{i.e.}, $R(g)_{\alpha\beta}=\delta_{\alpha\beta}$, $\forall g\in G$, the Lindbladian is said to have a \textit{strong symmetry}. The Hamiltonian term $\hat H$ is invariant under $\hat U_g$ if it satisfies  $\hat U_g^\dag \hat H \hat U_g = \hat H \iff [\hat H,\hat U_g] = \hat H$ -- unlike the jump operators $\hat L_\alpha$, there is no distinction between strong and weak symmetries for $\hat H$.

In the field theory representation, weak symmetries correspond to the invariance of the Keldysh path integral under symmetry transformations acting identically on both contours~\cite{Sieberer2023}
\begin{equation}
    \psi^\pm (\vecx) \rightarrow \mathcal{U}(g)\psi^\pm(\Lambda(g)\vecx) \quad \bar\psi^\pm(\vecx) \rightarrow \mathcal{U}^*(g)\bar\psi^\pm(\Lambda(g)\vecx)\,.
\end{equation}
Here we allow for both internal and external transformations: $\mathcal{U}(g)$ is a unitary matrix acting on the fields $\psi^\pm$, and $\Lambda(g)$ is a representation of the symmetry group acting on the base manifold. 

Indeed our problem features a weak symmetry of this type: It is invariant under the $\mathrm{Pin}(d)$ group. This group makes its appearance in Dirac theory~\cite{Zinn-Justin_book}, and its occurrence here is rationalized by the fact that the dark states considered also are ground states of Dirac Hamiltonians in $d$ dimensions. Relegating a brief review of the $\mathrm{Pin}(d)$ group to App.~\ref{app:weakPind1}, here we specify the action of $\mathrm{Pin}(d)$ on the $\psi^\pm$ spinors 
\begin{align}\label{eq:Pin_spinorfields}
    \begin{split}
    \psi^\pm(\vecx,t) &\to \exp(\frac{1}{4}\sum_{i<j}^d\alpha_{ij}\sigma_{ij})\psi^\pm(R\vecx,t),  \\
    \psi^\pm(\vecx,t) &\to \mi\gamma_{d+2}\gamma_i\ \psi^\pm(\dots,-x_i,\dots,t).
    \end{split}
\end{align}
Here $\sigma_{ij} = \frac{1}{2}[\gamma_i,\gamma_j]\,, i<j=1,\dots,d$ are the generators of the spin algebra in $d$ dimensions, $R\in \mathrm{SO}(d)$ is a rotation matrix and $\alpha_{ij}=-\alpha_{ij}$ is a real anti-symmetric matrix. The continuous transformations represent the $\mathrm{Spin}(d)$ group, together with the spatial reflections they form the $\mathrm{Pin}(d)$ group. 

In App.~\ref{app:weakPind2} we show that the continuum model discussed in Sec.~\ref{sec:dynamical_MF}, with almost unitary matrices given by Eq.~\eqref{eq:unitary_dirac}, is weak $\mathrm{Pin}(d)$ invariant. Therefore, this symmetry also constrains the full effective action. From the combination of chiral and weak $\mathrm{Pin}(d)$ symmetry, one finds the form (cf. App.~\ref{app:weak_symm_HDP} for a  derivation) 
\begin{align}\label{eq:FDS_matrices}
    \begin{split}
        D(\vecq) &= (\Delta_d + K_d\vecq^2)\mathbb{1} + \mi d_{d+1}\sum_i^d q_i\gamma_{d+1}\gamma_i \, ,\\
        H(\vecq) &= (\Delta_c + K_c\vecq^2)\mathbb{1} + \mi h_{d+1}\sum_i^d q_i\gamma_{d+1}\gamma_i\, ,\\
        P(\vecq) &= (\Delta_p + K_p\vecq^2)\gamma_{d+1} + p_d\sum_{i=1}^d q_i\gamma_i\,,
    \end{split}
\end{align}
with real-valued coefficients~\footnote{Note that the presence of the $d_{d+1}$ coefficient, while compatible with the symmetries of the microscopic model, renders $D$ non–positive semi-definite over a certain momentum range $q$, and can be set to zero by hand to ensure stability of the dark state.}.

These constraints systematically eliminate relevant directions that would otherwise require fine-tuning to reach the quantum critical point: Without any symmetry, Hermiticity alone allows for $3N^2$ independent relevant directions. $\mathrm{Spin}(d)$ invariance restricts these matrices to combinations of (pseudo-)scalar and (pseudo-)vector terms, reducing this number to $3 \times 6$. Imposing full $\mathrm{Pin}(d)$ symmetry eliminates pseudo-scalars and pseudo-vectors, yielding $3 \times 4$ directions. Finally, chiral symmetry further reduces this to just $6$  relevant directions. Not all relevant directions are linearly independent: For example, one can rotate to a basis where the $P$ matrix is diagonal reduces the number of independent directions by 1, which yields $6-1=5$ independent relevant directions.

However, with $5$ relevant parameters that need to be tuned to reach the critical point, the latter would still require tremendous fine-tuning. However, rephrasing the purity of the dark state as an additional emergent symmetry, in the next section, we find that the topological pure-to-pure state transition has only a single relevant direction, consistent with the microscopic model, which is tuned to criticality using $\theta$.

We emphasize a key technical outcome of this section. The symmetry principles identified here—most notably the weak (S)Pin$(d)$ symmetry together with the fundamental symmetries of the Keldysh action—provide a systematic framework for constructing mesoscopic field theories for pure–to–pure state transitions based solely on symmetry considerations. Following an approach analogous to App.~\ref{app:weak_symm_HDP}, this framework enables a controlled enumeration of all relevant perturbations beyond the single–particle description.

The construction developed here applies more broadly, beyond the microscopic models of Sec.~\ref{sec:MicroModel}. Rather than depending on a particular choice of jump operators or on detailed properties of the dark states across the transition, it captures the universal structure near criticality. From this perspective, the microscopic model simply serves to provide a physical intuition for the pure to pure state transition, the associated single particle gaps, and the underlying bulk phases. The topology of the underlying phases enters only indirectly: while the transition may be topological in origin, the mesoscopic action focuses on the critical region and need not encode the bulk topological properties. We will return to this point in Sec.~\ref{sec:Discussion}, where we discuss the relevance of topological terms for the bulk critical phenomena. Finally, extensions to other symmetry classes follow straightforwardly by selecting different combinations of Fock–space symmetries: Time-reversal($T$), charge-conjugation $(C)$, and/or chiral $(S)$ symmetry~\cite{Altland1997,Altland2021}.

\subsection{Fermionic dark state symmetry}\label{sec:FDS}

We now introduce a new symmetry principle that ensures the purity of fermionic stationary states—the \textit{fermionic dark state symmetry} (FDS). Heuristically derived from the fermionic dark state condition (FDC) for Gaussian states, the FDS can then be imposed at the level of the interacting action, and we construct explicit examples. In particular, the Yukawa term Eq.~\eqref{eq:Yukawa_action_1} respects the FDS. If present on the level of the bare, interacting action, it constrains the full renormalized action, much like a Ward identity. In particular, it guarantees the purity of the fermion state, in the sense $\Gamma(\mathbf{q})^2 = \mathbb{1}$ for the full covariance matrix. 

\subsubsection{Gaussian theory}\label{sec:Gaussianth}
For the derivation of the statement above, recall the FDC Eq.~\eqref{eq:FDC}, which implies the existence of a basis in which $H$, $D$, and $P$ are simultaneously diagonal. Rotating the fermion fields as $\psi \to \chi = U\psi$, we identify the FDS as the symmetry of the action under
\begin{align}\label{eq:dark_state_symmetry}
\begin{split}
    \mathcal{T}_F: \quad \chi(\vecx,t)&\rightarrow T_F \chi(\vecx,-t), \\ 
     \quad \bar{\chi}(\vecx,t)&\rightarrow \bar T_F \bar{\chi}(\vecx,-t), \quad \mi\rightarrow-\mi,
\end{split}
    \end{align}
where the transformation matrices $T_F$ and $\bar T_F$ act in contour and band space as 
\begin{align}\label{eq:dark_state_symmetry_matrix}
    T_F = \mqty(-\mathbb{1} & 0 \\ F & \mathbb{1})\,,\quad \bar T_F = \mqty(-\mathbb{1} & 0 \\ -F & \mathbb{1})\,. 
\end{align}
In the Gaussian problem, $F_{ij} = \text{sign} (\pi_i) \delta_{ij}$, with $\pi_i$ the eigenvalues of $P$. Thus, $F$ is Hermitian and involutive: $F = F^\dagger$, $F^2 = \mathbb{1}$. 
The FDS is an anti-unitary symmetry that squares to the identity, $\mathcal T_F^2 =1$. Note the relative minus sign in the off-diagonal sectors: The symmetry  relies on the fact that the Grassmann variables $\chi,\bar\chi$ are independent and not related to each other by complex conjugation, as is the case for bosonic fields. 

To connect to the FDC (cf. Sec.~\ref{sec:dynamical_MF}), we consider a general Gaussian action in the diagonal $\chi$ basis. Applying the symmetry transformation, one finds that the invariance of the Gaussian action requires 
\begin{equation}
    \begin{split}
        2P=\{F^{-1},D\}+\mi[F^{-1},H], \\ 
        F^{-1}HF=H\,,\ F^{-1} DF=D.
    \end{split}
\end{equation}
Purity follows from $F^2=\mathbb{1}$ if $[F,H]=[F,D]=0 \implies [D,P]=0$, which imposes the simultaneous diagonalizability of $F, H$, and $D$ and consequently of $P$. 

One can now lift the above discussion in the diagonal basis to an arbitrary basis, by inserting identities in the form of unitary transformations $U$ in the action \text{before} applying the complex conjugation. 
Consequently, in the new basis, we find that the Gaussian action 
\begin{equation}
    S_0 =\mi\int_{t} \bar{\psi}(t)\mqty(0 & \mi\partial_t - \tilde{H}-\mi \tilde{D}\\ \mi\partial_t - \tilde{H} + \mi \tilde{D} & 2\mi \tilde{P}) \psi(t)\,,
\end{equation}
with the matrices $\tilde{H}=UHU^\dag,\tilde{D}=UDU^\dag$ and $\tilde{P}=UPU^\dag$, is invariant under $\tilde{T}_F=U T_F U^\dagger$ and $\tilde{\bar{T}}_F=U^* \bar{T}_F U^\top$ followed by complex conjugation. 

Consequently, we have shown that a Gaussian action with a well-defined band index describes a pure state if it is invariant under \eqref{eq:dark_state_symmetry} and $F=D^{-1}P$ is a diagonal involution.

FDS has two central implications. We state them here, in the context of the Gaussian theory, but demonstrate below that they also hold for interacting systems equipped with FDS.\\
(a) Non-thermal fluctuation-dissipation relations. FDS imposes a rigid relation between correlation and response functions, analogous to fluctuation-dissipation relations in equilibrium systems. Specifically, the fermion two-point functions satisfy
\begin{align}
    [G^R,F] =[G^A,F]=0
\end{align}
and therefore
\begin{align}\label{eq:ds_gk}
    G^K (\omega
    ) = F(G^R (\omega) -G^A (\omega))\,,
\end{align}
where we omit all indices on the matrices (band and momentum). At equal times, $G^R(t,t) - G^A(t,t) = -\mi\mathbb{1}$, see Eq.~\eqref{eq:operatorgreen}, implying that the fermionic covariance matrix $\Gamma = G^K(t,t)$ obeys $\Gamma^2 = \mathbb{1}$: the fermion state is pure.

\noindent (b) FDS constrains the critical theory. Specifically for the model discussed here, FDS together with the weak $\mathrm{Pin}(d)$ symmetry~\eqref{eq:Pin_spinorfields} and chiral symmetry~\eqref{eq:chs}, excludes all but one tunable single-particle gap. To show this, we start from Eq.~\eqref{eq:ds_gk} and the parametrization $G^K = -G^R P^K G^A$, where again $P^K$ is the noise matrix, and $P^{R,A} = \left[G^{R/A}\right]^{-1}$. Inversion gives
\begin{equation}\label{eq:invfdr}
    P^K(\omega) = F (P^R(\omega) - P^A(\omega))\,,
\end{equation}
with $P^{R/A/K}$ as in Eq.~\eqref{eq:bare_action} determined by the matrices $H, D$ and $P$. These are constrained by the microscopic symmetries to the form~\eqref{eq:FDS_matrices}.
Rotating to the basis $\chi = U^\dag \psi$, where $P\sim\tau_z$ is diagonal and $H\rightarrow UHU^\dag$ and $D\rightarrow UDU^\dag$ take a similar form to Eq.~\eqref{eq:FDS_matrices} up to pre-factors, FDS then requires that $U[D,P]U^\dag=0$, which implies
\begin{equation}
     \mi d_{d+1}\left(p_d \vecq^2 \gamma_{d+1} - (p_{1} + p_{2}\vecq^2)\sum_{i=1}^d q_i\gamma_i \right) = 0\,.
\end{equation}
Since $\gamma_i\ \forall\ i=1,\dots,d$ and $\gamma_{d+1}$ are linearly independent, for arbitrary values of $p_1,p_2$ and $p_d$, the only consistent solution with $P(\mathbf{q})\neq 0$ is given by $d_{d+1} = 0$. Thus, the most general mass matrix (up to unitary transformations) reads $D(\omega =\vecq=0) =\Delta_d \mathbb{1}$, leaving a single dissipative gap scale $\Delta_d$ to tune to the critical point. Eq.~\eqref{eq:invfdr} further implies that $P^K(\omega =\vecq=0)$ must share this scale which locks $\Delta_d=\Delta_p$. Similarly, for the Hamiltonian term, the condition $[F,H]=0$ requires $h_{d+1}=0$.
Hence, the Hamiltonian mass merely defines a rotating frame and does not affect tuning to criticality. Consequently, only a single fine-tuning $\Delta_d \to 0$ is required to realize criticality, protected by the combination of chiral symmetry, weak $\mathrm{Pin}(d)$ symmetry, and FDS. 

Summarizing, the most general form of $H, D$ and $P$ matrices, compatible with the microscopic symmetries and FDS, reads:
\begin{align}
    \begin{split}
        D(\vecq) &= (\Delta_d + K_d\vecq^2)\mathbb{1}\, ,\\
        H(\vecq) &= (\Delta_c + K_c\vecq^2)\mathbb{1}\, ,\\
        P(\vecq) &= D(\vecq)\tau_z\, ,\\
    \end{split}
\end{align}
where we have used that in the basis of the topological fields $\chi$, chiral symmetry requires $P\sim\tau_z=\sigma_z\otimes\mathbb{1}_{N/2}$. The prefactor follows from observing that a consequence of the FDS conditions is the locking $P^2 = D^2$ (since $F^2 = \mathbb{1}$).

\subsubsection{Interacting theories}

Now we construct interaction terms that are invariant under FDS. To focus on the leading contributions in a derivative expansion, we will rely here on the canonical power counting introduced in Sec.~\ref{sec:treelevel} below. Following the discussion of the most important examples, we provide the general framework and interpretation for the FDS of an interacting theory.

\paragraph*{Yukawa interaction ---}
We remind the reader that this vertex was already introduced in Sec.~\ref{sec:diffusivemodes}, where we noted that a minimal interaction term of the conserved density field with spinors takes the form of a Yukawa coupling~\eqref{eq:Yukawa_action_1}. To emphasize its compliance with the symmetry principles identified above, we repeat the construction of the Yukawa vertex using only the microscopic symmetries (i)-(v), and FDS.

Retaining only relevant terms to leading order in a derivative expansion, the interactions take the form of a Yukawa term
\begin{align}\label{eq:Yukawa_action_general}
	S_Y[\chi,\bar{\chi},n_c] =\!\int_{\vecx,t} n_c(\vecx,t) \bar{\chi}^\top(\vecx,t) \mqty(0 & A\\ A^\dag & 2\mi B )\chi(\vecx,t)\,,
\end{align}
which is constrained only by Hermiticity (i) and probability conservation (ii), which requires $B=B^\dag$. To proceed, we need to first distinguish the transformation behavior of $n_c$ under chiral, $\mathrm{Pin}(d)$, and FDS symmetry. We distinguish two cases 
\begin{itemize}
    \item[(a)] $n_c$ is a conserved density given by Eq.~\eqref{eq:density_bosonic}, associated to an independent bosonic degree of freedom, coupled to fermionic modes as shown in Sec.~\ref{sec:diffusivemodes}.
    \item[(b)] $n_c(\vecx,t)=\bar{\psi}_c(\vecx,t)\psi_c(\vecx,t)$ is the local fermionic density (cf. (iv)). 
\end{itemize} 
In (a), the density field $n_c$ is invariant under chiral symmetry (cf. (iii)) and transforms as a scalar under weak $\mathrm{Pin}(d)$ symmetry (cf. (v)). Under FDS, the density fluctuations $\phi$ (cf. Eq.~\eqref{eq:Hydro_action}) transform as $\phi(\mathbf{x},t)\to\phi(\mathbf{x},-t)$. Thus, the matrix structure for the Yukawa coupling is identical to the structure of the inverse bare Green's function at zero momentum and frequency (cf. Eq.~\eqref{eq:bare_propagators}) 
\begin{align}\label{eq:Yukawa_action}
	S_Y[\chi,\bar{\chi},n_c] =\!\int_{\vecx,t} n_c(\vecx,t) \bar{\chi}^\top(\vecx,t) \mqty(0 & g\mathbb{1}\\ g^*\mathbb{1} & 2\mi g_d\tau_z )\chi(\vecx,t)\,,
\end{align}
where the Yukawa couplings $g\in\mathbb{C}$ and $g_d=-\Im(g)\in\mathbb{R}$. In particular, the locking of the dissipative coupling $g_d$ to $\Im(g)$ is a consequence of the FDS. Moreover, since the noise field $\theta_q(\vecx,t)$ can only enter the action through its derivatives, as a consequence of the conservation law, only the classical field $n_c(\vecx,t)$ couples to fermions. More precisely, a scaling analysis (cf. Sec.~\ref{sec:treelevel}) shows that interactions with $\theta_q(\vecx,t)$ are irrelevant in all spatial dimensions. 

For case (b), however, the composite density field has odd parity under chiral symmetry,  $n_c\rightarrow -n_c$. Formally, a Yukawa interaction term of the form~\eqref{eq:Yukawa_action} obtains by a substitution $\mathbb{1}\rightarrow\tau_z$ in the retarded/advanced sectors and $\tau_z\rightarrow\mathbb{1}$ in the Keldysh sector.
However, as discussed in (iv), the fermionic local density is pinned to $1/2$ as a consequence of chiral symmetry. Hence, all couplings of fermionic bilinears with the density mode $n_c$ cannot be generated near the critical point.

\paragraph*{Quartic fermion terms ---}
We present examples of quartic fermion interaction vertices compatible with $U_q(1)$ charge conservation, weak Pin symmetry, chiral symmetry, and FDS in 2 and 3 spatial dimensions. These terms are irrelevant near the upper critical dimension and can therefore be neglected in the upcoming RG analysis, but serve to illustrate the constraints resulting from FDS. We focus only on vertices without any derivative couplings.

In $2+1$ dimensions,  quartic interactions are marginal. In the fundamental representation ($j=1/2$) of the spin group Spin($3$) $\simeq \mathrm{SU}(2)$, the spinor fields are given by $\chi^\sigma(\vecx,t)=(\chi^\sigma_\uparrow(\vecx,t),\chi^\sigma_\downarrow(\vecx,t))^\top$, with $\sigma$ denoting the contour index. Quartic interactions that satisfy FDS include density-density and current-current terms such as 
\begin{align}
    \begin{split} 
    &\text{density-density:}\\
    &\mi\left(\bar\chi(\vecx,t)^\top(\sigma^z\otimes\mathbb{1}_2)\chi(\vecx,t)\right)\left(\bar\chi(\vecx,t)^\top(\sigma^z\otimes\mathbb{1}_2)\chi(\vecx,t)\right)\,,\\
    &\text{current-current }(i,j=1,2):\\
    &\begin{cases}
    \mi\left(\bar\chi(\vecx,t)^\top(\sigma^y\otimes\gamma_i)\chi(\vecx,t)\right)\left(\bar\chi(\vecx,t)^\top(\sigma^y\otimes\gamma_i)\chi(\vecx,t)\right)\,,\\
    \left(\bar\chi(\vecx,t)^\top(\sigma^y\otimes\gamma_i)\chi(\vecx,t)\right)\left(\bar\chi(\vecx,t)^\top(\sigma^y\otimes\gamma_j)\chi(\vecx,t)\right)\, \\
    \end{cases}
    \end{split}
\end{align}
($i\neq j$), where $\chi = (\chi^+,\chi^-)^\top$ and $\bar \chi =(\bar \chi^+,\bar \chi^-)^\top $ with $\sigma^{z,y}$ acting on the contour space and $\gamma_i$ on the band space. 

In $3+1$ dimensions, due to spin group Spin($4$)$\simeq$ SU($2$)$\times$ SU($2$), the spinor representation corresponds to the $(1/2,0)\oplus(0,1/2)$ representation of SU($2$)$\times$ SU($2$). Concretely, the 4 component spinor fields can be written as $\chi^\sigma=(\chi^\sigma_U,\chi^\sigma_L)$, where $\chi_{U/L}$ are two component spinor fields $\chi_{U/R}^\sigma=(\chi^\sigma_{U/R,\uparrow},\chi^\sigma_{U/R,\downarrow})^\top$. With these, we can write, for example
\begin{align}
    \begin{split} 
    &\text{density-density:}\\
    &\mi\left(\bar\chi(\vecx,t)^\top(\sigma^z\otimes\mathbb{1}_4)\chi(\vecx,t)\right)\left(\bar\chi(\vecx,t)^\top(\sigma^z\otimes\mathbb{1}_4)\chi(\vecx,t)\right)\,,\\
    &\text{current-current }(i=1,2,3):\\
    &\begin{cases}
    \mi\left(\bar\chi(\vecx,t)^\top(\sigma^y\otimes\gamma_i)\chi(\vecx,t)\right)\left(\bar\chi(\vecx,t)^\top(\sigma^y\otimes\gamma_i)\chi(\vecx,t)\right)\,,\\
    \mi\left(\bar\chi(\vecx,t)^\top(\sigma^y\otimes\gamma_5)\chi(\vecx,t)\right)\left(\bar\chi(\vecx,t)^\top(\sigma^y\otimes\gamma_5)\chi(\vecx,t)\right)\,,\\
    \end{cases}
    \end{split}
\end{align}
where in the last line we have exploited the existence of a non-trivial chirality matrix $\gamma_5= \prod_{j=1}^{4} \gamma_j$ in even dimensions to construct a pseudo-scalar current-current interaction.

\paragraph*{General framework ---} For interacting theories, we promote the simultaneous diagonalizability of $H, D$, and $P$ to the existence of a well-defined band index quantum number that uniquely labels each state. A sufficient condition for the existence of a well-defined band index quantum number is the existence of some weak symmetry of the Lindbladian. This principle is general; we will exemplify how it is realized in our specific problem.

We start by briefly reminding the reader of the connection between unitary symmetries and quantum numbers in the context of closed quantum systems. A Hamiltonian $\hat H$ is said to have a symmetry if and only if it commutes with some symmetry generator(s) $\hat U$. For unitary (abelian) symmetries, this leads to a symmetry decomposition of the Hilbert space $\mathcal{H} = \bigoplus_{j} \mathcal{H}^{(j)}$ into irreducible subspaces, labeled by the distinct eigenvalues (or charges) $j$ of the symmetry operator $\hat U$, and the Hamiltonian acting irreducibly within each block $\mathcal{H}^{(j)}$. In the basis which diagonalizes $\hat H$ and $\hat U$ simultaneously, the energy eigenstates of $\hat H$ are labeled by the quantum number $j$. 
For non-abelian groups, $j$ labels the irreducible representations of the symmetry group $G$ on $\mathcal{H}$. The complete set of good quantum numbers $\mathcal{J}$ is defined by the eigenvalues (or the so-called weights of the representation) of the (maximal) mutually commuting set of symmetry generators. Hence, given a unitary symmetry of the Hamiltonian, all energy eigenstates $\ket{\mathcal{J}}$ of $\hat H$ are labeled by the set of quantum numbers $\mathcal{J}$.

In open quantum systems, the existence of weak symmetries plays a similar role and allows one to assign a well defined index/quantum number to the fields on the Keldysh contour. Consider the unitary super-operator $\mathcal{U}_g$, associated with the unitary operator $\hat U_g$, for some $g\in G$
\begin{align}
    \begin{split}
        \mathcal{U}_g:\ &\mathcal{H}\otimes\mathcal{H}^*\rightarrow \mathcal{H}\otimes\mathcal{H}^*,\\
        &\hat A \rightarrow\hat{U}_g \hat A \hat{U}_g^\dag,
    \end{split}
\end{align}
where $\mathcal{H}$ denotes the Fock space and $\mathcal{H}^*$ its dual space \textit{i.e.} $\mathcal{H}\otimes\mathcal{H}^*$ denotes the vector space of linear operators acting on the Fock space. The Lindbladian $\mathcal{L}$ transforms in the adjoint representation under the action of $\mathcal{U}_g$,
\begin{equation}
    \mathcal{L}(\rho)\rightarrow (\mathcal{U}_g\mathcal{L}\mathcal{U}_g^\dag)(\rho) = \hat U_g \mathcal{L}(\hat U_g^\dag\rho \hat U_g) \hat U_g^\dag,
\end{equation}
hence the weak symmetry condition~\eqref{eq:weaksymm} 
\begin{equation}
    \mathcal{U}_g \mathcal{L}\mathcal{U}_g^\dag = \mathcal{L} \implies [\mathcal{U}_g,\mathcal{L}] = 0
\end{equation}
is equivalent to the statement that the symmetry ($\mathcal{U}_g$) commutes with the generator of the dynamics ($\mathcal{L}$).

However, a crucial distinction from closed systems is that weak symmetries of the Lindbladian induce a decomposition of the operator space $\mathcal{H} \otimes \mathcal{H}^*$ into a direct sum of irreducible subspaces, each labeled by the eigenvalues of $\hat{U}_g \otimes \hat{U}_g^\dagger$ \cite{deGroot2022,Buca2012}. Within each subspace, the Lindbladian acts irreducibly.

For the present model, this structure is most transparent in the basis of the fields $\chi$ corresponding to the annihilation operators $\hat{x}$ of the topological dark state, where the Hermitian matrix $P \sim \tau_z$ is diagonal, and the transformed matrices $\sim U H U^\dag$ and $\sim U D U^\dag$, which govern the spectral properties, take a similar form as in Eq.~\eqref{eq:FDS_matrices} up to pre-factors. 

An emergent weak $U(1)$ symmetry appears in the long wavelength theory near the dark state, which in this basis is implemented as
\begin{align}
\mathcal{U}_z(\theta) = \exp(i \theta \tau_z)\,.
\end{align}
In the basis of $\psi = U^\dag\chi$ (cf. Eq.~\eqref{eq:coherent_sup_x}, Eq.~\eqref{eq:WfU}), the symmetry is specified via $\tau_z \to U^\dag \tau_zU$; regardless, it is a global symmetry generated by a single transformation parameter $\theta$. Invariance under this symmetry enforces $d_{d+1} = h_{d+1} = 0$, thereby ensuring the simultaneous diagonalizability of the retarded and Keldysh Green’s functions, $G^R$ and $G^K$, as anticipated (cf. discussion below Eq.~\eqref{eq:invfdr}).

More generally, in $d=3$, the microscopic symmetries decompose the local Hilbert space into two blocks corresponding to the fundamental representations of the two $SU(2)$ factors in Spin($4$). The mesoscopic action then exhibits an emergent weak global $U(1)$ symmetry, which fully diagonalizes these blocks into states with definite "magnetic" spin quantum number, labeled by the eigenvalues of $\tau_z$. In the field-theoretic language, the resulting decomposition of the operator space imposes that the renormalized matrices $H$, $D$, and $P$ remain simultaneously diagonalizable.

It is important to realize that the generality of this argument, relying only on symmetries, ensures that the  implications of FDS discussed above in Sec.~\ref{sec:Gaussianth} apply to the full action.
The presence of a weak symmetry ensures that the band index remains a good quantum number for the full renormalized action, enabling the decomposition of the operator space into one-dimensional blocks labeled by this index. In this basis, the FDS retains the form~\eqref{eq:dark_state_symmetry} with $T_F$, $\bar T_F$ given by Eq.~\eqref{eq:dark_state_symmetry_matrix}, and $F$ diagonal with eigenvalues $\pm 1$. In any other basis, $F$ is similar to a diagonal matrix with these eigenvalues. As a result, the non-thermal fluctuation-dissipation relation~\eqref{eq:ds_gk} continues to hold for the full interacting two-point functions, with the same interpretation of $F$ as a generalized distribution function.
Moreover, the symmetry constraints derived in the quadratic theory—such as the forms of $H$, $D$, and $P$ in Eq.~\eqref{eq:FDS_matrices}—apply identically to the full renormalized propagators. In particular, it reduces the number of independent RG directions. For the model introduced in Sec.~\ref{sec:dynamical_MF}, FDS as well as weak $\mathrm{Pin}(d)$ and chiral symmetry restrict the theory to a single relevant direction at the critical point, which is obtained by tuning the renormalized $\Delta_d\to 0$.

We emphasize the analogy with thermal equilibrium at zero temperature. In the present context, the FDS takes over the role of energy as the ordering principle, replacing it with the band index. As a result, the conservation of band index plays a role analogous to that of energy conservation in thermal equilibrium. Accordingly, the function $F$ assumes the role played by the zero-temperature distribution function $F_{\mathrm{th}}(\omega) = \text{sign}(\omega - \mu)$, with chemical potential $\mu \in \mathbb{R}$. In this sense, the FDS generalizes the notion of zero temperature to  finite-dimensional representations of arbitrary (including non-abelian) groups.

\textit{Summary ---} FDS consists of two components, an emergent weak global symmetry that ensures the simultaneous diagonalizability of the full renormalized correlation and response functions, and an anti-unitary symmetry that fixes the eigenvalues of the distribution function $F$. 

The low-energy effective action near the critical point, including all non-irrelevant terms just below $d_c=4$ spatial dimensions, consistent with symmetries (i)-(v) and FDS, is given by
\begin{align}\label{eq:effective_action}
    S= S_F + S_B + S_Y\,,
\end{align}
with $S_B$ and $S_Y$ defined in Eqs.~\eqref{eq:Hydro_action} and~\eqref{eq:Yukawa_action} as well as
\begin{align} \label{eq:fermion_greens_fct}
\begin{split}
    S_F[\chi,\bar{\chi}] &=\int_{\vecq,\omega}\bar{\chi}^\top(\omega,\vecq)G_F^{-1}\chi(\omega,\vecq)\,,\\
    G_F^{-1}&=\mqty(0 & Z\omega\mathbb{1}-K(\vecq)\\Z^*\omega\mathbb{1} -K^\dagger(\vecq) & 2\mi P(\vecq))\,,
\end{split}
\end{align}
where 
\begin{align} \label{eq:DH}
\begin{split}
    K(\vecq) &=H(\vecq)+\mi D(\vecq) \\&= \left[\left(\Delta_c +\mi  \Delta_d\right) + \left(K_c + \mi K_d\right)\vecq^2\right]\mathbb{1}\,, \\
    P(\vecq) &= D(\vecq)\tau_z\,.
\end{split}
\end{align}
This is a slight generalization compared to Eq.~\eqref{eq:matrices} since we accommodate for all terms that can be generated during the RG flow. In the above formulas, the Yukawa coupling $g$ and the fermionic wave function renormalization $Z$ are allowed to be complex; similarly, the matrix $K(\vecq)$ has Hermitian and anti-Hermitian contributions in general, with the Hermitian part corresponding to a Hamiltonian; its most general form is $H(\vecq)=\left(\Delta_c + K_c \vecq^2\right)\mathbb{1}$. All these couplings have to be taken into account in a systematic, symmetry-based RG analysis; the initial values for the RG flow are taken from the 'pre-renormalized' mesoscopic action Eq.~\eqref{eq:effective_action}, which we have derived from the microscopic model in Sec.~\ref{sec:MicroModel}. 

\section{Renormalization group analysis}\label{sec:RG}

\subsection{Tree level}\label{sec:treelevel}
As a first step, we discuss the tree-level RG scaling, or canonical power counting, filling the gap of the last section. Consider a rescaling of the coordinates $\vecx,t$ and fields by a parameter $b>1$ as follows~\cite{kamenev_book}
\begin{align}
\begin{split}
    &\vecx\rightarrow b\vecx\,,\quad t\rightarrow b^zt\,,\\
    &\psi_{c,q}\rightarrow b^{\xi_{c,q}}\psi_{c,q}\,,\quad n_c\rightarrow b^{\zeta_c}n_c\,,\quad \theta_q\rightarrow b^{\zeta_q}\theta_q\,,
\end{split}
\end{align}
where the exponents $\xi_{c,q},\zeta_{c,q}$ are the scaling dimensions of the fields and $z$ is the dynamical exponent. The bare scaling exponents and dimensions of the fermionic fields are determined from the action $S_F[\chi,\bar{\chi}]$ at the critical point. As customary, we fix the bare scaling dimension of the kinetic coefficients $Z$, $K_c$ and $K_d$ to zero. Scale invariance of the action at the critical point then determines the dynamical exponent $z$ and the bare scaling dimensions of the fields to be
\begin{align}\label{eq:qscalingfermions}
	z=2\,,\quad \xi_{c,q} = - d/2\,.
\end{align}
The scaling analysis for the bosonic action in Eq.~\eqref{eq:Hydro_action} follows analogously. We fix the bare scaling dimension of $D$ and $T$ to zero and arrive at the same relations as above, with the replacement $\xi_{c,q}\rightarrow\zeta_{c,q}$ and $z\to z_B$, {\it i.e.\/},
\begin{align}
	z_B=2\,,\quad \zeta_{c,q} = - d/2\,.
\end{align}
The identical scaling of classical and quantum fields is a typical feature of quantum critical scaling solutions at equilibrium or out of equilibrium~\cite{Sieberer2023}, but also of classical diffusive hydrodynamic modes -- the common feature is a canonical scaling of the noise level $P^K\sim P^{R,A}$. This canonical scaling also appears in the directed percolation problem~\cite{kamenev_book}. We will come back to further parallels between these problems in Sec.~\ref{sec:percolation}. 

Combining the above, the renormalized mass and Yukawa coupling become
\begin{align}
	\Delta_{c,d}' = b^{2}\Delta_{c,d}\,,\  g' = b^{2-d/2}g\,.
\end{align}
Consequently, the Yukawa coupling is relevant for $d<d_c=4$. On the other hand, vertices involving $\theta_q$ must include two spatial derivatives or a time derivative, and therefore scale as $b^{-d/2}$. Thus, they are irrelevant in all spatial dimensions $d$, validating our previous claim. As the Yukawa vertex with a single classical field is the only relevant coupling, it follows that the fermionic modes do not renormalize the diffusion constant $D$ and the boson temperature $T$. Hence, the dynamical exponent $z_B$ and scaling dimensions of the boson field $\zeta_{c,q}$ remain pinned to their bare values. 

\subsection{One-loop flow equations}\label{sec:RG_flow}
\begin{figure}[t!]
 \input{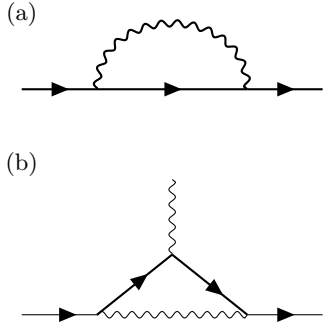}
\caption{Feynman diagrams for the one-loop RG flow equations. Fermions correspond to straight lines and bosons to wavy lines. The Feynman diagram renormalizing  the mass and kinetic coefficients are shown in (a), and (b) denotes the correction to the Yukawa vertex.} 
\label{Fig:Loop_diagrams}
\end{figure}

In this section, we analyze the one-loop corrections to the parameters of the effective action Eq.~\eqref{eq:effective_action} in an expansion around $\epsilon = 4 - d$. A detailed derivation of the flow equations using the momentum-shell RG is relegated to App.~\ref{app:RG_eq}. Here we examine the fixed points and their stability of the corresponding beta functions for the renormalized couplings. In addition to the Gaussian fixed point, where the Yukawa coupling vanishes ($g = 0$), leading to a decoupling of the boson and fermion sectors, we identify a nontrivial, infrared stable Wilson-Fisher fixed point for $d < 4$. At this fixed point, the mass $\Delta$ and kinetic coefficient $K$ are purely imaginary, reflecting their dissipative nature. Interestingly, the Yukawa coupling is real at this fixed point ($g \in \mathbb{R}$), so that at the fixed point the Yukawa term would be represented by a cubic Hamiltonian in an operatorial language. Furthermore, the scaling at this fixed point modifies both dynamical and static exponents, yielding $z \neq 2$ and $\nu\neq1/2$, and provides a fermion anomalous dimension, such that  $\xi_{c,q}\neq -d/2$.

Following the scaling analysis discussed in the last section, we first introduce the dimensionless parameters 
\begin{align}\label{eq:tildeg}
    \tilde{g} = \sqrt{\frac{\Omega_d}{(2\pi)^d}T}g \Lambda^{-\epsilon/2}\,,\quad  \delta = \Lambda^{-2}\Delta\,,
\end{align}
where $\Lambda$ is the ultraviolet momentum cutoff, and $\Omega_d = \text{Area}(S^{d-1})$. At one-loop level, these parameters satisfy the flow equations
\begin{align}
\begin{split}
\partial_l\delta&=\left[2+\Tilde{g}^2\left(\frac{1}{K_{\text{eff}}}\right)^2\right]\delta\,,\\
    \partial_l\tilde{g}&=\beta(g) = \left[\frac{\epsilon}{2}+\Tilde{g}^2\left(\frac{1}{K_\text{eff}}\right)^2\right]\tilde{g}\,,   
\end{split}
\end{align}
where $K_\text{eff} = K_c + \mi (K_d + D)$ is an effective kinetic coefficient, involving both the complex  fermionic ($K = K_{c} + \mi K_d$) and real bosonic ($D$) kinetic coefficients. The corresponding diagrams are presented in Fig.~\ref{Fig:Loop_diagrams}. Due to the fermionic dark state symmetry, $g_d$ remains locked to $g_d=-\Im(g)$ under RG transformations. The fixed points in the complex plane are given by $\beta(g^*) = 0$, leading to
\begin{align}\label{eq:fixed_point}
    g^* = 0\,,\quad (g^*)^2 = -\frac{\epsilon}{2}K_\text{eff}^2\,.
\end{align}

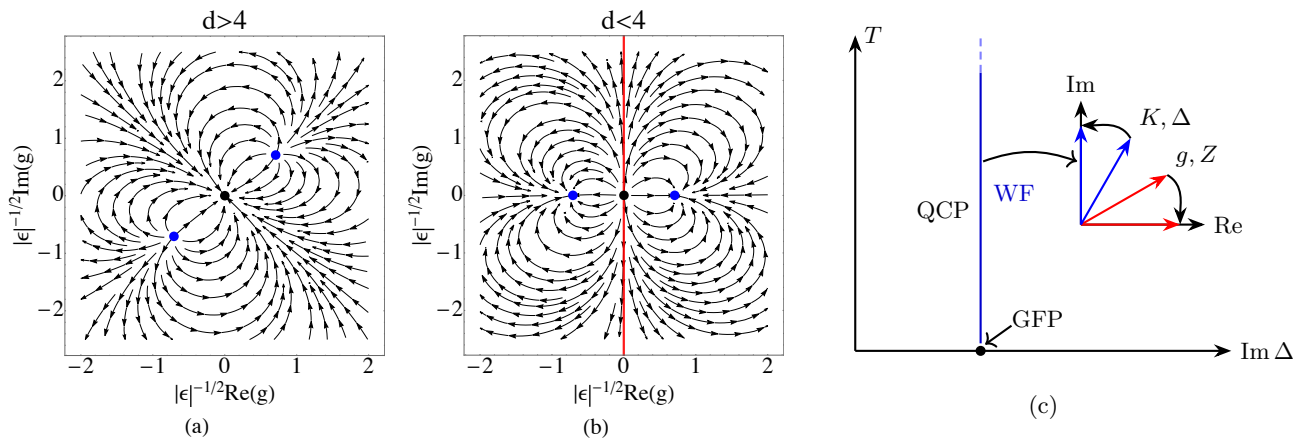
\begin{figure*}[t!]
    \begin{minipage}[b!]{0.28\textwidth}
    \centering
        \includegraphics[width=\textwidth]{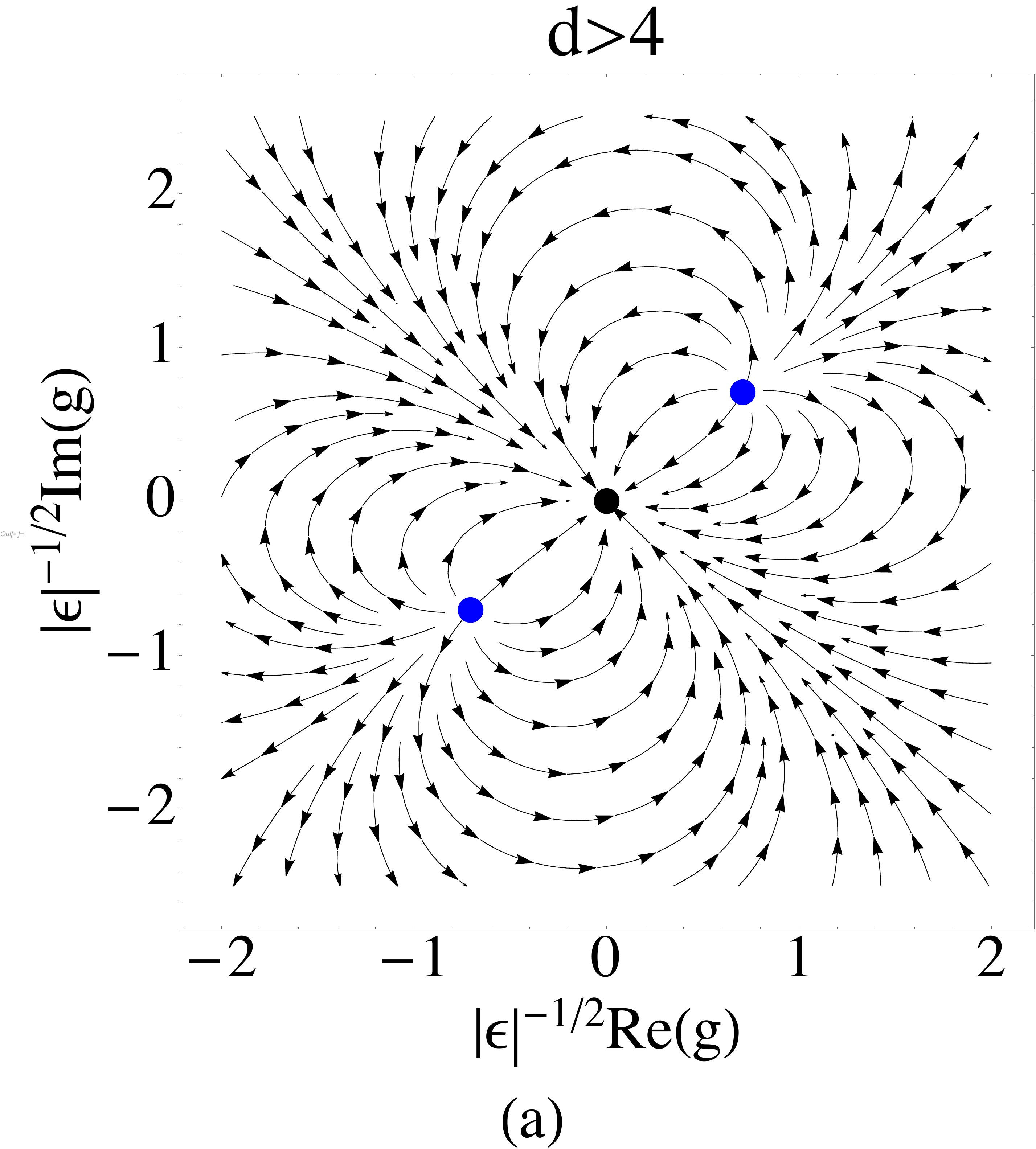}
    \end{minipage}
    \hspace{2pt}
    \begin{minipage}[b!]{0.28\textwidth}
    \centering
        \includegraphics[width=\textwidth]{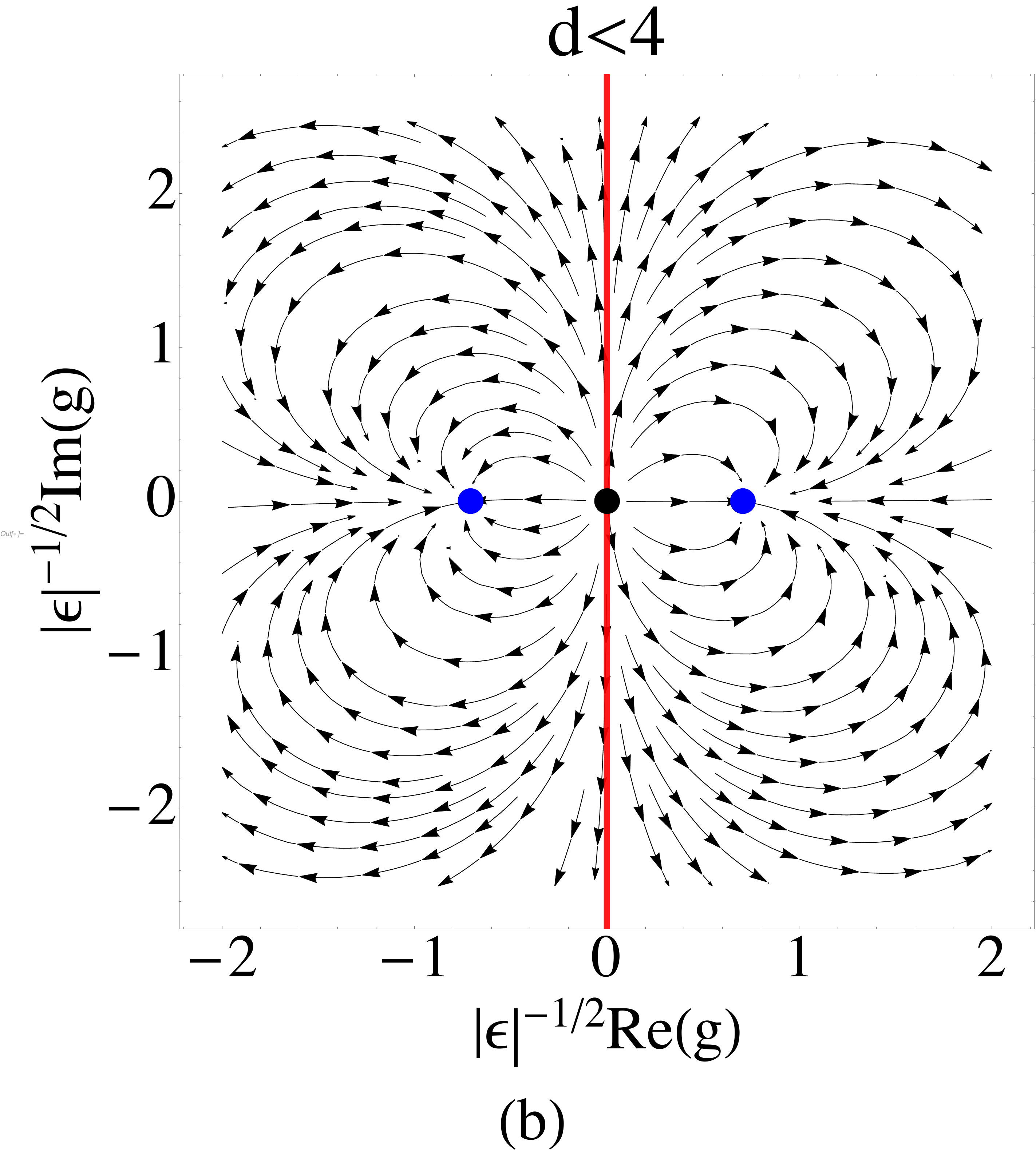}
    \end{minipage}
    \hspace{2pt}
    \begin{minipage}[b!]{0.4\textwidth}
    \centering
\definecolor{acidgreen}{rgb}{0.5,0.8,0.2}
\definecolor{burgundy}{RGB}{231, 0, 14}
\begin{tikzpicture}
\def\x{5}
\def\y{4.2}
    \draw[-Stealth,black, thick] (0,0) -- (\x,0) node[anchor=west] {$\Im \Delta$};
    \draw[-Stealth,black, thick] (0,0) -- (0,\y) node[anchor=west] {$T$};
    \path (\x/3-0.5,\y/2) node[below]   {QCP};
    \draw[thick,blue!80!black, thick] (\x/3,.1) -- (\x/3,\y-0.5);
    \draw[densely dashed, blue!50!white, thick] (\x/3,\y-0.5) -- (\x/3,\y);
    \path[thick,blue!80!black] (\x/3+0.05,\y/2) node[anchor=west] {WF};
    \draw[->,black, xshift=0.3pt, yshift=1pt, thick] (\x/3+0.4,0.3)--(\x/3+0.05,0.05);
    \node at (\x/3+0.3,0.2) [above right] {GFP};
    \filldraw[black, thick] (\x/3,0) circle (1.5pt);
    \node at (\x/2,-0.75) {{\small (c)}};

\draw[->, thick] (0.33*\x+0.05,0.6*\y) to [out=20,in=160] (0.6*\x-0.05,0.6*\y);

\def\s{0.33*\x}
\def\a{0.8}
\def\offx{0.6*\x}
\def\offy{0.4*\y}

\coordinate (a) at (0.87*\a*\s+\offx,0.5*\a*\s+\offy);
\coordinate (b) at (0.5*\a*\s+\offx,0.87*\a*\s+\offy);

\draw[-Stealth, thick] (0+\offx,0+\offy) -- (\s+\offx,0+\offy);
\draw[-Stealth, thick] (0+\offx,0+\offy) -- (0+\offx,\s+\offy);

\draw[-Stealth, red, thick] (0+\offx,0+\offy) -- (\a*\s+\offx,0+\offy);
\draw[-Stealth, blue, thick] (0+\offx,0+\offy) -- (0+\offx,\a*\s+\offy);

\draw[-Stealth, red, thick] (0+\offx,0+\offy) -- (a);
\draw[-Stealth, blue, thick] (0+\offx,0+\offy) -- (b);

\draw[-Stealth, thick] (a) to [out=330,in=90] node[auto] {} (\a*\s+\offx,0+\offy);
\draw[-Stealth, thick] (b) to [out=120,in=0] node[auto] {} (0+\offx,\a*\s+\offy);

\node at (\s+\offx,0+\offy) [right] {Re};
\node at (0+\offx,\s+\offy) [above] {Im};

\node at (a) [above right] {$g,Z$};
\node at (b) [above right] {$K, \Delta$};

\end{tikzpicture}
    \end{minipage}
    \caption{ (a,b) Plot for the beta functions for the complex Yukawa coupling $g$: (a) for $d>4$ and (b) for $d<4$ for $K_{\text{eff}} = K^*_{\text{eff}}$. In $d>4$ the Gaussian fixed point (black dot) is the only stable fixed point. In $d<4$, the Gaussian fixed point is unstable and there is an interacting (WF) fixed point (blue dots) with a single relevant direction (see main text). (c) Schematic of the phase diagram. The purely fermionic dark state model realizes a Gaussian quantum critical point (black dot in (b)), since the bosonic slow mode coupling to the fermions has vanishing noise level. Coupling additionally to $M\to\infty$ bosons at temperature $T>0$ activates the flow to the WF fixed point (blue in (b)), thus realizing a far-from-equilibrium interacting quantum critical point.
    The WF fixed point features both coherent and dissipative couplings, illustrated in the inset. At 1-loop order, the Yukawa coupling $g$ and the wave function renormalization $Z$ flow to a purely real fixed point. In contrast, the spectral mass and the kinetic coefficients become purely dissipative at the fixed point.}
    \label{Fig:beta_functions}
\end{figure*}

Linearizing $\beta(g^*)$ around the fixed points, we find that the Gaussian fixed point $(g=0)$ is unstable below the upper critical dimension $d_c=4$ and stable for $d>4$, while the opposite is true for the nontrivial fixed points given by $(g^*)^2 = -\frac{\epsilon}{2}K_\text{eff}^2$. The beta function for the Yukawa coupling $g$ is shown in Fig.~\ref{Fig:beta_functions}.

In Eq.~\eqref{eq:fixed_point} we observe that at the interacting fixed point the Yukawa coupling $g^*$ and the effective kinetic coefficient $K_\text{eff}$ enclose a right angle in the complex plane. Furthermore, the amplitudes satisfy the relation $|g^*|^2= \frac{\epsilon}{2}|K_\text{eff}|^2$.
At one-loop order, the flow equation of the kinetic coefficient $K_\text{eff}$ reads
\begin{align}
    \partial_l K_{\text{eff}} =\Tilde{g}^2(2\mi D)\left[1-\frac{4\mi D}{d}\frac{1}{K_{\text{eff}}}\right]\left(\frac{1}{K_{\text{eff}}}\right)^2\,.
\end{align}
At the Wilson-Fisher fixed point, this differential equation has the purely imaginary fixed point
\begin{align}
    K^*_\text{eff} = \frac{4\mi D}{d}\,,
\end{align}
from which we conclude that $g^*$ is purely real.
The anomalous dimension of the fermions is obtained from the corrections to the wavefunction renormalization coefficient $Z$. The one-loop correction is given by
\begin{align}
    \eta_Z= -\partial_l\ln Z = -\tilde{g}^2\left(\frac{1}{K_\text{eff}}\right)^2 + \mathcal{O}(\tilde{g}^4)\,.
\end{align}
At the nontrivial fixed point, this becomes $\eta_Z\approx\epsilon/2$, which, although the flow equations are complex, is purely real. Introducing the rescaled fields 
\begin{align}
    &\Bar{\psi}_{c,q} \rightarrow (Z)^{-1/2}\Bar{\psi}'_{c,q}\,, \quad \psi_{c,q} \rightarrow Z^{-1/2}\psi'_{c,q}\,,
 \end{align}
we see that the scaling dimension of the fermions is modified to
 \begin{align}
     \xi_c = \xi_q =\frac{\eta_Z-d}{2}= -\frac{d}{2}+\frac{\epsilon}{4}\,.
 \end{align}
The flow equations for the rescaled mass $\delta' = \delta/Z$ and the complex fermion kinetic coefficient $K' = K_\text{eff}/Z$ are
\begin{align}
\begin{split}
  \partial_l\delta' &= \left(2+\eta_Z + g^2\frac{1}{K^2}\right)\delta'\approx 2\delta'+O(\epsilon^2)\,,\\
  \partial_l K' & = \eta_Z K' + \frac{1}{Z}\Tilde{g}^2(2\mi D)\left[1-\frac{4\mi D}{d}\frac{1}{K_{\text{eff}}}\right]\left(\frac{1}{K_{\text{eff}}}\right)^2\\
  &\approx \left(\eta_Z  - \epsilon \right)K' = -\frac{\epsilon}{2}K'+O(\epsilon^2)\,.
\end{split}
\end{align}
In the second line, we linearized the equation around the nontrivial fixed point and kept terms only up to $O(\epsilon)$. The static exponent $\nu$, defined by the correlation length $\xi\sim\Delta^{-\nu}$ and the dynamical exponent is modified due to the anomalous scaling of $K'\sim b^{-\epsilon/2}\sim q^{\epsilon/2}$. Summarizing, we obtain the following critical exponents
\begin{align}\label{eq:crit_exponents}
\begin{split}
    \frac{1}{\nu} &= \partial_l\ln(\delta/K_\text{eff}) = 2+\frac{\epsilon}{2} +\mathcal{O}(\epsilon^2)\,,\\
    z &= 2 - \partial_l\ln K' = 2+\frac{\epsilon}{2}+\mathcal{O}(\epsilon^2)\,.
\end{split}
\end{align}
Finally, let us embed the purely self-interacting fermionic dark state problem into the fixed-point scenario established here. To this end, we note the form of the dimensionless coupling $\Tilde{g}\sim \sqrt{T}$, Eq.~\eqref{eq:tildeg}. Recall that $T$ is the temperature of the additional large $M$ boson degrees of freedom. In the limit $T\to0$, the dimensionless coupling is fine-tuned to zero, and so the Gaussian fixed point is physically realized, despite its relevant direction. This is the situation realized for the purely fermionic model: the effective temperature $T_{\text{eff}}$, or noise level, of the diffusive mode was found to vanish, and so the system remains at the Gaussian fixed point. From a more diagrammatic point of view, one can see that the absence of the bosonic noise level effectively decouples the bosons from the fermions (all diagrams with inner bosonic lines vanish), such that the system remains pinned at the Gaussian fixed point, see App.~\ref{app:RG_eq}. Conversely, any finite noise level of the diffusive mode will activate the flow to the Wilson-Fisher fixed point. This is illustrated in Fig.~\ref{Fig:beta_functions} (c). We emphasize the fact that the bosons reside at finite temperature is not in conflict with the quantum critical scaling solution manifestly realized for the fermions: In an out-of-equilibrium system, distinct subsystems do not need to see the same noise level and, in particular, do not need to thermalize.

In contrast to the out-of-equilibrium situation discussed above, thermal equilibrium requires all subsystems to share a common temperature and satisfy detailed balance. This imposes a thermal symmetry on the critical theory~\cite{Sieberer2023,sieberer2015prb}. Two limiting cases can be distinguished:\\
(i) If the fermions remain at $T=0$ (pure state), the same must be true for the bosonic bath. In this case, bosonic fluctuations vanish, fluctuation corrections to the Yukawa coupling are suppressed, and near upper critical dimension the system flows to a trivial Gaussian fixed point where bosons and fermions decouple. \\
(ii) If the bosonic bath is at finite temperature, thermal symmetry enforces that the fermions acquire the same temperature. The fermions then thermalize to a mixed state, and the purity gap is reduced below unity.\\
Thus, thermal equilibrium rules out the type of non-equilibrium critical behavior considered here.

\subsection{Discussion}\label{sec:Discussion}
We conclude this section with some discussion, also making contact with other critical and scaling phenomena.  As mentioned above, the fermionic dark state symmetry entails that there is only a single relevant direction at the critical point, as in a usual critical point: The symmetry stabilizes the fermionic quantum scaling solution Eq.~\eqref{eq:qscalingfermions}. Previously, quantum scaling solutions have been established in bosonic non-Markovian \cite{dalla2010quantum, DallaTorre2012PRB} and Markovian \cite{Marino2016a, Marino2016PRB} driven open quantum systems, but these correspond to bicritical points, where both the spectral (dissipative) and the noise gap have to be fine-tuned to zero simultaneously. 

Fermionic scaling solutions must be of the quantum type~\cite{Sachdev2023}, in- and out-of-equilibrium~\cite{Sieberer2023}.
It is worth comparing our scenario to quantum critical points involving gapless fermions, as appearing for magnetic quantum phase transitions in metals~\cite{Hertz1976,Millis1993,Loehneysen2007,Sachdev_2012}, or in Dirac-Yukawa systems~\cite{Zerf_2017,Boyack_2021,Herbut2024}. Of course, all these quantum phase transitions proceed in the equilibrium ground state.
Beyond this fundamental distinction, two more differences should be noted: First, the fermionic degrees of freedom in the problems above at $T=0$ equilibrium are generically gapless (living on a Fermi surface in a metal, or by symmetry and/or topology in the Dirac case), while the bosonic degrees of freedom are tuned to criticality -- opposite to our scenario, where the fermions are tuned to criticality while the collective bosons are gapless everywhere in parameter space. Second, the canonical fermion dynamical exponent is not $z=2$ in the cases above, due to anisotropies of the Fermi surface, or the linear dispersion of Dirac theory. 

Another class to compare to are phase transitions in topological insulators. Distinct from above and analogous to our scenario, these are topological phase transitions: the adjacent phases are characterized by topological invariants — nonlocal order parameters, instead of local order parameters indicating spontaneous symmetry breaking in conventional Landau transitions. Concretely, the topological-to-trivial insulator transition in $3+1$ dimensions is described in terms of a single massive Dirac fermion: While integrating out a gapped Dirac fermion generates a quantized axion response with coefficient $\theta=0,\pi$ depending on the sign of the mass, the $\theta$ term itself is not defined at the gap-closing point~\cite{Qi_2008}. Consequently, the critical theory at the transition is governed by massless Dirac fermions, and the topological response plays no dynamical role in the critical behavior, as shown explicitly in~\cite{Goswami2011QuantumCriticality,Isobe2012QuantumCriticalTI}. The same reasoning applies to our dissipative Dirac insulator setup: topology characterizes the non-local orders in the adjacent phases, but plays no role in the critical behavior.

Finally, it will be interesting to explore the connection of our findings to non-unitary fixed point actions in the Euclidean path integral framework~\cite{Giombi2020,katsevich2025,cardy2023,caetano2018,Gromov_2019}, which have been investigated recently in the context of coupled boson-fermion theories in~\cite{Klebanov_2023}. In particular, the ubiquitous $\mathcal{PT}$ symmetry of these works should be compared with the anti-unitary symmetry identified in our work.

\section{Comparison to equilibrium quantum criticality and directed percolation}\label{sec:percolation}
In this section we contextualize our results within the broader framework of critical phenomena. We have already noticed the similarity of our critical point with equilibrium quantum criticality and the non-equilibrium universality class of directed percolation on the level of canonical power counting in Sec.~\ref{sec:RG}. We now draw a further structural parallel between these critical points: All of them feature a protection by an anti-unitary $\mathbb{Z}_2$ symmetry involving the inversion of time, which ensures that a single fine-tuning is sufficient to reach criticality.

\textit{Relation to equilibrium quantum criticality ---}
In equilibrium, critical systems obey a thermal symmetry that enforces fluctuation-dissipation relations at all orders. This symmetry features a single parameter, the temperature $T$, which remains invariant under renormalization -- physically, all momentum shells in a Wilsonian RG procedure share the same temperature and maintain detailed balance. At zero temperature $(T=0)$ and near criticality, this symmetry thus protects the quantum critical nature of the state. (Of course, tuning temperature to absolute zero $T\to 0$ may be viewed as a form of physical fine-tuning, but not in the RG sense.) Consequently, quantum critical points require only a single fine-tuning parameter, distinguishing them from multicritical points. 
Thermal symmetry, like the case discussed above, combines (anti-unitary) quantum mechanical time reversal with an additional $\mathbb{Z}_2$ component. Defining a unitary matrix in band space $U \in U(N)$, for a system of complex fermions considered in this paper, it takes the form \cite{Sieberer2023}\footnote{There are two equivalent ways of formulating the action of thermal symmetry on the fields, cf. Ref.~\cite{sieberer2015prb}. Here, we chose the one that emphasizes the similarity to the action of time reversal on field operators.}
\begin{align}\label{eq:thermalsymm}
\mathcal{T}_\beta: \quad\psi^\sigma_\alpha(\vecx,t)&\to   \sigma U_{\alpha\beta}{\psi}^{-\sigma}_\beta(\vecx,-t-\mi\sigma\beta/2),  \quad \nonumber\\
\bar{\psi}^\sigma_\alpha(\vecx,t)&\to  \sigma U^*_{\alpha\beta}\bar\psi^{-\sigma}_\beta(\vecx,-t-\mi\sigma\beta/2)\,,\nonumber\\
\mi &\to - \mi,
\end{align}
which is to be compared with Eq.~\eqref{eq:dark_state_symmetry}, see also the discussion at the end of Sec.~\ref{sec:RG_flow}.

Despite this resemblance, a profound difference arises in the structure of correlation functions, relating to the non-equilibrium nature of the present setup. Thermal symmetry corresponds to a highly non-Markovian situation in the zero temperature limit, where the fluctuation-dissipation relation becomes
\begin{align}
    G^K(\omega,\vecq)=\text{sign}(\omega)\left(G^R (\omega,\vecq) -G^A (\omega,\vecq)\right)\,,
\end{align}
where the frequency-dependent prefactor comes with a unit matrix, and $\text{sign}(\omega)^2=1$,
guaranteeing that the state is pure. In contrast, the fermionic dark state symmetry respects the Markovian nature of the underlying Lindblad generator and leads to a frequency-independent relation
\begin{align}
    G^K (\omega,\vecq) = F\left(G^R (\omega,\vecq) -G^A (\omega,\vecq)\right)\,,
\end{align}
where $F$ is a constant matrix satisfying $F^2=\mathbb{1}$.
Consequently, the dark state is highly non-thermal and far from equilibrium.

\textit{Comparison with directed percolation ---}
A second key parallel exists with classical absorbing-state transitions, particularly those governed by the directed percolation universality class. The field theory for the latter is defined with
\begin{align}
    \mathcal{Z} &= \int \mathcal{D}(\phi_q,\phi_c) e^{\mi S[\phi_q,\phi_c]}, \quad\\\nonumber
    S[\phi_q,\phi_c] &= \int_{\vecx,t} [\phi_q(\partial_t + D\nabla^2 - r)\phi_c + g \phi_c\phi_q (\phi_c + \mi\phi_q) ],
\end{align}
with all parameters real, and real scalar bosonic fields. Here, we work in a real-time formulation to emphasize the parallel to our scenario, while often a Wick-rotated variant is presented, which obtains via $\phi_q \to \mi \phi_q$. This field theory is equipped with rapidity inversion symmetry. The latter shares two features with our fermionic dark state symmetry: First, rapidity inversion is often absent in microscopic models of directed percolation, such as those describing epidemic spreading or wetting processes, but emerges at the level of the directed percolation action. Second, rapidity inversion is again an anti-unitary $\mathbb{Z}_2$ parity symmetry, which includes inversion of time as an ingredient:
\begin{align}
    \mathcal{R}: \quad \phi_c(\vecx,t)\longleftrightarrow\mi\phi_q(\vecx,-t)\,, 
    \quad \mi \to - \mi.
\end{align}
 At first sight, the concrete implication of rapidity inversion on the directed percolation action might seem quite different to our case, namely, it excludes the presence of an additive noise level and thus enforces $G^K=0$ if not spontaneously broken. Conceptually, however, it parallels our scenario, see Fig.~\ref{fig:RGpicture}: Its primary consequence is the prohibition of an additive noise term, which is a relevant perturbation at the directed percolation fixed point. From an RG perspective, this mirrors the role of fermionic dark state symmetry, which prevents the activation of certain relevant couplings. 

Also in this case, there are fundamental differences: (i) Directed percolation describes bosonic instead of fermionic criticality. (ii)  Standard directed percolation lacks any conservation law coupled to the critical degrees of freedom, which is the case here. Variants of directed percolation including such a coupling have been considered in Refs.~\cite{Kree1989,VanWijland1998,Janssen2005,Doussal2015,Polovnikov_2022}. (iii) It is a classical transition, where quantum coherence and entanglement play no role, whereas our model lacks a classical counterpart. (iv) The absorbing state transitions in the directed percolation universality class proceed between a trivial vacuum (empty absorbing state) and a mixed state, whereas here we face a transition between two nontrivial vacua. Regarding this point and (iii), however, we refer again to Ref.~\cite{Thompson2024}, which establishes a quantum variant of bosonic directed percolation, yet with an empty dark state. 

\section{Conclusions and Outlook}\label{sec:outlook}

We have constructed and quantitatively analyzed a scenario of fermionic quantum criticality far from equilibrium. Our results establish a novel far-from-equilibrium universality class, where fermionic degrees of freedom in a pure state drive the RG flow at macroscopic scales. Unlike previously studied non-equilibrium quantum critical behavior, the critical point here emerges via a single fine-tuning. This is the consequence of an underlying emergent symmetry protecting the purity of the fermion quantum state.

Clearly, symmetry protection does not immediately equate to physical robustness. In fact, the difficulty in experimentally realizing the related directed percolation universality class stems from this very distinction. Nevertheless, directed percolation is central to non-equilibrium statistical mechanics precisely due to rapidity inversion symmetry without an equilibrium counterpart. Our work extends this symmetry protection concept to a fermionic quantum setting for the first time. The fact that this symmetry protects the purity of subsystems---such as the fermionic degrees of freedom in our fermion-boson theory---may make it significantly more useful than rapidity inversion in classical systems. It can serve as a design principle for constructing novel symmetry-protected pure state scenarios (some of which are discussed further below), potentially including implementations in quantum simulators with sufficient control to engineer many-body dynamics.

The dark state symmetry crucially involves the fermionic nature of the system. An open question is whether an analogous structure could exist for bosonic or spin systems with finite many-body excitation densities. Such a scenario would go beyond directed percolation, where the absorbing state is a product state devoid of excitations.

In fact, our findings may represent just a first instance of a broader class of quantum critical fermions far from equilibrium. Several extensions are worth exploring: First, it is conceivable that dark state transitions can occur between topological superfluids instead of insulating dark states. 
In this case, the critical fermions additionally couple to the Goldstone mode of the superfluid that exists across the transition. More technically, the present transition proceeds in a `strong to weak' symmetry broken scenario~\cite{Sieberer2023,Ogunnaike_2023,Huang_Lucas_2025}, where $U_c(1)\times U_q(1) \to U_c(1)$ on both sides of the transition, hosting a diffusive hydrodynamic mode throughout (where we note that we need an additional noisy diffusive mode to reach the Wilson-Fisher fixed point). A superfluid setting instead constitutes a `strong to nothing' broken case $U_c(1)\times U_q(1) \to \emptyset$, with an additional Goldstone mode for the broken $U_c(1)$ phase rotations; potentially, the Goldstone mode could replace the additional gapless diffusion mode. Second, replacing the $U(1)$ symmetry by larger symmetry groups such as $SU(N)$ or $Sp(N)$ will likely give rise to distinct non-equilibrium universality classes, with a large-$N$ limit as a handle for controlling critical exponents beyond dimensional expansion. Finally, a compelling direction is the construction of fermionic quantum directed percolation scenarios, for which the parallel to directed percolation is even more direct. This could be realized by putting the dark state Lindblad operators considered here into competition with a Hamiltonian featuring the same dark state as an unstable dynamical fixed point. This naturally leads to a fermionic quantum absorbing state transition, where the system transforms from a pure to a mixed state. Such problems are connected to measurement-induced phase transitions~\cite{Skinner2019,Li2019,Choi_2020,Gullans2020} in fermion systems~\cite{alberton2021enttrans,buchhold2021effective,TurkeshiZeroClick,fava2023,bao2021symmetry,chahine2023,Poboiko2023,Poboiko2024,Cao2019,Starchl2024,Klocke2023}, via the pre-selection concept~\cite{Buchhold2022}, according to which the criticality of measurement-induced transitions can be brought to the observable level using active feedforward \cite{Buchhold2022,Odea2022,sierant2023b,iadecola2023b,Piroli2023}. 

Our work thus paves the way to a plethora of novel critical phenomena far from equilibrium, whose ‘quantumness’ is ascertained by the very fact that fermions continue to dictate the universal physics up to the largest scales, a situation without a classical counterpart. 

\section*{Acknowledgments}
We thank M. Buchhold, R. Daviet, Z. Huang, A. Kamenev, I. Klebanov, T. Müller, A. Rosch, M. Scherer, L. Sieberer, F. Thompson, K. Weisenberger, and C. Zelle for discussions. This work was supported by the Deutsche Forschungsgemeinschaft (DFG, German Research Foundation) under Germany’s Excellence Strategy Cluster of Excellence Matter and Light for Quantum Computing (ML4Q) EXC 2004/1 390534769, and by the DFG Collaborative Research Center (CRC) 183 Project No. 277101999.

\appendix
\begin{widetext}
\section{Hubbard-Stratonovich decoupling: Fermion-boson theory }\label{app:HS}
We start from the microscopic fermionic action Eq.~\eqref{eq:micro_action} shown in Sec.~\ref{sec:MicroModel}. To access the relevant degrees of freedom, it is convenient to introduce the identity
\begin{align}\label{eq:HS}
    1=\int \mathcal{D}\eta \mathcal{D}\phi\, \text{exp}\left\{-\mi\,\text{Tr}\left[\left(\eta-X\right)\left(\phi-\Psi\right)\right]\right\}\,.
\end{align}
Inserting it in the functional integral \eqref{eq:Kpart} implements a Hubbard-Stratonovich transformation ~\cite{Trunin2021}: It locks the real bosonic fields $\eta$ and $\phi$ to the fermionic bilinears $X$ and $\Psi$ defined in Eq.~\eqref{eq:bilinears}. In particular, taking the saddle point conditions for variations with respect to $\eta$ and $\phi$ yields
\begin{align}\label{eq:locking}
    \langle \phi_{\alpha}^{\sigma\rho}(\vecx,t)\rangle=\langle \Psi_{\alpha}^{\sigma\rho}(\vecx,t)\rangle\,,\quad \langle \eta_{\alpha}^{\sigma\rho}(\vecx,t)\rangle=\langle X_{\alpha}^{\sigma\rho}(\vecx,t)\rangle\,.
\end{align}
Hence, $\phi$ relates to the total number of fermions, while $\eta$ detects the excitations above the dark state. At fixed particle number, these correspond to density waves.
The representation of unity in Eq.~\eqref{eq:HS} is not unique in general, but its particular choice here is guided by the knowledge that $\hat{x}_L^\dagger|\text{DS}\rangle=\hat{x}_U|\text{DS}\rangle=0$, which implies that $\langle \text{DS}|\hat X|\text{DS}\rangle$ vanishes. Furthermore, we require that the insertion of Eq.~\eqref{eq:HS} in the action cancels all terms involving more than two Grassmann fields. After the Hubbard-Stratonovich transformation, the microscopic action takes the form
\begin{align}\label{eq:HS_action}
\begin{split}
    S[\bar{\psi},\psi,\eta,\phi]=&S_0[\bar{\psi},\psi]+ S_B[\eta,\phi]+S_Y[\bar{\psi},\psi,\eta,\phi]\,,\\
    S_B[\eta,\phi]=&-\text{Tr}(\eta\phi)\,,\\
S_Y[\bar{\psi},\psi,\eta,\phi]=&\,\text{Tr}\left(\eta \Psi + \phi X\right)\,,
\end{split}
\end{align}
which is quadratic in the fermions. Thus, the original quartic coupling of fermions is replaced by a cubic (Yukawa) coupling between the Grassmann fields and the new bosonic variables. The density wave fields $\eta_{U/L}$ are coupled to the fields of the total density $\phi_{U/L}$, which are furthermore linearly dependent as they satisfy
\begin{align}
    \phi^{\sigma\rho}_{U}(\vecx,t)=1 - \phi^{\rho\sigma}_{L}(\vecx,t)\,,
\end{align}
which follows directly from the same relation satisfied by the fermionic bilinear $\Psi$.

\subsection{Dynamical mean field theory}\label{app:dynamical_MF}

To proceed, using the knowledge of the exact stationary state \eqref{eq:darkstate2}, we perform a saddle point approximation for the action \eqref{eq:HS_action}. To this end, we formally perform the Gaussian integral over the Grassmann fields to obtain the effective action in terms of $\eta$ and $\phi$,
\begin{align}\label{eq:HS_action_2}
    S[\phi,\eta]=-\Tr(\phi\eta)-\mi\Tr\ln(\mi  G^{-1})\,,
\end{align}
where the fermionic inverse Green’s function is given by
\begin{align}\label{eq:GF}
    G^{-1}[\phi,\eta]=(\sigma^z\otimes\mathbb{1}_N)\left(\mi \partial_t-hW^\dag W\right) + \Sigma[\phi,\eta]
\end{align}
with self-energy
\begin{align}
    \Sigma[\phi,\eta] = \left(\mi W^\dag\boldsymbol{\phi} W + \boldsymbol{\eta}\right)\,.
\end{align}
Here, and in the following, we define bold symbols as diagonal matrices in band space of the form 
\begin{align}\label{eq:boldsymbols}
    \begin{split}
    \boldsymbol{\phi}^{\sigma\rho}&= \text{diag}(\underbrace{\phi_U^{\rho\sigma},\dots,\phi_U^{\rho\sigma}}_{N/2},\underbrace{-\phi_L^{\sigma\rho},\dots,-\phi_L^{\sigma\rho}}_{N/2})\,, \\
    \boldsymbol{\eta}^{\sigma\rho} &=(c^{\sigma\rho}\eta_L^{\sigma\rho} - c^{\rho\sigma}\eta_U^{\rho\sigma})\otimes \mathbb{1}_{N}\,,
    \end{split}
\end{align}
where repeated indices are not summed over. 

Due to the quasi-local nature of the matrices $W(\vecx)$ (cf. Eq.~\eqref{eq:W_pos}), the matrix $W^\dag \boldsymbol{\phi} W$ shown in the self energy is given by the convolution
\begin{equation}
    (W^\dag\boldsymbol{\phi} W)(\vecx,\vecx';t) \equiv \int_\vecy W^\dag(\vecx-\vecy)\boldsymbol{\phi}(\vecy,t) W(\vecy-\vecx') .
\end{equation}
Taking the functional derivatives with respect to $\eta$ and $\phi$ we recover Eq.~\eqref{eq:locking} from the saddle point conditions
\begin{align}\label{eq:saddle_point_app}
   \frac{\delta S}{\delta\eta^{\sigma\rho}_\alpha} \overset{!}{=} 0 \implies &\phi^{\sigma\rho}_U = \Tr(\mi G^{\rho\sigma}c^{\rho\sigma})\,,\ \phi^{\sigma\rho}_L = -\Tr(\mi G^{\sigma\rho}c^{\rho\sigma}),\\
    \frac{\delta S}{\delta\phi^{\sigma\rho}_\alpha} \overset{!}{=} 0\implies&\eta^{\sigma\rho}_U = -\mi\Tr(\mi G^{\rho\sigma}W^\dag P^U W)\,,\ \eta^{\sigma\rho}_L = \mi\Tr(\mi G^{\sigma\rho}W^\dag P^L W)\,,
\end{align}
with $P^U = \text{diag}(\mathbb{1}_{N/2},0)$ and $P^L = \text{diag}(0,\mathbb{1}_{N/2})$ projecting onto the upper and lower bands, respectively. The trace runs over space-time indices and band indices. Here and in the remainder of the section, repeated indices are not summed over. Below, we demonstrate that the homogeneous ansatz solves the saddle point equations
\begin{align}\label{eq:saddle_point}
     \eta_\alpha(\vecx,t) = 0\,,\ \phi_\alpha(\vecx,t)=\phi_0\equiv\varphi_0 c^\top\,,\ \varphi_0\in\mathbb{R}^+\,,
\end{align}
where $\varphi_0$ satisfies (in the continuum limit)
\begin{align}
  \varphi_0 = N\varphi_0\int_{\omega,|\vecq|\leq\Lambda} \frac{\lambda(\vecq)}{(\omega-h\lambda(\vecq))^2 + \varphi_0^2\lambda(\vecq)^2}   \,,
\end{align}
which is obtained by carefully tracking the point-splitting of the fermionic fields in the time domain. The proof of this statement requires carefully tracking the point-splitting for the fermionic fields in the time domain (see below). To regularize the formally divergent momentum integral on the right-hand side, we introduce a UV cutoff $\Lambda$. The integral then yields a volume factor $\text{Vol}(S^d(\Lambda))$\,, where $S^d(\Lambda)$ denotes the $d$-dimensional sphere of radius $\Lambda$\,. 

For a realization on the lattice, this corresponds to the volume of the first Brillouin zone, which entails $\varphi_0=\frac{N}{2}a^{-d}\equiv \rho$, the total particle density. 

We now discuss the detailed evaluation of the saddle point equations ~\eqref{eq:saddle_point_app}. For this purpose, it is convenient to move to a Fourier representation of the composite fields
\begin{align}
    \phi(\omega,\vecq) &= \int_{\vecx,t} \phi(t,\vecx) e^{-\mi(\vecq\cdot\vecx -\omega t)}\,, \quad
    \eta(\omega,\vecq) = \int_{\vecx,t} \eta(t,\vecx) e^{-\mi(\vecq\cdot\vecx -\omega t)},\\
    \phi(t,\vecx) &= \frac{1}{L^d}\sum_{\vecq}\int_\omega \phi(\omega,\vecq) e^{\mi(\vecq\cdot\vecx -\omega t)}\,, \quad
    \eta(t,\vecx) = \frac{1}{L^d}\sum_{\vecq} \int_{\omega} \eta(\omega,\vecq) e^{\mi(\vecq\cdot\vecx -\omega t)}\,,
\end{align}
where $L$ denotes the linear dimension of the system. In the Fourier basis, the ansatz reads
\begin{equation}
    \phi_\alpha(\vecq,\omega) = \varphi_0 c^\top\delta_{\vecq,0}\delta(\omega)L^d\,,\quad \eta_\alpha(\omega,\vecq) = \eta_\alpha\delta_{\vecq,0}\delta(\omega)L^d = 0\,.
\end{equation}

\paragraph*{Contour off-diagonal terms ---}
The self-consistent equations for the contour off-diagonal sector read
\begin{align}\label{eq:SP_contour_off_diagonal}
    \phi^{+-}_U &= \sum_\veck\int_\omega\tr(\mi G_0^{-+}(\omega,\veck))c^{-+}= \varphi_0c^{-+}\,,\quad\phi^{+-}_L  = -\sum_\veck\int_\omega\tr(\mi G_0^{+-}(\omega,\veck))c^{-+}= \varphi_0c^{-+}\,,\\
    \eta^{-+}_U &= -\mi\sum_\veck\int_\omega\tr(\mi G_0^{+-}(\omega,\veck)W^\dag(\veck)P^U W(\veck))=0\,,\quad\eta^{-+}_L = \mi\sum_\veck\int_\omega\tr(\mi G_0^{-+}(\omega,\veck)W^\dag(\veck)P^L W(\veck))=0\,,
\end{align}
where $G_0$ is the bare Green's function obtained from evaluating $G$ on the mean field ansatz. 
The components $G_0^{+-} = G^<$ and $G_0^{+-}=G^>$ correspond to the physical particle and hole densities respectively \cite{kamenev_book}, evaluated in the stationary state, and are given by
\begin{align}\label{eq:greater_lesser_green_functions}
G_0^{+-}(\omega,\vecq) = U^\dag(\vecq)\frac{2\mi\varphi_0\lambda(\vecq)}{(\omega-h(\vecq))^2 + \varphi_0^2\lambda(\vecq)^2}P^L U(\vecq) ,\quad G_0^{-+}(\omega,\vecq) = -U^\dag(\vecq)\frac{2\mi\varphi_0\lambda(\vecq)}{(\omega-h(\vecq))^2 + \varphi_0^2\lambda(\vecq)^2} P^U U(\vecq)\,,
\end{align}
where we have used Eq.~\eqref{eq:WfU} to express $W(\vecq)$ in terms of the unitary $U(\vecq)$ and $\lambda(\vecq)$. Using this, we obtain
\begin{align}
    \varphi_0 &= N\sum_\vecq\int_\omega \frac{\varphi_0\lambda(\vecq)}{(\omega-h(\vecq))^2 + \varphi_0^2\lambda(\vecq)^2}\,.
\end{align}
Similarly, for the $\eta^{-+}_\alpha$ saddle point equations, it is straightforward to verify that the integrals are zero due to the vanishing trace over the band indices. 

\paragraph*{Contour-diagonal terms ---} For the contour diagonal terms, the saddle point equations read
\begin{align}
    \phi^{\sigma\sigma}_U & = \sum_\veck\int_\omega\tr(\mi G_0^{\sigma\sigma}(\omega,\veck))= \varphi_0\,,\quad \varphi^{\sigma\sigma}_L = -\sum_\veck\int_\omega\tr(\mi G_0^{\sigma\sigma}(\omega,\veck))= \varphi_0\,,\\
    \eta_U^{\sigma\sigma} &=-\mi \sum_\veck\int_\omega\tr(\mi G_0^{\sigma\sigma}(\omega,\veck)W^\dag(\veck)P^U W(\veck))=0\,,\quad \eta_L^{\sigma\sigma} = \mi\sum_\veck\int_\omega\tr(\mi G_0^{\sigma\sigma}(\omega,\veck)W^\dag(\veck)P^L W(\veck))=0\,,
\end{align}
with the diagonal components of the bare Green's function given by
\begin{equation}\label{eq:time_ord_green_functions}
    G^{++}(\omega,\vecq)= U^\dag(\vecq)\left(\frac{1}{\omega-h(\vecq)+\mi \lambda(\vecq)}P^U +  \frac{1}{\omega-h(\vecq)-\mi \lambda(\vecq)} P^L\right)U(\vecq)\,,\ G^{--}(\omega,\vecq) = -(G^{++}(\omega,\vecq))^\dag\,.
\end{equation}
The frequency integrals involving $G_0^{\sigma\sigma}$ are singular. At the operator level, this follows from the result that the contour diagonal Green's functions correspond to time (anti-)ordered Green's functions evaluated at equal times ~\cite{kamenev_book}, which is not well-defined. This is resolved by reinstating point-splitting in the full interacting fermionic action to properly account for operator ordering ~\cite{Sieberer2014}. Carefully tracking the point-splitting through the HS decoupling leads to the Yukawa action
\begin{equation}
    S_Y[\bar{\psi},\psi,\phi,\eta] = \mi\int_{t,t',\vecx}\Bar{\psi}(t,\vecx) W^\dag\boldsymbol{\phi}(t-t',\vecx) W\psi(t',\vecx) + \int_{t,t',\vecx}\Bar{\psi}(t,\vecx)\boldsymbol{\eta}(t-t',\vecx)\psi(t',\vecx)\,,
\end{equation}
with
\begin{align}\label{eq:eta_phi_time_reg}
\begin{split}
    \boldsymbol{\phi}^{\sigma\rho}(t-t')&=\text{diag}(\phi^{\rho\sigma}_U(t)\Lambda^{\sigma\rho}(t-t'),\dots,-\phi^{\sigma\rho}_L(t)\Lambda^{\rho\sigma}(t'-t),\dots)\,,\\
    \boldsymbol{\eta}^{\sigma\rho}(t-t')&=(c^{\sigma\rho}(t-t')\eta_L^{\sigma\rho}(t) - c^{\rho\sigma}(t'-t)\eta_U^{\rho\sigma}(t))\otimes \mathbb{1}_{N}\,,
\end{split}
\end{align}
where $c(t-t')$ is given by Eq.~\eqref{eq:c_matrix_timereg} and the matrix
\begin{equation}
\Lambda(t-t')= \mqty(\delta(t-(t'+0^+)) & 0 \\ \frac{1}{2}\left(\delta(t-(t'+0^+))+ \delta(t-(t'-0^+)) \right) & \delta(t-(t'-0^+)) )\,.
\end{equation}
For the saddle point equations, this leads to a regularization of the frequency integrals
\begin{align}\label{eq:SP_contour_diagonal}
\begin{split}
    \phi^{\sigma\sigma}_U & = \sum_\veck\int_\omega\tr(\mi G_0^{\sigma\sigma}(\omega,\veck))e^{-\mi\omega\sigma 0^+}= \varphi_0\,,\quad \varphi^{\sigma\sigma}_L = -\sum_\veck\int_\omega\tr(\mi G_0^{\sigma\sigma}(\omega,\veck))e^{\mi\omega\sigma 0^+}= \varphi_0\,,\\
    \eta_U^{\sigma\sigma} &=-\mi \sum_\veck\int_\omega\tr(\mi G_0^{\sigma\sigma}(\omega,\veck)W^\dag(\veck)P^U W(\veck))e^{\mi\omega\sigma 0^+}=0\,,\quad \eta_L^{\sigma\sigma} = \mi\sum_\veck\int_\omega\tr(\mi G_0^{\sigma\sigma}(\omega,\veck)W^\dag(\veck)P^L W(\veck))e^{-\mi\omega\sigma 0^+}=0\,.
\end{split}
\end{align}
The frequency integrals are now well-defined and can be evaluated using the residue theorem, which yields the same result for $\varphi_0$ as discussed above.

\section{Fate of the strong \texorpdfstring{$U(1)$}{} symmetry for the fermions}

In this section, we perform a fluctuation analysis beyond the saddle point solution and show that the gapless mode associated with the strong $U(1)$ symmetry is indeed frozen out. That is, it has zero noise level or effective temperature, which in turn implies that the Gaussian fixed point is not destabilized by the bosonic fluctuations.

As discussed in the main text, the microscopic action has a strong $U(1)$ symmetry associated with particle number conservation. We first review the emergence of hydrodynamic theories from this symmetry of the microscopic action.
Following this, we introduce a generalized phase-amplitude decomposition in the microscopic theory and perform a Wigner-Moyal expansion to derive the low-energy effective action for the hydrodynamic mode.

\subsection{Review of hydrodynamics from strong symmetries}\label{app:strong_symm_hydro}
The microscopic models of fermions described in Sec.~\ref{sec:Lindblad} exhibit a strong $U(1)$ symmetry (cf.  Eq.~\eqref{eq:Lcommutator}). In the Keldysh formalism, this corresponds to the action remaining invariant under independent $U(1)$ transformations, see Eq.~\eqref{eq:strong_sym}. This $ U_+(1)\times U_-(1)$ group includes a subgroup that acts identically on both contours, {\it i.e.\/}, $\theta^+=\theta^-$, referred to as a classical or weak $U_c(1)$ symmetry~\cite{Sieberer2023, Buca2012}. If broken spontaneously, this symmetry gives rise to a Goldstone mode. For the models discussed in Sec.~\ref{sec:MicroModel}, this is not the case. Yet, strong symmetries often influence the effective long-wavelength theory via Noether currents associated with conserved charges, which give rise to hydrodynamic modes. To see this explicitly, we consider an infinitesimal local transformation given by
\begin{align} 
    \psi^\pm(\vecx,t)\rightarrow \me^{\mi \theta^\pm(\vecx,t)}\psi^\pm(\vecx,t)\,.
\end{align}
The action then transforms as $S\rightarrow S + \delta S$, with
\begin{align}\label{eq:Noether_pm}
	\delta S = \int_{\mathbf{x},t} \ [ \partial_\mu\theta^+(\vecx,t) J^\mu_+(\vecx,t) - \partial_\mu\theta^-(\vecx,t) J^\mu_-(\vecx,t)]\,.
\end{align}
The Noether currents $J^\mu_\pm(\vecx,t)$ associated with the symmetry satisfy the continuity equation
\begin{align}\label{eq:conservation_law}
	\partial_\mu\langle J_\pm^\mu\rangle = 0\,.
\end{align}
To identify the relevant slow modes, we rewrite Eq.~\eqref{eq:Noether_pm} in terms of symmetric (under contour exchange; also referred to as classical) and antisymmetric (quantum) fields
\begin{align}\label{eq:deltaS}
   \delta S = \int_{\mathbf{x},t}[\  \partial_\mu\theta_c(\vecx,t) J^\mu_q(\vecx,t) + \partial_\mu\theta_q(\vecx,t) J^\mu_c(\vecx,t)]\,,
\end{align}
with
\begin{align}\label{eq:current}
\begin{split}
    J^\mu_c &= \frac{1}{2}\left(J^\mu_+ + J^\mu_-\right)\,,\quad J^\mu_q = \left(J^\mu_+ - J^\mu_-\right)\,,\\
    \theta_c &= \frac{1}{2}\left(\theta^+ + \theta^-\right)\,,\quad\theta_q = \left(\theta^+ - \theta^-\right)\,.
\end{split}
\end{align}
We have introduced here the parametrization $ U_+(1)\times U_-(1)\simeq  U_q(1)\times U_c(1)$, with strong and weak symmetry transformation parameters $\theta_{q,c}$ respectively. Due to probability conservation of the Keldysh action, expectation values of the quantum current $J_q$ vanish. Instead, the expectation value of the classical current $J_c$ is generally non-vanishing and observable, its zero component is associated with the conserved total particle number $N_c = \int_\vecx \langle J^0_c\rangle$, with $J^0_c$ defined in Eq.~\eqref{eq:current} and $J^0_\pm =\sum_\alpha\bar\psi^\pm_\alpha(\vecx,t)\psi^\pm_\alpha(\vecx,t)$. 
According to the Noether construction, the total action can depend on derivatives of the symmetry transformation parameter of the strong symmetry only, $S = S[\partial_\mu\theta_q]$, as a consequence of the global $U_q(1)$ symmetry, $\theta_q(\vecx,t) \to \theta_q(\vecx,t)+ \alpha_q$ with $\alpha_q\in\mathbb{R}$. 

The strong symmetry is not preserved under renormalization, physically corresponding to the system acting as its own bath. More precisely, the symmetry is broken according to $U_q(1)\times U_c(1) \to U_c(1)$ in our case (cf. Sec.~\ref{app:dynamical_MF}). However, as usual in spontaneous symmetry breaking \cite{Zinn-Justin_book}, the gapless character of the dynamics is preserved along the broken directions, \textit{i.e.} $\theta_q(\vecx,t) \to \theta_q(\vecx,t)+ \alpha_q$ remains intact. Thus, the effective action must depend on derivative terms only,  $S = S[\partial_\mu\theta_q]$.

This allows one to construct the hydrodynamics of the conserved charge, including irreversible terms~\cite{Sieberer2023,Huang_Lucas_2025}. The conserved charge density is conjugate to the symmetry transformation parameter, according to $e^{\pm\mi \theta_q \hat Q}$, with $\hat Q = \int_\vecx[\sum_\alpha\hat n_\alpha(\vecx)]$, where $\hat n_\alpha(\vecx) = \hat\psi^\dag_\alpha(\vecx)\hat\psi_\alpha(\vecx)$ (cf. Eq.~\eqref{eq:Charge}) applied to the $\pm$ branches in the Hamiltonian operator formalism. In the action language, this imposes the structure $\int_{\vecx,t}\theta_q \partial_t n_c$ to appear, with total charge density $n_c = J_c^0$ (summed over bands $\alpha$). Hydrodynamics can be expressed entirely in terms of these conjugate variables. For example, the spatial current must then take the form $J_c^i = J_c^i[n_c] \approx \partial_i n_c + ...$, to additionally comply with rotation invariance. Accounting additionally for the Hermiticity and probability conservation, the hydrodynamic action thus reads to leading order 
{\small
\begin{align}\label{eq:hydro_action}
	S_B[\phi] =\frac{1}{2} \int_{\vecx,t}  \Tilde{\phi}^\top(\vecx,t)\mqty(0 & -\partial_t - D\nabla^2\\ \partial_t - D\nabla^2 & -2 \mi\lambda\nabla^2) \Tilde{\phi}(\vecx,t)\,,
\end{align}}
\noindent where $D\in\mathbb{R}^+$ is the diffusion constant, and we have introduced $\Tilde{\phi}=(n_c(\vecx,t),\theta_q(\vecx,t))^\top$. Note that the field $n_c$ (and its conjugate noise field $\theta_q$) introduced here is not related to the diffusive modes introduced in Sec.~\ref{sec:diffusivemodes}. The bare action for the diffusive modes, although similar, describes density fluctuations of physically distinct fields: the former associated with the conserved fermionic density and the latter associated with the conserved bosonic density of the thermal bath.

The form $S[\partial_\mu \theta(q)]$ forbids a constant term in the Keldysh component, as such a term would generate a mass under RG flow. 
In Eq.~\eqref{eq:hydro_action} we have omitted higher-order terms, as they are irrelevant in all spatial dimensions (see power counting analysis in Sec.~\ref{sec:RG}). In thermal equilibrium, the noise coefficient $\lambda$ and the diffusion constant $D$ are related by the fluctuation-dissipation relation. Out of equilibrium, however, they are generally independent. 

The bosonic sector has an anti-unitary symmetry characteristic of diffusive actions
$\theta_q(\vecx,t) \to \theta_q(\vecx,-t) - \tfrac{\mi}{T_\text{eff}} n_c(\vecx,t), n_c(\vecx,t)\to n_c(\vecx,-t)$ with an effective temperature $T_\text{eff} = \lambda/D$.

We also note that a Burgers-type non-linearity $\sim \theta_q(\vecx,t)\vec v^\top\nabla n_c^2$ is ruled out by this symmetry, and independently by spatial inversion symmetry $\phi(\vecx,t) \to \phi(-\vecx,t)$ --  resulting from the underlying theory at the critical point 

\subsection{Fluctuation analysis and derivation of the hydrodynamic mode}\label{app:Hydro_mode}
Following this brief review of hydrodynamics as the relevant low-energy description of systems with $U(1)_q$ symmetry, in this section we provide a perturbative derivation of the low-energy effective theory for near-critical Class AIII dissipative insulators, as indicated in the main text. The derivation relies on an appropriate parameterization of the composite fields/bosonic fields in terms of a generalized phase-amplitude decomposition, followed by a gradient expansion in slowly varying derivatives of the conjugate field $\theta_q$.
However, we find that at 1-loop order, the noise coefficient $\lambda= 0$ or $T_\text{eff}=0$, implying that local density fluctuations are suppressed or "frozen" out (analogously to a zero temperature situation). Although this result is perturbative, we expect it to hold upon inclusion of hydrodynamic interactions, in accordance with chiral symmetry as per the discussion in the main text.

Since the total number of particles is conserved, the amplitude of $\boldsymbol{\phi}$ is fixed by the condition of half-filling. To distill the low-energy degrees of freedom, it is therefore convenient to introduce a generalized phase-amplitude decomposition based on the $U(1)\times U(1)$ symmetry of the microscopic fermionic action~\eqref{eq:micro_action}. The strong $U(1)$ symmetry (see Eq.~\eqref{eq:Lcommutator}) is recovered in the field theory formalism as an equivalent global $U(1)\times U(1)$ symmetry of the microscopic fermionic action~\eqref{eq:micro_action}, under which fermions transform as 
\begin{align}\label{eq:strong_sym}
\begin{split}
    g: \quad &\begin{pmatrix}
        \psi^+(\vecx,t)\\
        \psi^-(\vecx,t)
    \end{pmatrix}
   \rightarrow g\begin{pmatrix}
        \psi^+(\vecx,t)\\
        \psi^-(\vecx,t)
    \end{pmatrix}, \\
    &g =\text{diag}(\exp(\mi \theta^+), \exp(\mi \theta^-))\otimes\mathbb{1}_N\,,
\end{split}
\end{align}
{\it i.e.\/}, the action is invariant under independent $U(1)$ transformations of the fields on the two branches of the Keldysh contour~\cite{Sieberer2023}. The Hubbard-Stratonovich fields are locked to fermionic bilinears via Eq.~\eqref{eq:locking}, thus transforming in the adjoint representation
\begin{align}\label{eq:HS_symm_rep}
\begin{split}
    g: \quad \boldsymbol{\phi} &\rightarrow g\boldsymbol{\phi} g^{-1}\,,\ \boldsymbol{\eta} \rightarrow g\boldsymbol{\eta} g^{-1}\,.
\end{split}
\end{align}
To extract the long-wavelength degrees of freedom, we promote global $U(1)\times U(1)$ rotations to slowly varying local transformations, $g \rightarrow g(\vecx,t)$ (see e.g.~\cite{kamenev_book} for similar constructions in disorder physics).
This leads to the following parameterization of the low-energy fluctuations around the saddle point configuration
\begin{align}\label{eq:u(1)_transf_boson}
    \begin{split}
        g: \quad \boldsymbol{\phi}(\vecx,t) &\rightarrow g(\vecx,t)\boldsymbol{\phi_0} g^{-1}(\vecx,t)\,,\ 
        \boldsymbol{\eta}(\vecx,t) \rightarrow g(\vecx,t) \boldsymbol{\eta}(\vecx,t)g^{-1}(\vecx,t)\,,
    \end{split}
\end{align}
where
\begin{align}
    g(\vecx,t) =\text{diag}(\me^{\mi \theta^+(\vecx,t)} , \me^{\mi \theta^-(\vecx,t)})\otimes\mathbb{1}_N\,, 
\end{align}
and $\boldsymbol{\phi_0}$ has the same structure as $\boldsymbol{\phi}$ defined in Eq.~\eqref{eq:boldsymbols}.
This parametrization captures fluctuations of $\boldsymbol{\eta}$ and phase fluctuations of $\boldsymbol{\phi_0}$ while neglecting the gapped amplitude excitations. 

To derive a long-wavelength action, we first integrate out high-momentum fermions above an intermediate momentum scale $\mu$. This yields the action
\begin{align}\label{eq:HS_eft_intscale}
S[\bar\psi,\psi,\eta,\phi]&= S_0[\bar\psi,\psi] + S_B[\eta,\phi] + S_Y[\bar{\psi},\psi,\eta,\phi] \notag \ -\mi\Tr_\mu\ln((\sigma^z\otimes\mathbb{1}_N)(\mi\partial_t-h(\vecq)) + \Sigma[\phi, \eta ])\ +\dots \,,
\end{align}
where $\Tr_\mu$ denotes the integration over momenta $|\mathbf{q}|\geq\mu$. The first line corresponds to Eq.~\eqref{eq:HS_action} restricted to slow fermions with $|\vecq|\leq\mu$. In the second line, all irrelevant terms arising from integrating out the high-momentum fermions have been neglected (the irrelevance has been justified via canonical power counting developed in Sec.~\ref{sec:RG}). 

A comment is in order regarding the choice of the intermediate scale $\mu$. Within the topological phases, the fermions are gapped, and $\mu$ can be sent to zero. In contrast, at the critical point, fermions become gapless  (cf. Sec.~\ref{sec:MicroModel}), and we have to choose $\mu >0$ to avoid singularities. For technical convenience, we replace the IR cutoff $\mu$ by a soft regulator, implemented by restoring a finite bare dissipative gap and noise gap $\sim\mu^2$.

\textit{Gaussian action ---} The quadratic action for the slow fermions is obtained in a saddle point approximation for the Hubbard-Stratonovich fields 
\begin{align}
    \begin{split}
        S_F[\bar\psi,\psi] &= S_0[\bar\psi,\psi] + S_Y[\bar\psi,\psi,\phi=\phi_0,\eta=0]=\int_{\vecq,\omega}\bar{\psi}(\omega,\vecq) G_0^{-1}  \psi(\omega,\vecq)=\int_{\vecq,\omega}\bar\chi(\omega,\vecq) G_{F}^{-1} \chi(\omega,\vecq)\,,
    \end{split} 
\end{align}
where we have defined the Grassmann fields $\chi= U\psi,\bar\chi = \bar\psi U^\dagger$ with the unitary $U(\vecq)$ defined in Eq.~\eqref{eq:WfU}. The bare Green's function $G_0^{-1} = (\sigma^z\otimes\mathbb{1}_N) (\mi\partial_t- h(\vecq)) + \mi W^\dagger\boldsymbol{\phi}_0 W$ is given by Eq.~\eqref{eq:GF} evaluated on the saddle point solution.
Using Eq.~\eqref{eq:WfU} and $\lambda(\vecq)= \vecq^2 + \theta^2$, the bare inverse Green's function in the diagonal basis $G_F^{-1} = U(\vecq) G_0^{-1} U^\dag(\vecq)$ becomes
\begin{align}\label{eq:GFinv}
    G_{F}^{-1}=(\sigma^z\otimes\mathbb{1}_N)(\mi\partial_t-h(\vecq)) +\mi\boldsymbol{\phi}_0(\vecq^2 + \theta^2)\,.
\end{align}

\textit{Slow modes ---} The action for the slow bosonic modes is obtained from an expansion of the trace-log part of the action in slow variations of the $g$ and small fluctuations of $\boldsymbol{\eta}$. For convenience, we will set all coherent terms to zero and dissipative couplings to 1. It can be verified that the addition of compatible coherent couplings does not modify the results discussed below. Using the parametrization~\eqref{eq:u(1)_transf_boson} the action for the hydrodynamic mode takes the form
\begin{align}
\begin{split}
    S_{\text{hydro}}[g,\eta]=S_B[g\eta g^{-1},\varphi_0] -\mi\Tr_\mu\ln[&g^{-1}(\sigma^z\otimes\mathbb{1}_N)\mi\partial_t g + \mi g^{-1}W^\dag g\boldsymbol{\phi_0}g^{-1} W g + \boldsymbol{\eta}]\,.
\end{split}
\end{align}

Recalling that the operator $W$ acts to the right and $W^\dag$ acts to the left (cf. Eq.~\eqref{eq:W_pos}
), we obtain the following useful relations
\begin{align}
\begin{split}
    g^{-1} W g &= W + \mi g^{-1}\overrightarrow{\slashed{\nabla}}_\vecx g \,,\ g^{-1} W^\dag g = W^\dag - \mi g^{-1}\overleftarrow{\slashed{\nabla}}_\vecx g\,, 
\end{split}
\end{align}
and therefore
\begin{equation}\label{eq:trace_log_g}
  S_{\text{hydro}}[g,\eta]=S_B[\eta,\varphi_0]\!-\mi\Tr_\mu\ln\Big[G_0^{-1}  + g^{-1} (\sigma^z\otimes\mathbb{1}_N)\mi\partial_t g + \mi g^{-1}\overleftarrow{\slashed{\nabla}}g\boldsymbol{\phi_0} g^{-1}\overrightarrow{\slashed{\nabla}}g \notag -\left(W^\dag\boldsymbol{\phi}_0 g^{-1}\overrightarrow{\slashed{\nabla}}g - g^{-1}\overleftarrow{\slashed{\nabla}}g\boldsymbol{\phi}_0 W  \right) + \boldsymbol{\eta}\Big]\,.
\end{equation}
For the first term, we have used that the purely bosonic part of the action remains unchanged under the local $U(1)\times U(1)$ transformation, \textit{i.e.}, $S_B[g\phi g^{-1},g\eta g^{-1}]=S[\phi,\eta]$.

To derive the action of the long-wavelength hydrodynamic density fluctuations, we perform a \textit{Wigner-Moyal} expansion of the effective action~\eqref{eq:trace_log_g} to leading order in derivatives for the slowly varying fields  $g^{-1}\partial_t g$ and $g^{-1}\partial_i g$, as well as to first-order in the fluctuations of the order parameter field $\boldsymbol{\eta}$. The density field is then identified as a linear combination of the matrix elements of $\boldsymbol{\eta}$ with its conjugate phase field given by the antisymmetric combination of phases $\theta_q=\theta_+-\theta_-$. The transport coefficients are then evaluated by using the explicit form of the Green's functions~\eqref{eq:bare_propagators}. A diagrammatic representation of the relevant terms is shown in Fig.~\ref{fig:tracelog_diagrams}.  

\textit{Wigner-Moyal Expansion ---} are particularly useful when the fields have smooth/slow variations in the center coordinate, but rapidly vary with respect to the relative coordinate. By performing a Wigner transformation, one systematically eliminates/averages over these fast variations while preserving the slow variations in the field. We briefly review Wigner transformations and the Moyal product for reference. To this end, let us introduce a collective label for space-time coordinates $X=(\vecx,t)$  and frequency-momentum coordinates $Q=(\vecq,\omega)$. We then change coordinates to center and relative coordinates. The spatio-temporal Wigner transform for a space-time bi-local operator $O$ is then defined as follows \cite{kamenev_book}
\begin{align}\label{eq:WT1}
    O(X_1,X_2) &=\int_{Q} e^{-\mi(X_1-X_2)^\top\eta Q}\ O\left(X,Q\right)\,,\ \int_{Q} = \int\frac{\dd q}{(2\pi)^d}\frac{\dd \omega}{2\pi}\, ,
\end{align}
where $X=(t,\vecx)^\top = (X_1 + X_2)/2$ denotes the center coordinate and the metric has signature $\eta=\text{diag}(1,-\mathbf{1}_d)$. Similarly, we have also introduced a collective label for frequency-momentum coordinates $Q=(\vecq,\omega)$ dual to the difference coordinate $X_1-X_2$. The inverse Wigner transform is defined as
\begin{align}\label{eq:WT2}
     O(X,Q) &=\int_{\Delta X}e^{\mi \Delta X^\top\eta Q}\ O\left(X+\frac{\Delta X}{2},X -\frac{\Delta X}{2}\right)\,,\ \int_{\Delta X} = \int\dd t\dd x \, .
\end{align}
Expanding the trace-log formula~\eqref{eq:trace_log_g}, one encounters products of the form $\Tr((A \circ B)^n)\ n\geq 1$, where $\circ$ denotes a space-time convolution of two operators $A$ and $B$ i.e $(A\circ B)(X,Z) = \int_Y A(X,Y)B(Y,Z)$, where $X,Y,Z$ labels the space-time coordinates. In the Wigner basis, such convolutions are simplified via the Moyal product identity \cite{kamenev_book}
\begin{equation}\label{eq:WME}
    (A\circ B)(X,Q)\approx A(X,Q)B(X,Q) + \frac{\mi}{2}\{A,B\} + \hdots\,,
\end{equation}
where the Poisson bracket $\{A,B\} = \nabla_X A\cdot \nabla_Q B- \nabla_Q A\cdot \nabla_X B$ with $\nabla_X = (-\partial_t,\nabla_\vecx)^\top$ and $\nabla_Q = (\partial_\omega,\nabla_\vecq)^\top$ and the $\hdots$ comprise higher order derivative terms which we can neglect to leading order.
\begin{figure}[t!]
    \centering
    \input{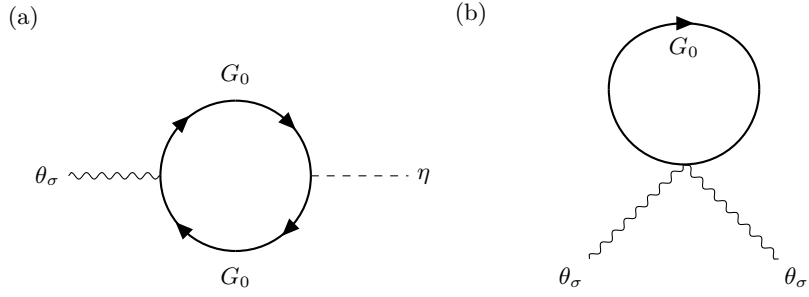}
    \caption{Diagrammatic representation of the generation of the effective bosonic action from the microscopic action~\eqref{eq:HS_action}.}
    \label{fig:tracelog_diagrams}
\end{figure}
\subsection{First order terms}
Expanding the trace log to first order, we obtain the action contribution
\begin{align}
    S^{(1)}[g,\eta] = -\mi\Tr(G_0 \left[g^{-1}(\sigma^z\otimes\mathbb{1})\mi \partial_t g + \mi g^{-1}\overleftarrow{\slashed{\nabla}}g\boldsymbol{\phi_0} g^{-1}\overrightarrow{\slashed{\nabla}}g
    -\left(W^\dag\boldsymbol{\phi}_0 g^{-1}\overrightarrow{\slashed{\nabla}}g - g^{-1}\overleftarrow{\slashed{\nabla}}g\boldsymbol{\phi}_0 W  \right) + \boldsymbol{\eta} \right] )\,.
\end{align}
We analyze each of the terms below. In the remainder of the section, repeated indices are summed over.

\paragraph{Noise terms ---} 
The leading contribution to the noise vertex arises from evaluating the term
\begin{equation}\label{eq:noise_1}
S^{(1)}_{\text{noise}}[g]=\Tr_\mu(G_0 g^{-1}\overleftarrow{\slashed{\nabla}}g\boldsymbol{\phi}_0 g^{-1}\overrightarrow{\slashed{\nabla}}g)=\int_{t_1,t_2}\tr_\mu(G_0^{\sigma\rho}(t_1,t_2)\gamma_j\boldsymbol{\phi}_0^{\rho\sigma}\gamma_k)\partial_j\theta_\rho(t_2)\partial_k\theta_\sigma(t_1)\,.
\end{equation}
The trace $\tr_\mu$ runs over spatial and band indices. In the remainder of the section, the traces without $\mu$ subscript refer to traces over the band indices. As in our discussion of dynamical mean field theory (cf. App.~\ref{app:HS}), traces involving time (anti-)ordered Green's functions are ill-defined at equal times, wherefore evaluating the trace requires a careful analysis. To proceed, we restore the point splitting arguments in the Keldysh action. This is implemented by a replacement of the matrix $c\delta(t-t')\rightarrow c(t,t')=c(\tau=t-t')$ (see Eq.~\eqref{eq:c_matrix}) in the stationary configuration $\boldsymbol{\phi}_0$, with
\begin{align}\label{eq:c_matrix_timereg}
	c(\tau) = \mqty(\delta(\tau+0^+) & 0 \\ -\left(\delta(\tau+0^+)+ \delta(\tau-0^+) \right) & \delta(\tau-0^+) )\,.
\end{align}
This reproduces the correct operator ordering in the fermionic bilinears (cf. Eq.~\eqref{eq:bilinears}), such that Eq.~\eqref{eq:noise_1} is uniquely expressed in terms of the physical particle and hole densities $\Ev{\hat{\psi}^\dag(t)\hat{\psi}(t)}$ and $\Ev{\hat{\psi}(t)\hat{\psi}^\dag(t)}$. Recall that the matrix $\boldsymbol{\phi}_0(\tau)$ is of the same form as Eq.~\eqref{eq:boldsymbols} with $\phi^{\sigma\rho}_U = \varphi_0 c^{\rho\sigma}(\tau)$ and $\phi^{\sigma\rho}_L= \varphi_0 c^{\rho\sigma}(-\tau)$, and $c(\tau)$ as defined in Eq.~\eqref{eq:c_matrix_timereg}. 
Performing this substitution, the contour diagonal terms can be simplified as follows
\begin{align}\label{eq:noise_CD}
&\int_{t_1,t_2}\tr_\mu(G_0^{\sigma\sigma}(t_1,t_2)\gamma_j\boldsymbol{\phi}_0^{\sigma\sigma}(t_2-t_1)\gamma_k)\partial_j\theta_\sigma(t_2)\partial_k\theta_\sigma(t_1)\notag\\
&=\varphi_0\int_t\tr_\mu(G_0^{\sigma\sigma}(t,t+\sigma 0^+) \gamma_j P^U \gamma_k)\partial_j\theta_\sigma(t+\sigma 0^+)\partial_k\theta_\sigma(t) -\varphi_0\int_t\tr_\mu(G_0^{\sigma\sigma}(t,t-\sigma 0^+) \gamma_j P^L \gamma_k)\partial_j\theta_\sigma(t-\sigma 0^+)\partial_k\theta_\sigma(t)\,\notag\\
&=\varphi_0\int_t\left[\tr_\mu(G_0^<(t,t)\gamma_jP^U\gamma_k) - \tr_\mu(G_0^>(t,t)\gamma_jP^L\gamma_k)\right]\partial_j\theta_\sigma(t)\partial_k\theta_\sigma(t)\,,
\end{align}
where $P^U=\text{diag}(\mathbb{1}_{N/2},0)$ and $P^L=\text{diag}(0,\mathbb{1}_{N/2})$. In the last line, we set the regularization to zero after replacing the time (anti-) ordered Green's functions by the physical densities
\begin{align*}
    G_0^{\sigma\sigma}(t,t+\sigma 0^+) &=   G_0^<(t,t+\sigma0^+)\,,\quad
	G_0^{\sigma\sigma}(t,t-\sigma 0^+) = G_0^>(t,t-\sigma0^+)\,.
\end{align*}
Similarly, for the contour off-diagonal terms we substitute $G^{+-}(t,t\pm 0^+) = G^<(t,t\pm 0^+)$ and $G^{-+}(t,t\pm 0^+)=G^>(t,t+\pm 0^+)$, and then set the regularization to zero, which yields 
\begin{align}\label{eq:noise_COD}
    -2\varphi_0\int_{t}\left[\tr_\mu(G^{<}_0(t,t)\gamma_j P^U \gamma_k) - \tr_\mu(G^{>}_0(t,t)\gamma_j P^L \gamma_k)\right]\partial_j\theta_-(t)\partial_k\theta_+(t)\,.
\end{align}
Putting together the ~\eqref{eq:noise_CD} and ~\eqref{eq:noise_COD}, the noise vertex takes the form
\begin{align}
    S^{(1)}_{\text{noise}}[\theta_q] = \varphi_0\int_{X}\left(\int_{Q}\left[\tr(G_0^<(Q)\gamma_j P^U \gamma_k) - \tr(G_0^>(Q)\gamma_j P^L\gamma_k)\right]\right)\partial_j\theta_q(X)\partial_k\theta_q(X)\,.
\end{align}
with $\theta_q(X) = \theta_+(X) - \theta_-(X)$. Using the explicit form of the bare Green's functions shown in Eq.~\eqref{eq:greater_lesser_green_functions} to evaluate the frequency integrals, we obtain
\begin{align}\label{eq:S_noise_first_order}
	S^{(1)}_{\text{noise}}[\theta_q] = \mi\lambda\int \nabla_\vecx\theta_q(\vecx,t) \cdot\nabla_\vecx\theta_q(\vecx,t),\quad \text{where} \quad \lambda = \mu^2\varphi_0 N\int_{|\vecq|\leq\Lambda} \frac{1}{q^2 + \mu^2}\,.
\end{align}

\paragraph{Tadpole diagrams ---} The action $S^{(1)}[g,\eta]$ also comprises terms linear in $\boldsymbol{\eta}$ given by
\begin{equation}
    S_B[\varphi_0,\eta]-\Tr_\mu(\mi G_0\boldsymbol{\eta}) = -\int_X\sum_{\alpha=U,L}\varphi_0c^{\sigma\rho}\eta^{\rho\sigma}_\alpha(X) - \int_X\int_{Q}\tr(\mi G_0^{\sigma\rho}(Q)\boldsymbol{\eta}^{\rho\sigma}(X))\,,
\end{equation}
which explicitly violate the $U(1)\times U(1)$ symmetry of the action. On the right-hand side, we have performed a Wigner transformation, and the trace only runs over band indices. However, using the saddle point equations derived in App.~\ref{app:HS}, we show below that these terms vanish.

First, consider the contour off-diagonal terms
\begin{align}
    &\int_X\varphi_0\sum_{\alpha=U,L}c^{-+}\eta^{-+}_\alpha(X) + \int_X\int_{Q}\left[\tr(\mi G_0^{<}(Q)c^{-+})\eta_L^{-+}(X) - \tr(\mi G_0^{>}(Q)c^{-+})\eta_U^{-+}(X)\right]\,.
\end{align}
Using the saddle point equations, Eq.~\eqref{eq:SP_contour_off_diagonal}, the traces over $G_0^{</>}(Q)$ can be replaced with the saddle point value $\varphi_0 c^{-+}$. This leads to a cancellation of the two terms. Similarly, for the contour diagonal terms, restoring the time regularization for the $\eta$ field using Eq.~\eqref{eq:eta_phi_time_reg}, we obtain
\begin{equation}
    \int_X\varphi_0\sum_{\alpha=U,L}c^{\sigma\sigma}\eta^{\sigma\sigma}_\alpha(X) + \int_X\int_Q\left[\tr(\mi G_0^{\sigma\sigma}(Q))e^{\mi\omega\sigma\epsilon}\eta_L^{\sigma\sigma}(X) - \tr(\mi G_0^{\sigma\sigma})e^{-\mi\omega\sigma\epsilon}\eta_U^{\sigma\sigma}\right].
\end{equation}
Using the saddle point equations, Eq.~\eqref{eq:SP_contour_diagonal} to replace the trace over the $G^{\sigma\sigma}(Q)$ with the saddle point value $\varphi_0$, 
we again find an exact cancellation with the corresponding terms from the bosonic action $S_B[\phi_0,\eta]$. Hence, all tadpole contributions to the action vanish, preserving the $U(1)\times U(1)$ symmetry of the action. 

\paragraph{Total derivative terms ---} Lastly, we consider the tadpole diagrams arising from total derivative terms given by
\begin{align}
    &\mi\Tr_\mu\left(G_0\left[g^{-1} (\sigma^z\otimes\mathbb{1}_N)\mi\partial_t g +W^\dag\boldsymbol{\phi}_0 g^{-1}\overrightarrow{\slashed{\nabla}}g - g^{-1}\overleftarrow{\slashed{\nabla}}g\boldsymbol{\phi}_0 W  \right]\right)\notag\\
    &=\sigma\int_Q\tr(\mi G^{\sigma\sigma}(Q))\int_X\mi\partial_t\theta_\sigma(X) + \int_Q\left[\tr(\mi G^{\sigma\rho}(Q)W^\dag(\vecq)\boldsymbol{\phi}_0^{\rho\sigma}\gamma_j) + \tr(\mi G^{\rho\sigma}(Q)\gamma_j\boldsymbol{\phi}_0^{\sigma\rho}W(\vecq))\right]\int_X\mi\partial_j\theta_\sigma(X)\,.
\end{align}
Although the second term appears to violate the spatial inversion symmetry of the action, evaluating the traces by performing a Wigner-Moyal expansion and using Eqs.~\eqref{eq:greater_lesser_green_functions} and \eqref{eq:time_ord_green_functions}, one finds that these terms vanish due to the trace over the band indices being zero. 
Note that the first order term (and all higher order ones) in the Moyal expansion do not contribute to the trace, since the coefficient of the first derivative term $\sim\int_Q \nabla_Q(G_0(Q) W^\dag(\vecq)\gamma_j) =0$.

\subsection{Second order terms}
For the second-order term in the trace-log expansion, it suffices to retain terms up to first order in the fluctuation field $\boldsymbol{\eta}$, and up to second order in derivatives of $g$
\begin{align}\label{eq:2ndorder_trace}
\begin{split}
    S^{(2)}[g,\eta]&\approx \mi\Tr_\mu\left(G_0\left[g^{-1} (\sigma^z\otimes\mathbb{1}_N)\mi\partial_t g  -\left(W^\dag\boldsymbol{\phi}_0 g^{-1}\overrightarrow{\slashed{\nabla}}g - g^{-1}\overleftarrow{\slashed{\nabla}}g\boldsymbol{\phi}_0 W  \right) \right]G_0\boldsymbol{\eta}\right)\\
    &+\frac{1}{2}\mi\Tr_\mu\left(G_0\left[g^{-1} (\sigma^z\otimes\mathbb{1}_N)\mi\partial_t g  \right]G_0\left[g^{-1} (\sigma^z\otimes\mathbb{1}_N)\mi\partial_t g\right]\right)\\
    &+\frac{1}{2}\mi\Tr_\mu\left(G_0\left(W^\dag\boldsymbol{\phi}_0 g^{-1}\overrightarrow{\slashed{\nabla}}g - g^{-1}\overleftarrow{\slashed{\nabla}}g\boldsymbol{\phi}_0 W  \right)G_0\left(W^\dag\boldsymbol{\phi}_0 g^{-1}\overrightarrow{\slashed{\nabla}}g - g^{-1}\overleftarrow{\slashed{\nabla}}g\boldsymbol{\phi}_0 W  \right)\right)\,.
\end{split}
\end{align} 
Due to spatial inversion symmetry, terms involving both spatial and temporal derivatives have been dropped. They only contribute higher-order derivative terms in the action, which are power-counting irrelevant. Higher order terms involving more than two $\theta$ fields have also been dropped, for the same reason (cf. Sec.~\ref{sec:RG}).

To evaluate the traces in $S^{(2)}[g,\eta]$, we rotate to the Keldysh basis by the transformation \cite{kamenev_book}
\begin{align}\label{eq:Keldysh_rotate}
    G_0\rightarrow R^\top G_0R = G_F \,,\  \boldsymbol{\eta}\rightarrow R^\top\boldsymbol{\eta}R =\mqty(\eta^V & \eta^A \\ \eta^R & \eta^K)\,,
\end{align}
with $R=(\sigma^x+\sigma^z)/\sqrt{2}$. The bare Green's function in the rotated basis has the standard causality structure and reads (cf. Eq.~\eqref{eq:bare_propagators})
\begin{align}
    G_F(Q) = \mqty(g^K(Q) & g^R(Q) \\ g^A(Q) & 0)=\mqty(-2\mi(\omega-K^\dag(\vecq))^{-1}W^\dag(\vecq)\tau_z W(\vecq)(\omega-K(\vecq))^{-1} & (\omega-K^\dag(\vecq))^{-1}\mathbb{1}_N \\ (\omega-K(\vecq))^{-1}\mathbb{1}_N & 0)\,.
\end{align} 
We also introduce a compact notation for the derivatives of $g$ appearing in the trace-log term
\begin{align}
    g^{-1}\overrightarrow{\slashed{\nabla}} g = \mi\slashed{\nabla}_\vecx\theta(X)\,,\quad g^{-1}\overleftarrow{\slashed{\nabla}} g = -\mi\slashed{\nabla}_\vecx\theta(X)\,,
\end{align}
where $\theta(X) = \text{diag}(\theta^+(X),\theta^-(X))\otimes \mathbb{1}_N$. As discussed previously, to evaluate the traces we perform a Wigner transformation (Eq.~\eqref{eq:WT1}), followed by a Moyal expansion (cf. Eq.~\eqref{eq:WME}) to identify the leading order contributions to the hydrodynamic action. 

\paragraph{Dynamical term ---} The dynamical term is obtained by performing a Wigner-Moyal expansion to leading order in time derivation of the following term
\begin{equation}
   S[\partial_t g,\eta] = \mi\Tr_\mu\left(G_0g^{-1} (\sigma^z\otimes\mathbb{1}_N)\mi\partial_t gG_0R^\top\boldsymbol{\eta}R\right)\,.
\end{equation}
To evaluate the right-hand side, we perform a Keldysh rotation and split this term into symmetric and antisymmetric contributions of the phase $\theta$
\begin{align}\label{eq:dynamical_term}
    S[\partial_t g,\eta] &= -\Tr_\mu\left(G_F(\sigma^x\otimes\mathbb{1}_N)G_FR^\top\boldsymbol{\eta}R\right)\mi\partial_t\theta_c - \frac{1}{2}\Tr_\mu\left(G_FG_FR^\top\boldsymbol{\eta}R\right)\mi\partial_t\theta_q\,,
\end{align}
where $\theta_c = (\theta_+ + \theta_-)/2$ and $\theta_q=\theta_+-\theta_-$.
To evaluate the terms in Eq.~\eqref{eq:dynamical_term}, we perform a Wigner-Moyal expansion to first order in time derivatives and zeroth order in space derivatives. For the first term, this yields
\begin{align}
\begin{split}
    \Tr\left(G_F (\sigma^x\otimes\mathbb{1}_N) G_FR^\top\boldsymbol{\eta}R\right)\mi\partial_t\theta_c&\approx\int_{X}\int_{Q} \tr\left(g^A(Q)g^A(Q)\eta^A(X)\right)\mi\partial_t\theta_c(X) + \int_{X}\int_{Q} \tr\left(g^R(Q)g^R(Q)\eta^R(X)\right)\mi\partial_t\theta_c(X)\\
   &+ \int_{X}\int_{Q} \tr\left(\left[g^R(Q)g^K(Q)+g^A(Q)g^K(Q)\right]\eta^V(X)\right)\mi\partial_t\theta_c(X)=0\,.
\end{split}
\end{align}
On the RHS, we have used that the bare Green's functions are translation invariant, and thus their Wigner transform depends only on frequency and momentum. The frequency integrals vanish due to the causality structure of the Green's functions. Similarly, for the second term in Eq.~\eqref{eq:dynamical_term}, one finds the following non-vanishing contributions 
\begin{align}
    -\frac{1}{2}\Tr_\mu(G_FG_FR^\top\boldsymbol{\eta}R)\mi\partial_t\theta_q&\approx -\frac{1}{2}\int_{X}\left[\int_{Q} \tr(g^A(Q)g^R(Q)(\eta^K(X) +\eta^V(X))) + \int_{Q} \tr(g^K(Q)g^K(Q)\eta^V(X))\right]\mi\partial_t\theta_q(X)\notag\\
    &=\frac{1}{4\varphi_0}\int_X\int_{\vecq\leq\Lambda}\frac{1}{(q^2 + \mu^2)}\tr(\eta^V(X)-\eta^K(X))\mi\partial_t\theta_q(X) \notag\\
    &=-\mi Z_t\int_X\theta_q(X)\partial_t\tr(\sigma^x\boldsymbol{\eta}(X))\,,\quad \text{with}\quad Z_t = \frac{1}{4\varphi_0}\int_{\vecq\leq\Lambda} \frac{1}{q^2 + \mu^2}.
\end{align}
In the last line, the trace runs over both contour and band indices. 
In the second line, we have performed the frequency integrals using the bare Green's functions ~\eqref{eq:bare_propagators} and in the final step we rotate back to the $\pm$ basis to find that
\begin{align}\label{eq:eta}
-\frac{\mi}{2} \text{tr}\langle\sigma^x\boldsymbol{\eta}(\vecx,t)\rangle=&\sum_{\alpha\in U} \langle \hat{x}^\dag_\alpha(\vecx,t)\hat{x}_\alpha(\vecx,t)\rangle - \sum_{\alpha\in L} \langle \hat{x}_\alpha(\vecx,t)\hat{x}^\dag_\alpha(\vecx,t) \rangle\,,
\end{align}
which indeed corresponds to the local particle density conserved by the strong $U(1)$ symmetry. Furthermore,  the above calculation shows that its conjugate field is given by the phase $\theta_q = \theta^+ - \theta^-$. 

\paragraph{Kinetic terms ---}

We now turn to the kinetic terms. Due to the spatial inversion symmetry of the microscopic model, we expect the leading kinetic contribution to take the form of a diffusion term as discussed in Sec.~\ref{app:strong_symm_hydro}. It arises from the following contribution to Eq.~\eqref{eq:trace_log_g},
\begin{align}
    \begin{split}
         S_\text{diff}[g,\eta] =& -\mi\Tr_\mu(G_0\underbrace{\left(W^\dag\boldsymbol{\phi}_0 g^{-1}\overrightarrow{\slashed{\nabla}}g - g^{-1}\overleftarrow{\slashed{\nabla}}g\boldsymbol{\phi}_0 W  \right)}_{\Gamma[\theta]}G_0\boldsymbol{\eta}).
    \end{split}
\end{align}

While the calculation of this term is more involved, it proceeds analogously to the dynamical term and yields the diffusive term $n_c\nabla^2\tilde\theta_q$ in the effective action.

To this end, we first perform a Keldysh rotation (cf. Eq.~\eqref{eq:Keldysh_rotate}) and express the components of the rotated matrix $R^\top\Gamma[\theta]R$ in terms of symmetric and anti-symmetric combinations of the phases
\begin{equation}\label{eq:Gamma3}
    R^\top\Gamma[\theta]R = \frac{\mi}{2}\varphi_0\mqty(\slashed{\nabla}\theta_q W -W^\dag \slashed{\nabla}\theta_q & 2\slashed{\nabla}\theta_q \gamma_{d+1}W \\
    2W^\dag\gamma_{d+1}\slashed{\nabla}\theta_q & W^\dag\slashed{\nabla}\theta_q-\slashed{\nabla}\theta_q W) + \mi\varphi_0\mqty(0 &-(W^\dag\slashed{\nabla}\theta_c+\slashed{\nabla}\theta_c W)\\ W^\dag\slashed{\nabla}\theta_c+\slashed{\nabla}\theta_c W & 2W^\dag\gamma_{d+1}\slashed{\nabla}\theta_c+\text{h.c} ).
\end{equation}
The traces are most conveniently performed by projecting onto each component of the matrix $R^\top\boldsymbol{\eta} R$. 

We first consider the projection onto the $\eta^K$ component, which yields the following contribution to the action
\begin{align}\label{eq:diff_keldysh}
    S_\text{diff} \ni\ &\frac{\varphi_0}{2}\tr_\mu(g^R\left(\slashed{\nabla}\theta_q W -W^\dag \slashed{\nabla}\theta_q\right) g^A\eta^K)).
\end{align}
As a reminder, the traces without $\mu$ subscript refer to traces over the band indices. To evaluate the traces, we perform a Wigner-Moyal expansion to first order in spatial derivatives and zeroth order in time derivatives. For the first term in Eq.~\eqref{eq:diff_keldysh} we obtain
{\small
\begin{align}
    \begin{split}
        \tr_\mu(g^R\slashed{\nabla}\theta_q W g^A\eta^K) &\approx\int_{X,Q}\tr\Bigg[\left(g^R(Q)\slashed{\nabla}_\vecx\theta_q(X)-\frac{\mi}{2}\nabla_\vecq g^R(Q)\cdot\nabla_\vecx(\slashed{\nabla}_x\theta_q(X))+\dots\right)\times\\
        &\quad\quad\quad\left(W(\vecq)g^A(Q)\eta^K(X)-\frac{\mi}{2}\nabla_\vecq(W(\vecq)g^A(Q))\cdot\nabla_\vecx\eta^K(X)+\dots \right)\Bigg]\\
        &\approx-\frac{\mi}{2}\int_{X,Q}\Big[\tr(g^R(Q)\slashed{\nabla}_\vecx\theta_q(X) \nabla_\vecq(W(\vecq)g^A(Q))\cdot\nabla_\vecx\eta^K(X))\\&\quad\quad\quad\quad\quad\;\;\;+\tr(\nabla_\vecq g^R(Q)\cdot\nabla_\vecx(\slashed{\nabla}_x\theta_q(X))W(\vecq)g^A(Q)\eta^K(X))\Big]\,.
    \end{split}
\end{align}
}
Since the coefficient of the linear derivative term $\partial_j\theta_q$ given by $\int_Q\tr(g^R(Q)\gamma_j W(\vecq) g^A(Q)\eta^K(X)) =\int_Q\tr(g^R(Q)\gamma_j \left(-\slashed{\vecq} + \mu\gamma_{d+1}\right) g^A(Q)\eta^K(X))$ vanishes, in the second line we have retained only terms involving up to two spatial derivatives. The vanishing of the coefficient of the linear derivative term is seen as follows: the first term is antisymmetric under $\vecq\rightarrow-\vecq$, leading to a vanishing angular integral over the loop momenta, 
Analogously, for the second term in Eq.~\eqref{eq:diff_keldysh} we obtain
{\small
\begin{align}
    \begin{split}
        \tr_\mu(g^RW^\dag \slashed{\nabla}\theta_q g^A\eta^K) &\approx\int_{X,Q}\tr\Bigg[\left(g^R(Q)W^\dag(\vecq)\slashed{\nabla}_\vecx\theta_q(X)-\frac{\mi}{2}\nabla_q (g^R(Q)W^\dag(\vecq))\cdot\nabla_\vecx(\slashed{\nabla}_\vecx\theta_q(X))+\dots\right)\times\\
        &\quad\quad\quad\left(g^A(Q)\eta^K(X)-\frac{\mi}{2}\nabla_q(g^A(Q))\cdot\nabla_\vecx\eta^K(X)+\dots \right)\Bigg]\\
        &\approx-\frac{\mi}{2}\int_{X,Q}\Big[\tr(\nabla_\vecq (g^R(Q)W^\dag(\vecq))\cdot\nabla_\vecx(\slashed{\nabla}_\vecx\theta_q(X)) g^A(Q) \eta^K(X))\\&\quad\quad\quad\quad\quad\;\;\;+\tr(g^R(Q)W^\dag(\vecq)\slashed{\nabla}_\vecx\theta_q(X)\nabla_\vecq(g^A(Q))\cdot\nabla_{\vecx}\eta^K(X))\Big]\,.
    \end{split}
\end{align}
}
Combining the two equations above, the terms involving derivatives of the Green's functions cancel out, and the only non-vanishing contribution to Eq.~\eqref{eq:diff_keldysh} reads
\begin{align}
    \frac{\varphi_0}{2}\tr_\mu(g^R\left(\slashed{\nabla}\theta_q W -W^\dag \slashed{\nabla}\theta_q\right) g^A\eta^K))&\approx-\frac{\mi}{4}\varphi_0\int_{X,Q}\Big[\tr(g^R(Q)\slashed{\nabla}_\vecx\theta_q(X)\nabla_\vecq W(\vecq) g^A(Q) \nabla_\vecx\eta^K(X))\notag\\
    &\quad\quad\quad\quad\quad\quad\;\;\;+\frac{\mi}{4}\varphi_0\int_{X,Q}\tr(g^R(Q)\nabla_\vecq W^\dag(\vecq)\cdot\nabla_\vecx(\slashed{\nabla}_\vecx\theta_q(X))g^A(Q)\eta^K(X))\Big] \notag\\
    &=-\frac{\mi}{4}\varphi_0\int_{X,Q}\tr(g^R(Q)\{\gamma_j,\gamma_i\} g^A(Q) \eta^K(X))\partial_j\partial_i\theta_q(X)\notag\\
    &=-\frac{\mi}{2}\varphi_0\int_{X,Q}\tr(g^R(Q) g^A(Q)\eta^K(X))\nabla_\vecx^2\theta_q(X)\,.
\end{align}
In going to the second line we used $\partial_{q_j} W(\vecq) = -\gamma_j$, and in the final step we used the Clifford relation $\{\gamma_i,\gamma_j\}=2\delta_{ij}$. 

Next, consider the projection onto the $\eta^{R}$ sector
\begin{align}
    &\frac{\varphi_0}{2}\tr_\mu(g^k\left(\slashed{\nabla}\theta_q W -W^\dag \slashed{\nabla}\theta_q\right)g^R\eta^R) + \varphi_0\tr_\mu(g^R W^\dag\gamma_{d+1}\slashed{\nabla}\theta_qg^R\eta^K)+\frac{\varphi_0}{2}\tr_\mu(g^R \left(W^\dag\slashed{\nabla}\theta_c+\slashed{\nabla}\theta_c W\right)g^R\eta^K)\,.    
\end{align}
The second and the third terms vanish due to the causality structure of the Green's functions, and thus the $\theta_c$ terms drop out from the action. To evaluate the first term, we perform a Wigner-Moyal expansion to leading order in spatial derivatives and zeroth order in time derivatives. The first-order spatial derivatives of $\theta_q$ vanish, and the second derivative terms are given by
\begin{align}
    \tr_\mu(g^k\left(\slashed{\nabla}\theta_q W -W^\dag \slashed{\nabla}\theta_q\right)g^R\eta^R) \approx&-\frac{\mi}{2}\varphi_0\int_{X,Q}\tr(\nabla_\vecq g^K(Q) \cdot\nabla_\vecx(\slashed{\nabla}_\vecx\theta_q(X)) W(\vecq) g^R(Q)\eta^R(X))\notag\\
    &+\frac{\mi}{2}\varphi_0\int_{X,Q}\tr(\nabla_\vecq (g^K(Q)W^\dag(\vecq)) \cdot\nabla_\vecx(\slashed{\nabla}_\vecx\theta_q(X)) g^R(Q)\eta^R(X))\notag\\
    &-\frac{\mi}{2}\varphi_0\int_{X,Q}\tr( g^K(Q) \slashed{\nabla}_\vecx\theta_q(X)\nabla_\vecq( W(\vecq) g^R(Q))\cdot\nabla_\vecx(\eta^R(X)))\notag\\
    &+\frac{\mi}{2}\varphi_0\int_{X,Q}\tr( g^K(Q) W^\dag(\vecq)\slashed{\nabla}_\vecx\theta_q(X) \nabla_\vecq g^R(Q)\cdot\nabla_\vecx(\eta^R(X)))\notag\,.
\end{align}
We illustrate the evaluation of one such term, consider
\begin{align}
    &\int_{X,Q}\tr(\partial_{q_i} g^K(Q)\gamma_jW(\vecq)g^R(Q)\eta^R(X))\partial_{x_i}\partial_{x_j}\theta_q(X)\notag\\
    &=-2\mi\int_{X,Q}\tr[\left((\partial_{q_i}g^R(Q)) W^\dag(\vecq) \gamma_{d+1} W(\vecq)g^A(Q) - g^R(Q)\gamma_i\gamma_{d+1} W(\vecq)g^A(Q) + \text{h.c} \right)\gamma_jW(\vecq)g^R(Q)\eta^R(X)]\partial_{x_i}\partial_{x_j}\theta_q(X)\notag\,.
\end{align}
Since the bare Green's functions $g^{R/A}(Q)$ and $\boldsymbol{\eta}(X)$ are diagonal in band space (cf. Eq.~\eqref{eq:boldsymbols} and Eq.~\eqref{eq:bare_propagators}), the trace over band space reduces to evaluating expressions such as $\tr(W^\dag(\vecq)\gamma_{d+1} W(\vecq)\gamma_jW(\vecq) )$, $\tr(\gamma_i\gamma_{d+1}W(\vecq)\gamma_jW(\vecq))$ and their Hermitian conjugates. These traces vanish as they involve an odd number of gamma matrices. An analogous argument applies to the remaining terms. Further, due to the conjugation symmetry of the action ensuring Hermiticity, the projection onto the $\eta^A$ sector also vanishes. Lastly, the projection onto the $\eta^V$ sector yields
{\small
\begin{align}
\begin{split}
    &\frac{\varphi_0}{2}\Big[-\tr_\mu(g^R\left(\slashed{\nabla}\theta_q W -W^\dag \slashed{\nabla}\theta_q\right)g^A\eta^V )+\tr_\mu(g^k\left(\slashed{\nabla}\theta_q W -W^\dag \slashed{\nabla}\theta_q\right)g^k\eta^V )+2\tr_\mu(g^k\slashed{\nabla}\theta_q \gamma_{d+1}W g^A\eta^V)+ 2\tr_\mu(g^R W^\dag\gamma_{d+1}\slashed{\nabla}\theta_q g^K\eta^V)\Big]\\
    &+\varphi_0\Big[\tr_\mu(g^R\left(W^\dag\slashed{\nabla}\theta_c +\slashed{\nabla}\theta_c W\right)g^K\eta^V) -\tr_\mu(g^K\left(W^\dag\slashed{\nabla}\theta_c +\slashed{\nabla}\theta_c W\right)g^A\eta^V) + 2\tr_\mu(g^R\left(W^\dag\gamma_{d+1}\slashed{\nabla}\theta_c +\text{h.c}\right)g^A\eta^V)\Big].
    \end{split}
\end{align}
}
The second line comprises terms with derivatives of $\theta_c$. However, these terms vanish since they always involve a trace over an odd number of gamma matrices. Evaluating the traces in the first line, we find a single non-vanishing contribution 
\begin{align}
    -\frac{\varphi_0}{2}\tr_\mu(g^R\left(\slashed{\nabla}\theta_q W -W^\dag \slashed{\nabla}\theta_q\right)g^A\eta^V) &\approx\mi\frac{\varphi_0}{2}\int_{X,Q}\tr(g^R(Q)g^A(Q)\eta^V(X))\nabla^2_\vecx\theta_q(X).
\end{align}
Summarizing, we have derived the following contribution to the action
\begin{equation}
    S_\text{diff} = \mi\frac{\varphi_0}{2}\int_{X,Q}\tr(g^R(Q)g^A(Q)(\eta^V(X)-\eta^K(X)))\nabla^2_\vecx \theta_q(X) = -D\int_X\theta_q(X)\nabla_\vecx^2n_c(X).
\end{equation}
In the last equality, we evaluated the frequency integral using the explicit form of the Green's functions and rotated to the $\pm$ basis. Note that we find the same combination of the fields $\eta_V,\eta_K$ as the conjugate field to $\theta_q$, which we previously identified as the density.

\paragraph{Corrections to noise coefficient $\lambda$ ---} Lastly, we consider the term
\begin{align}
    S^{(2)}_\text{noise} =\frac{\mi}{2}\Tr_\mu(G_0g^{-1} (\sigma^z\otimes\mathbb{1}_N)\mi\partial_t gG_0g^{-1} (\sigma^z\otimes\mathbb{1}_N)\mi\partial_t g) + \frac{\mi}{2}\Tr_\mu(G_0\Gamma[\theta]G_0\Gamma[\theta])\,,
\end{align}
in Eq.~\eqref{eq:2ndorder_trace} which, to leading order in Wigner-Moyal expansion, provides an additional contribution to the noise coefficient $\lambda$ (cf. Eq.~\eqref{eq:S_noise_first_order}).

To simplify the first term in $S^{(2)}_\text{noise}$, we perform a Keldysh rotation followed by a Wigner-Moyal expansion to leading order in time derivatives and zeroth order in spatial derivatives, resulting in 
\begin{align}
\begin{split}
     &\frac{\mi}{2}\int_{Q,X}\tr(G_F(Q)(\sigma^x\otimes\mathbb{1}_N)G_F(Q)(\sigma^x\otimes\mathbb{1}_N))(\partial_t\theta_c(X))^2+\frac{\mi}{2}\int_{Q,X}\tr(G_F(Q)G_F(Q))(\partial_t\theta_q(X))^2\\
    &+\mi\int_{Q,X}\tr(G_F(Q)(\sigma^x\otimes\mathbb{1}_N)G_F(Q))\partial_t\theta_q(X)\partial_t\theta_c(X)\,.
\end{split}
\end{align}
Evaluating the traces, we find that all of the above terms vanish due to the causality structure of the Green's functions. For the second term, we first use Eq.~\eqref{eq:Gamma3} to express the trace in terms of derivatives of $\theta_q$ and $\theta_c$. Followed by this, we perform a Wigner-Moyal expansion to leading order in spatial derivatives and zeroth order in time derivatives. The coefficients for $\partial_j\theta_c(X)\partial_k\theta_c(X)$ vanish due to the causality structure of the Green's functions. Moreover, the coefficients of the cross terms $\partial_j\theta_c(X)\partial_k\theta_q(X)$ are also zero, due to vanishing trace over the band indices. Hence, we find that the only non-vanishing contribution to the trace is given by
\begin{align}
    \begin{split}
        S_\text{noise}^{(2)}[\theta_q] =&\frac{\varphi_0^2}{8}\int_{Q,X}\tr(g^K(Q)(\slashed{\nabla}_j\theta_q(X)W -W^\dag\slashed{\nabla}_j\theta_q(X))g^K(\slashed{\nabla}_j\theta_q(X)W -W^\dag\slashed{\nabla}_j\theta_q(X)))\\
        &+\frac{\varphi_0^2}{2}\int_{Q,X}\tr(g^K(Q)(\slashed{\nabla}_j\theta_q(X)W -W^\dag\slashed{\nabla}_j\theta_q(X))g^RW^\dag\gamma_{d+1}\slashed{\nabla}_j\theta_q(X)))+\text{h.c.}\\
        &-\frac{\varphi_0^2}{4}\int_{Q,X}\tr(g^R(Q)(\slashed{\nabla}_j\theta_q(X)W -W^\dag\slashed{\nabla}_j\theta_q(X))g^A(\slashed{\nabla}_j\theta_q(X)W -W^\dag\slashed{\nabla}_j\theta_q(X))) \,.
        \end{split}
\end{align}
Evaluating the frequency integrals using~\eqref{eq:fermion_greens_fct}, we find that the corrections to the noise vertex at second order come exactly with a pre-factor $-\mi\lambda$, leading to a cancellation with the first order contribution~\eqref{eq:S_noise_first_order}. It is clear that higher-order terms in the trace-log expansion do not contribute to the noise vertex, thus leading to the conclusion that the Keldysh component for the hydrodynamic mode vanishes, indicating the absence of density fluctuations. 

It remains to verify if hydrodynamic interactions can lead to a renormalization of the noise such that it acquires a $\mathcal{O}(1)$ value. However, we do not expect this to be the case for Class AIII due to the presence of chiral symmetry, as indicated in the main text. 

\section{Weak \texorpdfstring{$\mathrm{Pin}(d)$}{} symmetry}
\label{app:weakPind}

\subsection{Brief review of \texorpdfstring{$\mathrm{Spin}(d)$}{} and \texorpdfstring{$\mathrm{Pin}(d)$}{} groups}\label{app:weakPind1}
As discussed in the main text, the Keldysh action for near-critical dissipative topological insulators with Dirac stationary states, in addition to chiral symmetry, is also constrained by the weak $\mathrm{Pin}(d)$ symmetry. Let us briefly review some basic facts about this group. 
The degrees of freedom are \emph{spinors}, which transform in the fundamental (complex) representation of the Spin$(d+1)$ group, the double cover of the rotation group $\mathrm{SO}(d+1)$. While the rotation group $\mathrm{SO}(d+1)$ acts on the base manifold $\mathbb{R}^{d+1}$ via global rotations, the double cover, Spin$(d+1)$ also acts on the spinors via unitary transformations generated by the spin algebra, which is spanned by the generators given in Eq.~\eqref{eq:spinalgebra} below \cite{Zinn-Justin_book}. The full orthogonal group $\mathrm{O}(d+1)$, in addition to rotations, also includes spatio-temporal reflections. The double cover of $\mathrm{O}(d+1)$ is the $\mathrm{Pin}(d+1)$ group, which also has a non-trivial action on the spinor fields, as described in Eq.~\eqref{eq:pinaction} below, with a subtle distinction between even and odd dimensions. 

\textit{Complex spinor representations ---} Spinor or spin representations are the irreducible representations of the Spin group realized over fermionic Fock space. In odd spacetime dimensions the spinor representation is the only irreducible representation, up to isomorphisms. These are known as Dirac spinors. However, in even dimensions, the spinor representation is not irreducible and can be decomposed into the direct sum of two representations, so called Weyl spinors, which are transformed into each other by the chirality matrix $\gamma_{d+2}$. 
With the inclusion of the full $\mathrm{Pin}(d+1)$ symmetry, we restrict our attention to theories with full parity symmetry and exclude those involving a single Weyl spinor~\cite{Zinn-Justin_book}. 

To this end, we first define the action of $\mathrm{Spin}(d+1)$ and $\mathrm{Pin}(d+1)$ on the spinor fields below and discuss its non-relativistic reduction. In Euclidean spacetime, spinors $\hat\psi(\vecx,t)$ transform in the fundamental representation of the $\mathrm{Spin}(d+1)$ group
\begin{equation}\label{eq:psi_spin_transf}
    \hat\psi_\alpha(\vec{X})\rightarrow \hat U_g \hat\psi_\alpha(\vec{X}) \hat U_g^\dag \equiv \sum_\beta \mathcal{U}_{\alpha\beta}(\alpha)\hat\psi_\beta(R\vec{X})\,,
\end{equation}
where $R\in \mathrm{SO}(d+1)$ is a rotation matrix acting on the coordinates $\vec{X} = (\vecx,t)$ and 
\begin{equation}\label{eq:spinalgebra}
\mathcal{U}(\alpha)=\exp(\frac{1}{4}\sum_{\mu<\nu}^{d+1}\alpha_{\mu\nu}\sigma^{\mu\nu})\,,
\end{equation}
is a unitary representation of Spin$(d+1)$ with $\sigma^{\mu\nu} = \frac{1}{2}[\gamma_\mu,\gamma_\nu]$ the generators of the spin algebra (\textit{i.e.}, the Lie algebra associated to the Spin group) and $\alpha_{\mu\nu}=-\alpha_{\nu\mu}$ a real antisymmetric matrix. 

In a non-relativistic setting, we restrict to the subgroup $\mathrm{Spin}(d)\subset\mathrm{Spin}(d+1)$ spanned by the generators $\sigma^{ij} = \frac{1}{2}[\gamma_i,\gamma_j]$ with $i<j=1,\dots,d$ which commutes with $\gamma_{d+1}$, \textit{i.e.}, $[\sigma^{ij},\gamma_{d+1}] = 0\, \forall\,i<j=1,\dots,d$. On the base manifold, these transformations preserve the temporal direction, and acts only on the spatial coordinates by rotations $R\in\mathrm{SO}(d)\subset \mathrm{SO}(d+1)$. The extension to the $\mathrm{Pin}(d)$ group follows similarly.

As discussed above, in even spacetime dimensions, the $\mathrm{Pin}(d)$ subgroup, also includes spatial reflections $x_i\rightarrow -x_i$, implemented as
\begin{align}\label{eq:pinaction}
    \begin{split}
        & \psi(\vecx,t)\rightarrow \Pi_i \psi(\dots,-x_i,\dots,t),\\
        & \bar\psi^\top(\vecx,t)\rightarrow\bar\psi^\top(\dots,-x_i,\dots,t)\Pi_i^\dag
    \end{split}
\end{align}
where the matrix $\Pi_i = \mi\gamma_{d+2}\gamma_i,\, i=1,\dots, d\,$. It is straightforward to verify that the transformation maps $\gamma_i\rightarrow-\gamma_i$ and leaves $\gamma_{d+1}$ and $\gamma_j\,,\ j\neq i$ invariant. Note that the transformation $\Pi_i^2 = \mathbb{1}$ with the definition above, thereby defining the Pin$^+(d)$ group. Alternatively, one can also define this without the $\mi$ factor, and the transformation then squares to $-1$, defining a Pin$^-(d)$ group. These groups are not isomorphic to each other; however, the natural choice here is the former since, the quadratic form given by the Euclidean metric is positive definite. The double cover of Pin$^\pm(d)$ in both cases is the orthogonal group O$(d)$.

In odd spacetime dimensions, since $\gamma_{d+2}\sim\mathbb{1}$ is trivial, a (spatial-) parity transformation for the spinor fields can be implemented by
\begin{equation}
    \psi(\vecx) \rightarrow \gamma_{d+1}\psi(-\vecx) \quad \bar\psi^\top(\vecx)\rightarrow \bar\psi^\top(-\vecx)\gamma_{d+1}
\end{equation}
which sends $\gamma_i\rightarrow -\gamma_i$ while leaving $\gamma_{d+1}$ invariant and squares to $\mathbb{1}$.

\subsection{Weak symmetry of the continuum action}\label{app:weakPind2}
For near-critical dissipative topological insulators with Dirac stationary states discussed in Sec.~\ref{sec:dynamical_MF}, we now show that the jump operators \eqref{eq:jumpop2}, with almost unitary matrices of the form \eqref{eq:unitary_dirac}, transform non-trivially under the action of the $\mathrm{Pin}(d)$ group (cf. Eq.~\ref{eq:weaksymm_jump}), and that the microscopic action~\eqref{eq:micro_action} is invariant under this transformation. We then specify the necessary and sufficient conditions for the Gaussian theory to be $\mathrm{Pin}(d)$ invariant in terms of the $H,D$ and $P$ matrices.

Recall that under the action of the $\mathrm{Spin}(d)$ group, the field operators $\hat \psi$ transform as~\eqref{eq:psi_spin_transf}. This implies that the $\hat x$ operators are charged under the action of the symmetry group, \textit{i.e.}, they transform as a singlet under the group action
\begin{align}\label{eq:x_spin_transf}
    \begin{split}
        \hat x_{\alpha\in U}(\vecx) &\to \sum_{\beta} \mathcal{U}_{\alpha\beta}(\alpha) \hat{x}_\beta(R\vecx) = \sum_{\beta\in U} \mathcal{V}_{\alpha\beta}(\alpha)\hat x_{\beta}(R\vecx)\,,
    \end{split}
\end{align}
and similarly for the lower band. Here we have used that in the basis where $\gamma_{d+1} = \tau_z$, the generators $\mathcal{U}(\alpha)$ take a block diagonal form: $\mathcal{U}(\alpha) =\mathbb{1}_2\otimes \mathcal{V}(\alpha)$, which follows from $[\sigma^{ij},\gamma_{d+1}] = 0$.
A sufficient condition for the operator $\hat x$ to carry $\mathrm{Spin}(d)$ charge is that $W$ transforms as a (pseudo-)scalar. In particular, for $W(\vecq) = \alpha_\mu(\vecq)\gamma_\mu$, the spatial components $\alpha_i$ for $i=1,\dots,d$ are required to transform as a (pseudo-)vector and the temporal component $\alpha_{d+1}$ as a (pseudo-)scalar. 

Together with Eq.~\eqref{eq:psi_spin_transf} and Eq.~\eqref{eq:x_spin_transf}, the quadratic Lindblad operators $L^{(X)}_{\alpha,\beta}$ transform as a rank-$2$ tensor,
\begin{align}
    L^{(U)}_{\alpha,\beta}(\vecx) &\rightarrow \sum_{\alpha'\in I,\beta'\in U}\mathcal{V}^*_{\alpha\alpha'}(\alpha) L^{(U)}_{\alpha'\beta'}(R\vecx) \mathcal{V}_{\beta\beta'}(\alpha),\\
    L^{(L)}_{\alpha,\beta}(\vecx) &\rightarrow \sum_{\alpha'\in I,\beta'\in L}\mathcal{V}_{\alpha\alpha'}(\alpha) L^{(L)}_{\alpha'\beta'}(R\vecx) \mathcal{V}^*_{\beta\beta'}(\alpha)\,,
\end{align}
under the action of $\mathrm{Spin}(d)$. Since the Lindblad equation involves a summation over both upper and lower bands, invariance of the Lindblad equation follows from the unitarity of $\mathcal{V}(\alpha)$. Invariance under the full $\mathrm{Pin}(d)$ group follows immediately since the Dirac operator~\eqref{eq:unitary_dirac} transforms as a scalar under reflections. 
In the field theory representation, as mentioned in the main text, weak symmetries act trivially on the Keldysh contour and only act in band space (cf. Eq.~\eqref{eq:Pin_spinorfields}). Thus, the invariance of the Lindblad equation under the weak $\mathrm{Pin}(d)$ symmetry implies that  the microscopic action~\eqref{eq:micro_action} is invariant under the transformation~\eqref{eq:Pin_spinorfields}. Note that the time derivative term in the action is invariant under the group action.

\textit{Weak symmetry conditions on $H,D$ and $P$ ---} The weak $\mathrm{Pin}(d)$ symmetry of the microscopic theory imposes constrains on the $H,D$ and $P$ matrices. Under the transformation~\eqref{eq:Pin_spinorfields}, the Gaussian action is invariant under the weak $\mathrm{Spin}(d)$ symmetry, if and only if the matrices $H,D,P$ satisfy
\begin{align}\label{eq:spinsgauss}
\begin{split}
    &\mathcal{U}(\alpha)^\dag H(R^\top\vecq) \mathcal{U}(\alpha) =  H(\vecq), \\
    &\mathcal{U}(\alpha)^\dag D(R^\top\vecq) \mathcal{U}(\alpha) =  D(\vecq), \\
    &\mathcal{U}(\alpha)^\dag P(R^\top\vecq) \mathcal{U}(\alpha) =  P(\vecq). 
\end{split}
\end{align}
Invariance under the full $\mathrm{Pin}(d)$ group then leads to the following constraints
\begin{align}
    \begin{split}
        &\Pi_i^\dag H(\dots,-q_i\dots) \Pi_i =  H(\vecq), \\
        &\Pi_i^\dag D(\dots,-q_i\dots) \Pi_i =  D(\vecq), \\
        &\Pi_i^\dag P(\dots,-q_i\dots) \Pi_i =  P(\vecq). \quad \forall\ i=1,\dots,d
    \end{split}
\end{align}

It is now straightforward to verify that the Gaussian action~\eqref{eq:bare_action} is invariant under $\mathrm{Pin}(d)$. Using the relations $\mathcal{U}^\dag(\alpha)\gamma_{d+1} \mathcal{U}(\alpha) = \gamma_{d+1}$ and 
\begin{equation}
    \Pi_i^\dag \gamma_{d+1}\Pi_i = \gamma_{d+1}\,,\ \Pi_i^\dag \gamma_j\Pi_i = 
        \begin{cases}
            -\gamma_i & j=i \\
            \gamma_j & j\neq i
        \end{cases}.
\end{equation}
one can indeed verify that $D$, $H$, and $P$ as derived in Eq.~\eqref{eq:D_P_effective} satisfies the above relations, implying that the Gaussian action~\eqref{eq:bare_action} is invariant under $\mathrm{Pin}(d)$.

\section{Gaussian actions in Class AIII with weak \texorpdfstring{$\mathrm{Pin}(d)$}{} symmetry}\label{app:weak_symm_HDP}

In this appendix, we discuss the construction of the $H,D$, and $P$ matrices, using only the weak $\mathrm{Pin}(d)$ symmetry and chiral symmetry in even spacetime dimensions. To proceed, recall that any $N\times N$, with ($=2^{D}$), Hermitian matrix can be expanded in an orthogonal basis provided by the generators of the Clifford algebra. Concretely, the basis set $\{\Gamma^A\}$, where $A$ denotes a collective index for tensors of rank $r_A\leq D$, contains the $\mathbb{1}$ matrix, the Clifford generators $\gamma_\mu$ and all possible totally antisymmetric products of the Clifford generators $\sim\gamma^{[\mu_1}\dots\gamma^{\mu_k]}$, $k\leq D$. It is straightforward to verify that the basis elements are orthogonal with respect to the trace, \textit{i.e.}, $D^{-1}\tr(\Gamma^A \Gamma^B) = \delta_{AB}$. Thus any Hermitian matrix $X=H,D,P$ can be written as
\begin{equation}
    X(\vecq) = \frac{1}{D}\sum_A\tr(X(\vecq)\Gamma^A)\Gamma^A\,.
\end{equation}
To leading order in a gradient expansion, we retain only terms up to second order in spatial derivatives. It is then sufficient to consider the subset of the basis elements $\{\Gamma^A\}$ which transform as scalars or vectors under $\mathrm{Pin}(d)$. Concretely, the minimal set of basis elements is given by
\begin{equation}
    \mathcal{B} = \{\mathbb{1},\gamma_{d+2},\gamma_{i},\frac{\mi}{2}[\gamma_{d+1},\gamma_j],\mi\gamma_{d+2}\gamma_j\}\ \text{with }j=1,\dots,d+1 \,.
\end{equation}
The identity matrix and $\gamma_{d+1}$ are invariant under $\mathrm{Pin}(d)$, whereas $\gamma_{d+2}$ and $\gamma_{d+2}\gamma_{d+1}$ transform as axial scalars. Similarly for $i=1,\dots, d$ the Clifford generators $\gamma_i$ and $\gamma_{d+1}\gamma_i$ transforms as a vector, whereas $\gamma_{d+2}\gamma_i$ transforms as an axial vector under $\mathrm{Pin}(d)$. 

Although we retain all terms up to two spatial derivatives, we have dropped all basis elements $\Gamma^A$ with $r_A\geq 2$. For example, for any contraction with $r_A=2$ tensors of the form $\sim\mi[\gamma_i,\gamma_j]$ $\forall i< j=1,\dots,d$ vanishes, since $\partial_i\partial_j$ is symmetric in $i,j$. This generalizes straightforwardly to higher-order tensor structures.

As detailed in (iii) in Sec.~\ref{sec:symmetries}, the invariance of the action under chiral symmetry requires that $\gamma_{d+2}$ commutes with $D,H$ and anti-commutes with $P$, which yields the following expansion for $H,D$ and $P$
\begin{align}
    \begin{split}
        D(\vecq) &= d_0(\vecq)\mathbb{1} + d_{d+2}(\vecq)\gamma_{d+2} + \mi \sum_{i=1}^d d_i(\vecq)\gamma_{d+1}\gamma_i\, ,\\
        H(\vecq) &= h_0(\vecq)\mathbb{1} + h_{d+2}(\vecq)\gamma_{d+2} + \mi \sum_{i=1}^d h_i(\vecq)\gamma_{d+1}\gamma_i\, ,\\
        P(\vecq) &= \sum_{i=1}^d p_i(\vecq)\gamma_i + p_{d+1}(\vecq)\gamma_{d+1} + \mi \sum_{i=1}^{d+1}p'_i(\vecq)\gamma_{d+2}\gamma_i\,.
    \end{split}
\end{align}
where, the Hermitian property of the matrices imply that the coefficients are real functions of $\vecq$. We proceed to discuss the transformation behavior of the coefficient functions such that $H,D$, and $P$ satisfy Eq.~\eqref{eq:spinsgauss}. 

\paragraph{(Axial-) Scalar coefficients ---} Invariance under $\mathrm{Spin}(d)$ requires that the coefficients $d_0(\vecq),d_{d+2}(\vecq),h_0(\vecq),h_{d+2}(\vecq),p_{d+1}(\vecq)$ and $p'_{d+1}(\vecq)$ must transform trivially under rotations, \textit{i.e.}
\begin{align}
    \begin{split}
        &d_0(R^\top\vecq) \overset{!}{=} d_0(\vecq)\,,\ d_{d+2}(R^\top\vecq) \overset{!}{=} d_{d+2}(\vecq)\, ,\\
        &h_0(R^\top\vecq) \overset{!}{=} h_0(\vecq)\,,\ h_{d+2}(R^\top\vecq) \overset{!}{=} h_{d+2}(\vecq)\, ,\\
        &p_{d+1}(R^\top\vecq) \overset{!}{=} p_{d+1}(\vecq)\,,\ p'_{d+1}(R^\top\vecq) \overset{!}{=} p'_{d+1}(\vecq)\,,
    \end{split}
\end{align}
where the last line follows from the fact that the generators of the spin algebra $\mathcal{U}(\alpha)$(cf. Sec.~\ref{sec:symmetries}) commute with $\gamma_{d+1}$. 
Invariance under the full $\mathrm{Pin}(d)$ group requires that the $D,H$ and $P$ are invariant under spatial reflections generated by $\Pi_i$. Recall that
\begin{equation}
    \Pi_i^\dag \gamma_{d+1}\Pi_i = \gamma_{d+1}\,,\ \Pi_i^\dag \gamma_j\Pi_i = 
        \begin{cases}
            -\gamma_i & j=i \\
            \gamma_j & j\neq i
        \end{cases}.
\end{equation}
Hence, the full $\mathrm{Pin}(d)$ symmetry requires that $d_0(\vecq),h_0(\vecq)$ and $p_{d+1}(\vecq)$ are even functions of $\vecq$. Combined with rotational invariance, one can write $d_0,h_0,p_{d+1}$ in a gradient expansion 
\begin{align}
    \begin{split}
        d_0(\vecq) &= \Delta_d + K_d\vecq^2 + \hdots \\
        h_0(\vecq) &= \Delta_c + K_h\vecq^2 +\hdots \\
        p_{d+1}(\vecq) &= \Delta_p + K_p\vecq^2 +\hdots\,.
    \end{split}
\end{align}
However, $d_{d+2},h_{d+2},p'_{d+1}$ must transform as an axial scalar under spatial reflections 
\begin{align}
    \begin{split}
        d_{d+2}(\dots,-q_i\dots) \overset{!}{=} -d_{d+2}(\vecq)\,,\ h_{d+2}(\dots,-q_i\dots) \overset{!}{=} -h_{d+2}(\vecq)\,,\ p'_{d+1}(\dots,-q_i\dots) \overset{!}{=} -p'_{d+1}(\vecq)\ \forall i\,,
    \end{split}
\end{align}
which follows noting that $\Pi_i^\dag\gamma_{d+2}\Pi_i = -\gamma_{d+2}$. This condition forbids an additional chiral mass term $\sim d_{d+2}$ in the $D(\vecq)$ matrix. Combined with rotational invariance, this implies that the coefficients $d_{d+2}(\vecq)=h_{d+2}(\vecq)=p'_{d+1}(\vecq) =0$. 

\paragraph{(Axial-) Vector coefficients ---} Recall, that under the adjoint action of the $\mathrm{Spin}(d)$ group via the unitary $\mathcal{U}(\alpha)$, the Clifford generators
\begin{equation}
    \gamma_i\rightarrow \mathcal{U}(\alpha)^\dag\gamma_i\mathcal{U}(\alpha) = \sum_{j=1}^d\gamma_jR_{ji}\ \text{with } R \in \mathrm{SO}(d).
\end{equation}
This implies that the coefficients $p_i(\vecq),d_i(\vecq),h_i(\vecq)$ and $p_i'(\vecq)$, $i=1, ... ,d$ must transform as a vector under  rotations,\textit{i.e}
\begin{align}
\begin{split}
    d_i(R^\top\vecq) \overset{!}{=} R_{ij}d_j(\vecq)\,,\ h_i(R^\top\vecq) \overset{!}{=} R_{ij}h_j(\vecq)\,,\\
    p_i(R^\top\vecq) \overset{!}{=} R_{ij}p_j(\vecq)\,,\ p'_i(R^\top\vecq) \overset{!}{=} R_{ij}p'_j(\vecq)\,.
\end{split}
\end{align}
Furthermore, spatial reflection symmetry requires that $p_i(\vecq)$ is an odd function of $\vecq$ 
    \begin{align}
        \begin{split}
            p_i(\dots,-q_i\dots) &\overset{!}{=} -p_i(\vecq)\,,\ p_j(\dots,-q_j\dots) \overset{!}{=} p_j(\vecq)\ \forall j\neq i\,,        
        \end{split}
    \end{align}
similarly for $h_i(\vecq)$ and $d_i(\vecq)$. In contrast, $p'_i(\vecq)$ transforms as an axial vector
 \begin{equation}
   \begin{split}
            p'_i(\dots,-q_i\dots) &\overset{!}{=} p'_i(\vecq)\,,\ p'_j(\dots,-q_j\dots) \overset{!}{=} -p'_j(\vecq)\ \forall j\neq i\,.  
        \end{split}     
 \end{equation} 
 Hence, to leading order in gradient expansion, for $d>2$, $p_i'(\vecq)=0$ (in $d=2$ one can write down a term of the form $p_i'(\vecq) = \epsilon_{ij}q_j$, which transforms as an axial vector) must vanish and
\begin{align}
    \begin{split}
        &d_i(\vecq) \sim d_{d+1} q_i + \hdots\,,\ h_i(\vecq) \sim h_{d+1} q_i + \hdots\,,\ p_i(\vecq) \sim p_{d} q_i + \hdots\,,
    \end{split}    
\end{align}
with $p_d, h_{d+1}, d_{d+1}\in\mathbb{R}$. Summarizing, we find that the $D,H$ and $P$ matrices compatible with both chiral and weak Pin symmetry are of the form
\begin{align}
    \begin{split}
        D(\vecq) &= (\Delta_d + K_d\vecq^2)\mathbb{1} + \mi d_{d+1}\sum_i^d q_i\gamma_{d+1}\gamma_i\, ,\\
        H(\vecq) &= (\Delta_c + K_c\vecq^2)\mathbb{1} + \mi h_{d+1}\sum_i^d q_i\gamma_{d+1}\gamma_i\, \\
        P(\vecq) &= (\Delta_p + K_p\vecq^2)\gamma_{d+1} + p_d\sum_{i=1}^d q_i\gamma_i\,,
    \end{split}
\end{align}
as stated in the main text.
\section{One-loop flow equations for the effective action}\label{app:RG_eq}

In this appendix, we provide a detailed calculation of the 1-loop RG flow equations discussed in Sec.~\ref{sec:RG}. The bare Green's functions for
\begin{enumerate}
	\item \textbf{Fermions}
	\begin{align}
    \begin{split}
		\Ev{\psi_{c,s}(\omega,\vecq)\Bar{\psi}_{c,s'}(\omega',\vecq')} &= \mi g_{ss'}^K(\omega,\vecq;\omega',\vecq') = \frac{2\lambda(\vecq)}{\left(\omega-h(\vecq)\right)^2 + \lambda(\vecq)^2}(\tau_z)_{ss'}\delta(\omega - \omega')\delta(\vecq-\vecq'),  \\
		\Ev{\psi_{c,s}(\omega,\vecq)\Bar{\psi}_{q,s'}(\omega',\vecq')} &= \mi g_{ss'}^R(\omega,\vecq;\omega',\vecq') = \frac{\mi }{\omega - h(\vecq) + \mi\lambda(\vecq)}\delta_{ss'}\delta(\omega - \omega')\delta(\vecq-\vecq'), \\
		\Ev{\psi_{q,s}(\omega, \vecq)\Bar{\psi}_{c,s'}(\omega', \vecq')} &= \mi g_{ss'}^A(\omega,\vecq;\omega',\vecq') =\frac{\mi }{\omega - h(\vecq) - \mi\lambda(\vecq)}\delta_{ss'} \delta(\omega - \omega')\delta(\vecq-\vecq').
    \end{split}
	\end{align}
	\item \textbf{Bosons}
	\begin{align}
		\Ev{n_c(\omega,\vecq)n_c(\omega',\vecq')} &= \mi G^K(\omega,\vecq;\omega',\vecq') = \frac{2 TD q^2}{\omega^2 + D^2q^4}\delta(\omega + \omega')\delta(\vecq + \vecq')\,,
	\end{align}
\end{enumerate}
and their graphical representations are shown in Fig.~\ref{fig:propagators}, while the relevant Yukawa vertices are depicted in Fig.~\ref{fig:Yukawa-vertices}.
\begin{figure}
    \centering
     \input{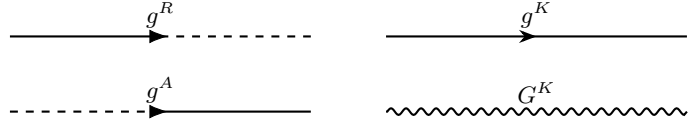}
    \caption{Diagrammatic representation of the bare fermion and boson propagators.}
    \label{fig:propagators}
\end{figure}

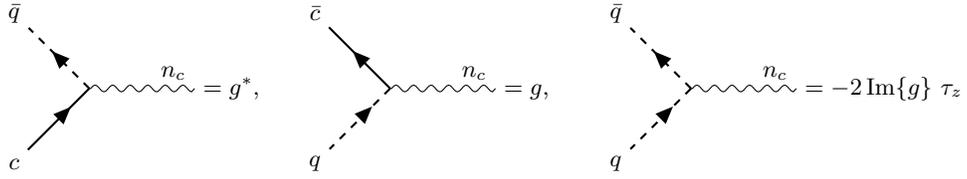
\begin{figure}
    \centering
     \begin{center}
	\begin{tikzpicture}
    \def\y{4}
		\centering
		\begin{feynman}
			\vertex (a1) at (0,0);
			\vertex (b1) at (1.4,0);
			\vertex (c1) at (-1,1) {$\bar{q}$};
            \vertex (d1) at (-1,-1) {$c$};
			\diagram*{
				(a1) -- [photon] (b1),
				(a1)-- [charged scalar, thick] (c1),
                (a1)-- [anti fermion,thick] (d1)
			};
			\end{feynman}
            \node[right] at (b1) {$= g^*,$};
            \node[above left] at (b1) {$n_c$};
		\begin{feynman}
			\vertex (a2) at (\y,0);
			\vertex (b2) at (1.4+\y,0);
			\vertex (c2) at (-1+\y,1) {$\bar{c}$};
            \vertex (d2) at (-1+\y,-1) {$q$};
			\diagram*{
				(a2) -- [photon] (b2),
				(a2)-- [fermion, thick] (c2),
                (a2)-- [anti charged scalar,thick] (d2)
			};
			\end{feynman}
            \node[right] at ($(b2)-(0,0.05)$) {$= g,$};
            \node[above left] at (b2) {$n_c$};
            \begin{feynman}
			\vertex (a3) at (0+2*\y,0);
			\vertex (b3) at (1.4+2*\y,0);
			\vertex (c3) at (-1+2*\y,1) {$\bar{q}$};
            \vertex (d3) at (-1+2*\y,-1) {$q$};
			\diagram*{
				(a3) -- [photon] (b3),
				(a3)-- [charged scalar, thick] (c3),
                (a3)-- [anti charged scalar,thick] (d3)
			};
			\end{feynman}
            \node[right] at (b3) {$= -2 \Im{g}\ \tau_z$};
            \node[above left] at (b3) {$n_c$};
	\end{tikzpicture}
\end{center}
    \caption{Feynman rules for the Yukawa vertices.}
    \label{fig:Yukawa-vertices}
\end{figure}

\subsection{Structure of one-loop corrections} 
 Define the supervector $\xi = (n_c, \theta_q,\psi_c,\psi_q)^\top,\Bar{\xi} = (n_c, \theta_q,\Bar{\psi}_c,\Bar{\psi}_q)^\top$. The generating function of one-loop diagrams is given by
 \begin{align*}
        \Gamma_{1-\text{loop}}&= \mi \Str\ln(\mi S^{(2)}[\Bar{\psi},\psi,n_c,\theta_q])= \mi \Str\ln(\mi G^{-1}_0 + \mi X[\Bar{\psi},\psi,n_c]) \\
        &\approx \mi \Str\ln(\mi G_0^{-1})+\Str(\mi  G_0X) +\frac{\mi }{2}\Str(\mi G_0X\mi G_0X) -\frac{1}{3}\Str(\mi G_0X\mi G_0X\mi G_0X) + \hdots\,,
    \end{align*}
where $S^{(2)}$ is the second variation of the action with respect to the supervector fields $\xi,\Bar{\xi}$. The supertrace $\Str$ is defined as
\begin{equation}
    \Str(X) = \Tr(X_{BB}) - \Tr(X_{FF})\,,
\end{equation}
where $BB(FF)$ refers to the boson-boson (fermion-fermion) block, respectively. The traces $\Tr$ are over Keldysh, orbital, and spacetime indices. The inverse bare Green's function $G^{-1}_0$ in the notation above has the block structure:
\begin{align}
    G^{-1}_0 = \mqty(\dmat{G^{-1}_{B},G^{-1}_{F}})\,,
\end{align}
where the definitions of $G^{-1}_{B}$ corresponds to the boson sector, and $G^{-1}_{F}$ corresponds to the fermion sector as given given in action Eq.~\eqref{eq:effective_action} and Eq.~\eqref{eq:fermion_greens_fct}. The fluctuation matrix $X[\Bar{\psi},\psi,n_c]$ has the structure
\begin{equation}
    X[\bar\psi,\psi,n_c] = \mqty( X_{BB} & X_{BF}\\X_{FB} & X_{FF})\,,
\end{equation}
where the matrices $X_{\alpha\beta}$ are given by
\begin{align}
    X_{BB} =\mathbf{0}\,,\quad X_{FF} = A(g,g_d) n_c\,,\quad X_{FB} = \mqty(A(g,g_d)\psi & 0)\,,\quad X_{BF} = \mqty(\bar{\psi}^\top A(g,g_d) \\ 0)\,,
\end{align}
where
\begin{equation}
    A(g,g_d) =\mqty(0 & g\mathbb{1}\\ g^*\mathbb{1} & 2\mi g_d\tau_z).
\end{equation}
We proceed with the computation of 1-loop corrections to the self-energy and the Yukawa couplings.
\subsection{Self-energy corrections}
At one-loop order, the retarded and the noise vertex functions are given by (see Fig.~\ref{fig:self-energy}) 

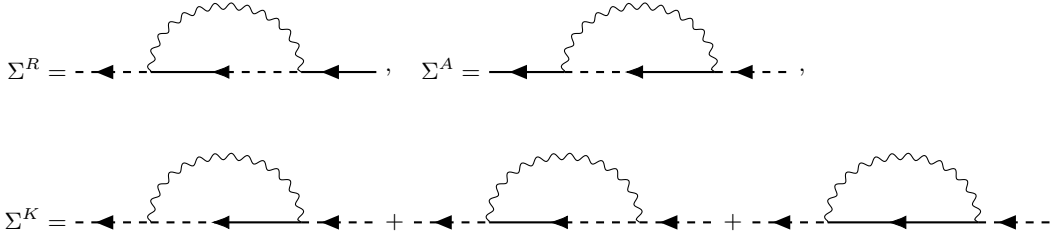
\begin{figure}
    \centering
     \begin{tikzpicture}
\def\y{4.5}
\def\z{2}
		\centering
		\begin{feynman}
			\vertex (a4) at (0,0);
			\vertex (b4) at (1,0);
			\vertex (c4) at (3,0);
            \vertex (d4) at (4,0);
			\diagram*{
				(a4) -- [anti charged scalar,thick] (b4),
				(c4)-- [anti fermion, thick] (d4),
                (b4)-- [photon,,half left] (c4)
			};
			\end{feynman}
            
            \coordinate (m4) at ($(b4)!0.5!(c4)$);
            \draw[thick] (b4) -- (m4);
            \draw[thick, dashed] (m4) -- (c4);
            \draw [arrows={-Latex[length=7pt]}] ($(m4)+(0,0)$) -- ($(m4)+(-0.2,0)$);
            \node[left] at ($(a4)+(0,0.05)$) {$\Sigma^R = $};
            \node[right] at (d4) {,};

		\begin{feynman}
			\vertex (a5) at (1+\y,0);
			\vertex (b5) at (2+\y,0);
			\vertex (c5) at (4+\y,0);
            \vertex (d5) at (5+\y,0);
			\diagram*{
				(a5) -- [anti fermion,thick] (b5),
				(c5)-- [anti charged scalar, thick] (d5),
                (b5)-- [photon,half left] (c5)
			};
			\end{feynman}
            
            \coordinate (m5) at ($(b5)!0.5!(c5)$);
            \draw[thick,dashed] (b5) -- (m5);
            \draw[thick] (m5) -- (c5);
            \draw [arrows={-Latex[length=7pt]}] ($(m5)+(0,0)$) -- ($(m5)+(-0.2,0)$);
            
            \node[left] at ($(a5)+(0,0.05)$) {$\Sigma^A = $};
            \node[right] at (d5) {,};

		\begin{feynman}
			\vertex (a1) at (0,-\z);
			\vertex (b1) at (1,-\z);
			\vertex (c1) at (3,-\z);
            \vertex (d1) at (4,-\z);
			\diagram*{
				(a1) -- [anti charged scalar,thick] (b1),
				(c1)-- [anti charged scalar, thick] (d1),
                (b1)-- [photon,,half left] (c1)
			};
		\end{feynman}
           
            \coordinate (m1) at ($(b1)!0.5!(c1)$);
            \draw[thick,dashed] (b1) -- (m1);
            \draw[thick] (m1) -- (c1);
            \draw [arrows={-Latex[length=7pt]}] ($(m1)+(0,0)$) -- ($(m1)+(-0.2,0)$);
            
            \node[left] at ($(a1)+(0,0.05)$) {$ \Sigma^K = $};
            \node[right] at (d1) {$+$};

		\begin{feynman}
			\vertex (a2) at (0+\y,-\z);
			\vertex (b2) at (1+\y,-\z);
			\vertex (c2) at (3+\y,-\z);
            \vertex (d2) at (4+\y,-\z);
			\diagram*{
				(a2) -- [anti charged scalar,thick] (b2),
				(c2)-- [anti charged scalar, thick] (d2),
                (b2)-- [photon,,half left] (c2)
			};
		\end{feynman}
           
            \coordinate (m2) at ($(b2)!0.5!(c2)$);
            \draw[thick] (b2) -- (m2);
            \draw[thick,dashed] (m2) -- (c2);
            \draw [arrows={-Latex[length=7pt]}] ($(m2)+(0,0)$) -- ($(m2)+(-0.2,0)$);
            
            \node[right] at (d2) {$+$};
            
        \begin{feynman}
			\vertex (a3) at (0+2*\y,-\z);
			\vertex (b3) at (1+2*\y,-\z);
			\vertex (c3) at (3+2*\y,-\z);
            \vertex (d3) at (4+2*\y,-\z);
			\diagram*{
				(a3) -- [anti charged scalar,thick] (b3),
				(c3)-- [anti charged scalar, thick] (d3),
                (b3)-- [photon,,half left] (c3)
			};
		\end{feynman}
           
            \coordinate (m3) at ($(b3)!0.5!(c3)$);
            \draw[thick] (b3) -- (m3);
            \draw[thick] (m3) -- (c3);
            \draw [arrows={-Latex[length=7pt]}] ($(m3)+(0,0)$) -- ($(m3)+(-0.2,0)$);
            
\end{tikzpicture}
    \caption{The retarded, advanced, and Keldysh contributions to the self-energy.}
    \label{fig:self-energy}
\end{figure}

 \begin{align*}
	   \Gamma_{\bar{\psi}_q\psi_c}(\omega,\vecp) =&\omega-h(\vecp)+\mi\lambda(\vecp) + \mi {g^*}^2\mathbb{1}\int_{\nu,q} \mi g^R(\nu,\vecq) \mi G^K(\omega-\nu,\vecp-\vecq)+ \mathcal{O}(2 \text{ loops}),\\
        \Gamma_{\bar{\psi}_c\psi_q}(\omega,\vecp) =&\omega-h(\vecp)-\mi \lambda(\vecp) + \mi g^2\mathbb{1}\int_{\nu,q} \mi g^A(\nu,\vecq) \mi G^K(\omega-\nu,\vecp-\vecq)+ \mathcal{O}(2 \text{ loops}),\\
	   \Gamma_{\bar{\psi}_q\psi_q}(\omega,\vecp) =&2\mi \lambda(\vecp)\tau_z + \mi \left[|g|^2\int_{\nu, q} \mi g^K(\nu,\vecq) \mi G^K(\omega-\nu,\vecp-\vecq) + 2\mi g_d\tau_z\left({g^*} \mi g^R(\nu,\vecq)+ g \mi g^A(\nu,\vecq)\right)\mi G^K(\omega-\nu,\vecp-\vecq) \right] \\
       &+ \mathcal{O}(2 \text{ loops}).
    \end{align*}
 
Evaluation of the frequency integrals yields
\begin{align*}
    \Gamma_{\bar{\psi}_c\psi_c}(\omega,\vecp) &=0,\\
	\Gamma_{\bar{\psi}_q\psi_c}(\omega,\vecp) &=\left(\omega-h(\vecp)+\mi \lambda(\vecp)\right)\mathbb{1} - T{g^*}^2\int\frac{\dd^d q}{(2\pi)^d}\ \frac{1}{\omega-h(\vecq)+\mi \left(\lambda(\vecq)+D\norm{\vecq-\vecp}^2\right)}\mathbb{1},\\
	\Gamma_{\bar{\psi}_c\psi_q}(\omega,\vecp) &=\left(\omega-h(\vecp)-\mi \lambda(\vecp)\right)\mathbb{1} - Tg^2\int\frac{\dd^d q}{(2\pi)^d}\ \frac{1}{\omega-h(\vecq)-\mi\left(\lambda(\vecq)+D\norm{\vecq-\vecp}^2\right)}\mathbb{1},\\
	\Gamma_{\bar{\psi}_q\psi_q}(\omega,\vecp) &= 2\mi \lambda(\vecp)\tau_z - 2\mi T\Im\left[(2\mi g_d g^*+|g|^2)\int\frac{\dd^d q}{(2\pi)^d}\ \frac{1}{\omega-h(\vecq)+\mi \left(\lambda(\vecq)+D\norm{\vecq-\vecp}^2\right)} \right]\tau_z.
\end{align*}
 
It is straightforward to verify that for $g_d=-\Im(g)$ (cf. Eq.~\eqref{eq:Yukawa_action} constrained by FDS), 
renders the vertex functions for the retarded and the noise equal (up to an overall $2\mi $ pre-factor) at one-loop order,  \textit{i.e.}
\begin{align}
2\mi \Im\Gamma_{\bar{\psi}_q\psi_c}(\omega,\vecp) =\tau_z\Gamma_{\bar{\psi}_q\psi_q}(\omega,\vecp)\eval_{g_d=-\Im g}.
\end{align}
Moreover, both the causality structure and probability conservation of the Keldysh action are preserved since $\Gamma_{\bar{\psi}_0\psi_c}(\omega,\vecp)=0$ and  $\Gamma_{\bar{\psi}_q\psi_c}(\omega,\vecp) = \Gamma^\dag_{\bar{\psi}_c\psi_q}(\omega,\vecp)$.\\ 
To evaluate the momentum integrals, we use the standard prescription of momentum shell integration according to Wilsonian RG. This is done by first extracting the leading order UV singular contributions near $d_c=4$, followed by computing the momentum integrals over the shell $\Lambda/b<|\vecq|<\Lambda$, where $b$ is the scaling parameter chosen such that $b>1$. 

\paragraph{Spectral and noise gap ---}
We first note the additive shift of the bare dissipative and noise gap due to fluctuations 
	\begin{align*}
	\Im\Gamma_{\bar{\psi}_q\psi_c}(0,0) =\frac{1}{2\mi }\Gamma_{\bar{\psi}_q\psi_q}(0,0)\eval_{g_d=-\Im g}&=\Delta_d+T\frac{\Omega_d}{(2\pi)^d}\Im\left[{g^*}^2\int_{\Lambda/b}^\Lambda \dd q\ \frac{q^{d-1}}{q^2}\left(\frac{1}{\overline{K}_\text{eff}}\right) + \mathcal{O}(g^4)\right],
	\end{align*} 
where we have defined $K_\text{eff} = K_c + \mi (K_d+D)$ and $\Omega_d/(2\pi)^d=(2^{d-1}\pi^{d/2}/\Gamma(d/2))^{-1}$. The loop integral shown above has a quadratic UV divergence near $d=4$ and is independent of bare spectral mass $\Delta_d$, which effectively shifts the bare value of the critical point at $\Delta_d=0$. The quadratic divergence is removed by
introducing a counter term $\Delta_{T}$ in the bare action and defining the distance from the critical point $\Tilde{\Delta}=\Delta-\Delta_T$, which vanishes at the critical point. After subtracting the quadratic divergences, the leading order UV singular part near $d=4$, for the mass in the R/A sectors is given by 
\begin{align}\label{eq:mass_corrections}
-\Gamma_{\bar{\psi}_c\psi_q}(0,0) = \delta\left[1+T\frac{\Omega_d}{(2\pi)^d}\left(g^2\left(\frac{1}{K_\text{eff}}\right)^2\int_{\Lambda/b}^\Lambda\dd q \frac{q^{d-1}}{q^4}\right)+\mathcal{O}(g^4)\right],\quad \delta = \Delta_c+ \mi \Delta_d.
\end{align}
 
\paragraph{Kinetic terms and wavefunction renormalization ---}
We also encounter finite momentum and frequency corrections at one-loop order, which provide a renormalization of the kinetic coefficient and the wave-function renormalization. The latter can be obtained by computing the frequency derivative of the vertex functions at zero external momenta
 
\begin{align}
\begin{split}
\partial_\omega \Gamma_{\bar{\psi}_c\psi_q}(\omega,0)\eval_{\omega=0}&=1+T\frac{\Omega_d}{(2\pi)^d} g^2\left(\frac{1}{K_\text{eff}}\right)^2\int_{\Lambda/b}^\Lambda\dd q \frac{q^{d-1}}{q^4} + \mathcal{O}(g^4) ,\label{eq:wavefunction_renorm}\\
\frac{1}{2\mi }\partial_\omega \Gamma_{\bar{\psi}_q\psi_q}(\omega,0)\eval_{\omega=0}&=-\Im\left[T\frac{\Omega_d}{(2\pi)^d}g^2\left(\frac{1}{K_\text{eff}}\right)^2\int_{\Lambda/b}^\Lambda\dd q \frac{q^{d-1}}{q^4} + \mathcal{O}(g^4)\right].
\end{split}
\end{align}
 Due to the complex-valued nature of the couplings, the one-loop correction to the wavefunction renormalization is complex. As a consequence of the locking of the vertex functions for $g_d=-\Im g$, this leads to a non-zero frequency dependence of the Keldysh self-energies, which was not present at the level of the bare theory. For the momentum corrections, we first note that due to the symmetry of the loop integrals, all odd powers of $p$ vanish. Hence, the leading order correction to the fermion diffusion term is given by 
 \begin{align} \label{eq:kinetic_renormalization}
    -\partial^2_p\Gamma_{\bar{\psi}_c\psi_q}(0,0)&=K_c+\mi K_d+Tg^2\partial_p^2\eval_{p=0}\left[ \int \frac{\dd^d q}{(2\pi)^d}\frac{1}{-h(\mathbf{q})-\mi\left(\lambda(\vecq)+ D\norm{\mathbf{q}-\mathbf{p}}^2\right)}\right]+ \mathcal{O}(g^4)\notag \\
    &=K+Tg^2\frac{\Omega_d}{(2\pi)^d}\left[2\mi D -\frac{2}{d}(2\mi D)^2\left(\frac{1}{K_\text{eff}}\right)\right]\left(\frac{1}{K_\text{eff}}\right)^2\int_q\dd q \frac{q^{d-1}}{q^4}\,.
\end{align} 
The last line is obtained by expressing $\mathbf{p}=p\hat{n}$ and using
\begin{align}
  \int_q (\mathbf{q}\hat{n})^2g(\norm{\vecq}^2) = \frac{1}{d}\frac{\Omega_d}{(2\pi)^d}\int \dd q\  q^{d+1} g(q^2) .
\end{align}
\subsection{Corrections to the Yukawa vertex}

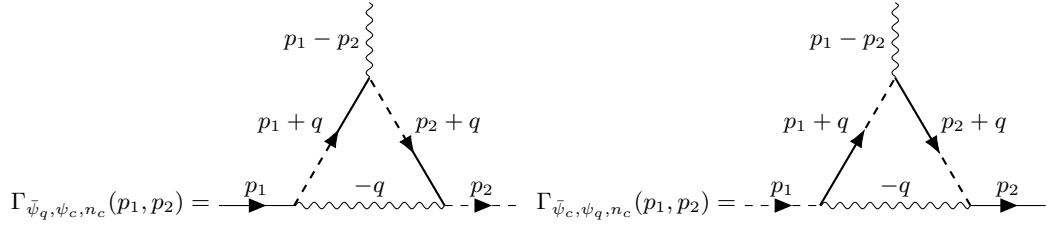
\begin{figure}
    \centering
     \begin{center}
	\begin{tikzpicture}
		\centering
		\begin{feynman} 
			\vertex (a) at (0,0);
			\vertex (b) at (2,0);
			\vertex (c) at (1,1.7);
            \vertex (f1) at (-1,0);
			\vertex (f2) at (3,0);
			\vertex (f3) at (1,2.7);
			\diagram*{
				(a) -- [boson,edge label=$-q$] (b),
				(f1) -- [ fermion,edge label=$p_1$] (a), (c) -- [boson,edge label=$p_1-p_2$] (f3), (b) -- [charged scalar,edge label=$p_2$] (f2)
			};
			\end{feynman}

            \coordinate (m) at ($(b)!0.5!(c)$);
            \draw[thick] (b) -- (m);
            \draw[thick,dashed] (m) -- (c);
            \draw [arrows={-Latex[length=7pt]}] ($(b)!0.5!(c)$)--($(b)!0.4!(c)$);

            \coordinate (m2) at ($(a)!0.5!(c)$);
            \draw[thick,dashed] (a) -- (m2);
            \draw[thick] (m2) -- (c);
            \draw [arrows={-Latex[length=7pt]}] ($(a)!0.5!(c)$)--($(a)!0.6!(c)$);

            \node[above right] at (m) {$p_2+q$};
            \node[above left] at (m2) {$p_1+q$};
            \node[left] at (f1) {$\Gamma_{\bar{\psi}_q, \psi_c, n_c}(p_1,p_2)=$};
	\end{tikzpicture}
    \begin{tikzpicture}
        \begin{feynman}
			\vertex (a) at (0,0);
			\vertex (b) at (2,0);
			\vertex (c) at (1,1.7);
            \vertex (f1) at (-1,0);
			\vertex (f2) at (3,0);
			\vertex (f3) at (1,2.7);
			\diagram*{
				(a) -- [boson,edge label=$-q$] (b),
				(f1) -- [charged scalar,edge label=$p_1$] (a), (c) -- [boson,edge label=$p_1-p_2$] (f3), (b) -- [fermion,edge label=$p_2$] (f2)
			};
			\end{feynman}

            \coordinate (m) at ($(b)!0.5!(c)$);
            \draw[thick,dashed] (b) -- (m);
            \draw[thick] (m) -- (c);
            \draw [arrows={-Latex[length=7pt]}] ($(b)!0.5!(c)$)--($(b)!0.4!(c)$);

            \coordinate (m2) at ($(a)!0.5!(c)$);
            \draw[thick] (a) -- (m2);
            \draw[thick,dashed] (m2) -- (c);
            \draw [arrows={-Latex[length=7pt]}] ($(a)!0.5!(c)$)--($(a)!0.6!(c)$);

            \node[above right] at (m) {$p_2+q$};
            \node[above left] at (m2) {$p_1+q$};
            \node[left] at (f1) {$\Gamma_{\bar{\psi}_c, \psi_q, n_c}(p_1,p_2)=$};
    \end{tikzpicture}
\end{center}
    \caption{Triangle diagram contributions to the Yukawa vertex renormalization in the retarded and advanced sectors.}
    \label{fig:RA-triangles}
\end{figure}

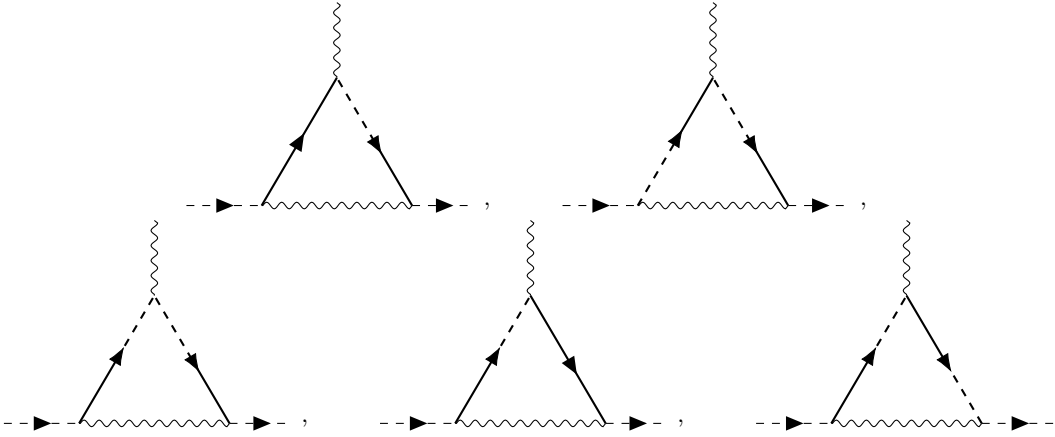
\begin{figure*}
    \centering
     \begin{center}
	\begin{tikzpicture}
		\centering
		\begin{feynman} 
			\vertex (a) at (0,0);
			\vertex (b) at (2,0);
			\vertex (c) at (1,1.7);
            \vertex (f1) at (-1,0);
			\vertex (f2) at (3,0) {,};
			\vertex (f3) at (1,2.7);
			\diagram*{
				(a) -- [boson] (b),
				(a)-- [fermion, thick] (c),
				(f1) -- [charged scalar] (a), (c) -- [boson] (f3), (b) -- [charged scalar] (f2)
			};
			\end{feynman}

            \coordinate (m) at ($(b)!0.5!(c)$);
            \draw[thick] (b) -- (m);
            \draw[thick,dashed] (m) -- (c);
            \draw [arrows={-Latex[length=7pt]}] ($(b)!0.5!(c)$)--($(b)!0.4!(c)$);

            \begin{feynman} 
			\vertex (a) at (5,0);
			\vertex (b) at (7,0);
			\vertex (c) at (6,1.7);
            \vertex (f1) at (4,0);
			\vertex (f2) at (8,0) {,};
			\vertex (f3) at (6,2.7);
			\diagram*{
				(a) -- [boson] (b),
				(f1) -- [charged scalar] (a), (c) -- [boson] (f3), (b) -- [charged scalar] (f2)
			};
			\end{feynman}

            \coordinate (m) at ($(b)!0.5!(c)$);
            \draw[thick] (b) -- (m);
            \draw[thick,dashed] (m) -- (c);
            \draw [arrows={-Latex[length=7pt]}] ($(b)!0.5!(c)$)--($(b)!0.4!(c)$);

            \coordinate (m2) at ($(a)!0.5!(c)$);
            \draw[thick,dashed] (a) -- (m2);
            \draw[thick] (m2) -- (c);
            \draw [arrows={-Latex[length=7pt]}] ($(a)!0.5!(c)$)--($(a)!0.6!(c)$);
	\end{tikzpicture}
    \begin{tikzpicture}
    \centering
        \begin{feynman} 
			\vertex (a) at (0,0);
			\vertex (b) at (2,0);
			\vertex (c) at (1,1.7);
            \vertex (f1) at (-1,0);
			\vertex (f2) at (3,0) {,};
			\vertex (f3) at (1,2.7);
			\diagram*{
				(a) -- [boson] (b),
				(f1) -- [charged scalar] (a), (c) -- [boson] (f3), (b) -- [charged scalar] (f2)
			};
			\end{feynman}

            \coordinate (m) at ($(b)!0.5!(c)$);
            \draw[thick] (b) -- (m);
            \draw[thick,dashed] (m) -- (c);
            \draw [arrows={-Latex[length=7pt]}] ($(b)!0.5!(c)$)--($(b)!0.4!(c)$);

            \coordinate (m2) at ($(a)!0.5!(c)$);
            \draw[thick] (a) -- (m2);
            \draw[thick,dashed] (m2) -- (c);
            \draw [arrows={-Latex[length=7pt]}] ($(a)!0.5!(c)$)--($(a)!0.6!(c)$);

            \begin{feynman} 
			\vertex (a) at (5,0);
			\vertex (b) at (7,0);
			\vertex (c) at (6,1.7);
            \vertex (f1) at (4,0);
			\vertex (f2) at (8,0) {,};
			\vertex (f3) at (6,2.7);
			\diagram*{
				(a) -- [boson] (b) -- [anti fermion,thick] (c),
				(f1) -- [charged scalar] (a), (c) -- [boson] (f3), (b) -- [charged scalar] (f2)
			};
			\end{feynman}

            \coordinate (m2) at ($(a)!0.5!(c)$);
            \draw[thick] (a) -- (m2);
            \draw[thick,dashed] (m2) -- (c);
            \draw [arrows={-Latex[length=7pt]}] ($(a)!0.5!(c)$)--($(a)!0.6!(c)$);

            \begin{feynman} 
			\vertex (a) at (10,0);
			\vertex (b) at (12,0);
			\vertex (c) at (11,1.7);
            \vertex (f1) at (9,0);
			\vertex (f2) at (13,0);
			\vertex (f3) at (11,2.7);
			\diagram*{
				(a) -- [boson] (b),
				(f1) -- [charged scalar] (a), (c) -- [boson] (f3), (b) -- [charged scalar] (f2)
			};
			\end{feynman}

            \coordinate (m) at ($(b)!0.5!(c)$);
            \draw[thick,dashed] (b) -- (m);
            \draw[thick] (m) -- (c);
            \draw [arrows={-Latex[length=7pt]}] ($(b)!0.5!(c)$)--($(b)!0.4!(c)$);

            \coordinate (m2) at ($(a)!0.5!(c)$);
            \draw[thick] (a) -- (m2);
            \draw[thick,dashed] (m2) -- (c);
            \draw [arrows={-Latex[length=7pt]}] ($(a)!0.5!(c)$)--($(a)!0.6!(c)$);
    \end{tikzpicture}
\end{center}
    \caption{Triangle diagram contributions to the Yukawa vertex renormalization in the Keldysh sector.}
    \label{fig:K-triangles}
\end{figure*}

The relevant diagrams for the Yukawa vertex (see Fig.~\ref{fig:Yukawa-vertices}) renormalization are shown in Figs.~\ref{fig:RA-triangles} and~\ref{fig:K-triangles} which correspond to the following integrals,
 
{\small
\begin{align}
		\Gamma_{\bar{\psi}_c \psi_c n_c}(0,0) =& 0\,,\\
        \begin{split}
		\Gamma_{\bar{\psi}_q\psi_cn_c}(0,0) =& g^* \mathbb{1} - {g^*}^3 \int_{\nu,\vecq} (\mi g^R(\nu,\vecq)^2\mi G^K(-\nu,-\vecq) + \mathcal{O}(g^5) \\=& g^* \left[ 1 + {g^*}^2T\frac{\Omega_d}{(2\pi)^d}\left(\frac{1}{\overline{K}_\text{eff}}\right)^2\int\dd q \frac{q^{d-1}}{q^4} +\mathcal{O}(g^4)\right]\mathbb{1}\,,
        \label{eq:YukawaA_correction}
        \end{split}\\
		\Gamma_{\bar{\psi}_c \psi_q n_c}(0,0) =& g\mathbb{1} - g^3 \int_{\nu,\vecq} (\mi g^A(\nu,\vecq))^2\mi G^K(-\nu,-\vecq) = g\left[ 1 + g^2T\frac{\Omega_d}{(2\pi)^d}\left(\frac{1}{K_\text{eff}}\right)^2\int\dd q \frac{q^{d-1}}{q^4}+\mathcal{O}(g^4) \right]\mathbb{1} \label{eq:YukawaR_correction}\,,\\
        \begin{split}
            \Gamma_{\bar{\psi}_q \psi_q n_c}(0,0) =& 2\mi g_d\tau_z - 2\int_{\nu,q}\Bigg[\frac{1}{2\mi }|g|^2g \mi g^K(\nu,\vecq)\mi g^A(\nu,\vecq)+\frac{1}{2\mi }|g|^2 {g^*} \mi g^K(\nu,\vecq)\mi g^R(\nu,\vecq) +\\
	       	&+g_d\tau_z \Big({g^*}^2 \mi g^R(\nu,\vecq)^2 + g^2 \mi g^A(\nu,\vecq)^2 + |g|^2\mi g^R(\nu,\vecq)\mi g^A(\nu,\vecq)\Big)\Bigg]\mi G^K(-\nu,-\vecq)\\
            =&2\mi g_d\tau_z - 2\mi \Re\left[\int_{\nu,q}\mi g^R(\nu,\vecq)\left(\mi g^R(\nu,\vecq)(2g_d-\mi g){g^*}^2 + \mi g^A(\nu,\vecq)(g_d+\Im g)|g|^2\right)\mi G^K(-\nu,-\vecq) \right]\tau_z\,.
        \end{split}
\end{align}}
 
The first line ensures that the loop corrections do not spoil the probability conservation of the action. Additionally, we note that the loop correction to $g_d$ is anti-Hermitian; thus, the one-loop correction also preserves the Hermiticity symmetry of the action, Eq.~\eqref{eq:hermiticity}. 
Moreover, it is straightforward to verify that setting $g_d=-\Im g$, the vertex function in the last line gives the same result as the imaginary part of $\Gamma_{\bar{\psi}_q\psi_cn_c}(0,0,0)$. This implies the locking of the vertex functions at the 1-loop level
\begin{align}
    \Im\Gamma_{\bar{\psi}_q\psi_cn_c}(\omega_1,\vecp_1;\omega_2,\vecp_2) = \frac{1}{2\mi } \Gamma_{\bar{\psi}_q \psi_q n_c}(\omega_1,\vecp_1;\omega_2,\vecp_2)\tau_z\,,
\end{align}
which is a consequence of the FDS discussed in the main text.
\end{widetext}
\bibliography{bibliography}
\end{document}


\begin{tikzpicture}
\def\pt{(1,0)}
\tikzset{snake it/.style={decorate, decoration=snake}}
    \begin{feynman}
        \foreach \x in {0,1,3,5,6}{
            \vertex (a\x) at (\x,0);
            \vertex (b\x) at (\x,1.2);
            \vertex (d\x) at (\x,1.2);
        }
			\diagram*{
				(a0) -- [fermion ,thick] (a1),
				(a1)-- [ fermion, thick] (a3),
                (a3)-- [ fermion, thick] (a5),
                (a5)-- [ fermion, thick] (a6),
                (a1)-- [scalar, thick] (d1),
                (a3)-- [scalar, thick] (d3),
                (a5)-- [scalar, thick] (d5),
                (d1) --[charged boson, thick] (d3),
                (d3) --[charged boson, thick] (d5),
			};
    \end{feynman}
    \foreach \x in {1,3,5}{
        \filldraw[black] (a\x) circle (1pt) node[anchor=north,black, node font =\large] {$g$};
        \filldraw[black] (b\x) circle (1pt);
    };
    \draw[snake it, thick,,decoration={amplitude=1.3,segment length=7},radius=2.8] (d5) arc[start angle=45,end angle=120] -- (d1);

    \draw[thick,-{Latex[length=2.5mm]}] (3,2.08) -- (2.8,2.08) ;

    \path (7,1) node[black, node font=\Large] {$\sim \frac{g^3}{N\sqrt{NM}}$};

\end{tikzpicture}